\pdfoutput=1
\documentclass[cernpreprint ,atlasdraft=false, UKenglish, texmf, orcidlogo]{atlasdoc}
\usepackage{amsmath}
\usepackage{atlaspackage}
\usepackage{atlasbiblatex}

\usepackage{atlasphysics}
\graphicspath{{logos/}{figures/}}

\usepackage{ANA-STDM-2018-41-PAPER-defs}

\AtlasTitle{Measurement of
double-differential charged-current Drell--Yan cross-sections at high transverse masses in $pp$ collisions at
$\sqrt{s} = 13$ TeV with the ATLAS detector}

\AtlasAbstract{
This paper presents a first measurement of the cross-section for the
charged-current Drell--Yan process $pp\rightarrow
W^{\pm} \rightarrow \ell^{\pm} \nu$ above the resonance region, where
$\ell$ is an electron or muon. The measurement is performed for
transverse masses, $m_{\text{T}}^{\text{W}}$, between $200$ GeV and
$5000$ GeV, using a sample of $140$~fb$^{-1}$ of $pp$ collision data
at a centre-of-mass energy of $\sqrt{s}$ = 13 TeV collected by the
ATLAS detector at the LHC during 2015--2018.  The data are presented
single differentially in transverse mass and double differentially in
transverse mass and absolute lepton pseudorapidity.  A test of lepton
flavour universality shows no significant deviations from the Standard
Model.  The electron and muon channel measurements are combined to
achieve a total experimental precision of 3\% at low
$m_{\text{T}}^{\text{W}}$.  The single- and double differential
$W$-boson charge asymmetries are evaluated from the measurements.  A
comparison to next-to-next-to-leading-order perturbative QCD
predictions using several recent parton distribution functions and
including next-to-leading-order electroweak effects indicates the
potential of the data to constrain parton distribution functions. The
data are also used to constrain four fermion operators in the Standard
Model Effective Field Theory formalism, in particular the lepton-quark
operator Wilson coefficient $c_{\ell q}^{(3)}$.}

\AtlasRefCode{STDM-2018-41}

\PreprintIdNumber{CERN-EP-2025-032}

\AtlasJournalRef{\JHEP 07 (2025) 026}
\AtlasDOI{10.1007/JHEP07(2025)026}

\AtlasCoverEgroupAnalysisTeam{atlas-stdm-2018-41-analysis-team@cern.ch}

\hypersetup{pdftitle={ATLAS document},pdfauthor={The ATLAS Collaboration}}

\begin{document}

\maketitle

\tableofcontents

\newcommand \figurepath{figures}

\newpage
\section{Introduction}
%

%
\label{sec:intro}

The Drell--Yan process~\cite{DrellYan} of lepton pair production
in hadronic interactions is a powerful tool in understanding the
nature of partonic interactions and hadronic structure, and it can be a probe of
the electroweak sector of the Standard Model.
It has been fundamental in developing perturbative quantum chromodynamics (QCD) with calculations
now available at next-to-next-to-next-to-leading-order (N$^3$LO)
accuracy~\cite{Duhr:2020sdp, Duhr:2021vwj, Chen:2022lwc,Campbell:2023lcy}.

Cross-section measurements from the Large Hadron Collider (LHC) of inclusive neutral-current
($pp \rightarrow Z/\gamma^* \rightarrow \ell^+\ell^-$)
and charged-current ( $pp \rightarrow W^\pm \rightarrow \ell^\pm \nu$) Drell--Yan processes
at centre-of-mass energies of
$\sqrt{s}=5.02$~\TeV, $13$~\TeV\ and at 13.6~\TeV\ have recently been published by the
ATLAS~\cite{STDM-2018-17, STDM-2023-16},
CMS~\cite{cmscollaboration2024measurementinclusivecrosssections,CMS-SMP-18-012,CMS-SMP-17-010}
and
LHCb~\cite{LHCbZ13TeV}   %
collaborations. These measurements, performed at the resonant peaks
where event yields are very large and
background contributions are extremely small, reach high
experimental accuracy. They provide new constraints on the parton
distribution functions (PDFs) of the proton and offer insights into
the initial-state QCD radiation dynamics in vector boson production~\cite{STDM-2018-17}.
They are however restricted in the kinematic range of
partonic momentum fraction $x$ and four-momentum transfer
$Q=m_{\ell\ell}$, where $m_{\ell\ell}$ denotes the invariant mass of the dilepton pair.

Measurements extending to lower or higher invariant masses have been
performed in the neutral-current channel only~\cite{STDM-2011-41, STDM-2012-10,
STDM-2014-06, CMS-EWK-10-007,CMS-SMP-13-003, CMS-SMP-14-003,CMS-SMP-17-001}
and access wider $x$ and $Q$ kinematic regions.  At low
$Q$ the measurements are sensitive to QCD resummation effects. At
large $Q$ they are sensitive to new physics beyond the Standard Model
and also provide new constraints on the poorly known high-$x$
anti-quark PDFs.

The charged-current process is highly complementary to the neutral-current process. The charge-separated
cross-sections for $W^+$ and $W^-$ production depend on different flavour
combinations, while the $Z$ production depends on same-flavour combinations.
The dominant partonic contributions are $(u,\bar{d})$ and $(c,\bar{s})$ for $W^+$, and
$(d,\bar{u})$ and $(s,\bar{c})$ for $W^-$ production.

Since the final-state neutrino is not observed, the dilepton invariant
mass cannot be experimentally determined, and the transverse mass \mtw
is used instead, defined by $\mtw=\sqrt{2p^{\ell}_{\textrm{T}}
p^{\nu}_{\textrm{T}} (1-\cos\Delta\phi(\ell,\nu))} $. Here
$p^{\ell}_{\textrm{T}}$ is the charged lepton transverse momentum,
$p^{\nu}_{\textrm{T}}$ is the neutrino or missing transverse momentum,
and $\Delta\phi(\ell,\nu)$ is the azimuthal angle between them.

To date no cross-section measurements of the charged-current process above the $W$-boson
resonance region have been made. This kinematic region is of
particular interest as at the highest transverse masses accessible at
the LHC, the observed \mtw spectrum in the decay to electron or muon final states is sensitive to new physics.
Such effects could manifest as a localised resonance \cite{EXOT-2018-30,CMS-EXO-19-017},
a violation of lepton flavour universality or an
enhancement or suppression of the continuum spectrum.   %
Modifications to the
Standard Model predictions can be described in terms of an Effective
Field Theory (EFT). In this formalism the Standard Model Lagrangian is
extended with new
operators which are suppressed by an energy cut-off scale. %
The charged-current Drell--Yan process is sensitive to operators which modify the couplings between fermions and the $W$-boson,
and operators which allow for a direct interaction between four fermions without a mediator.
The effects of four-fermion operators grow quadratically with energy,
so the measurement of charged-current Drell--Yan at high \mtw therefore provides strong sensitivity to such effects.
This measurement is thus expected to provide electroweak precision tests that surpass LEP in sensitivity~\cite{Farina:2016rws}.

This article reports a first measurement of the cross-section for the process $pp\rightarrow \ell
\nu + X$ above the $W$-boson resonance production region.
The measurements are performed using a sample of $140$~fb$^{-1}$ of $pp$ collision data
at
$\sqrt{s}=13$~\TeV\
collected
by the ATLAS detector at the LHC
in both electron and muon channels.
The double-differential measurement is reported as a
function of \mtw\ and absolute charged-lepton
pseudorapidity $|\eta|$. The data are also presented as single-differential
cross-sections ${\textrm d}\sigma/{\textrm d}\mtw$.
The data cover the kinematic region of $200 \leq \mtw \leq 5000$~\GeV\ and access
partonic momentum fractions from $x\sim 10^{-2}$ up to $x\sim 1$.
%


%
\section{ATLAS detector}
%


%
%
%
%

The ATLAS detector~\cite{PERF-2007-01} at the LHC covers nearly the entire solid angle around the collision point.\footnote{\AtlasCoordFootnote}
It consists of an inner tracking detector surrounded by a thin superconducting solenoid, electromagnetic (EM) and hadronic calorimeters,
and a muon spectrometer incorporating three large superconducting air-core toroidal magnets.

The inner-detector system (ID) is immersed in a \qty{2}{\tesla} axial magnetic field
and provides charged-particle tracking in the range \(|\eta| < 2.5\).
The high-granularity silicon pixel detector covers the vertex region and typically provides four measurements per track,
the first hit generally being in the insertable B-layer (IBL) installed before Run~2~\cite{ATLAS-TDR-19,PIX-2018-001}.
It is followed by the SemiConductor Tracker (SCT), which usually provides eight measurements per track.
These silicon detectors are complemented by the transition radiation tracker (TRT),
which enables radially extended track reconstruction up to \(|\eta| = 2.0\).
The TRT also provides electron identification information
based on the fraction of hits (typically 30 in total) above a higher energy-deposit threshold corresponding to transition radiation.

The calorimeter system covers the pseudorapidity range \(|\eta| < 4.9\).
Within the region \(|\eta|< 3.2\), electromagnetic calorimetry is provided by barrel and
endcap high-granularity lead/liquid-argon (LAr) calorimeters,
with an additional thin LAr presampler covering \(|\eta| < 1.8\)
to correct for energy loss in material upstream of the calorimeters.
Hadronic calorimetry is provided by the steel/scintillator-tile calorimeter,
segmented into three barrel structures within \(|\eta| < 1.7\), and two copper/LAr hadronic endcap calorimeters.
The solid angle coverage is completed with forward copper/LAr and tungsten/LAr calorimeter modules
optimised for electromagnetic and hadronic energy measurements respectively.

The muon spectrometer (MS) comprises separate trigger and
high-precision tracking chambers measuring the deflection of muons in a magnetic field generated by the superconducting air-core toroidal magnets.
The field integral of the toroids ranges between \num{2.0} and \qty{6.0}{\tesla\metre}
across most of the detector.
Three layers of precision chambers, each consisting of layers of monitored drift tubes, cover the region \(|\eta| < 2.7\),
complemented by cathode-strip chambers in the forward region, where the background is highest.
The muon trigger system covers the range \(|\eta| < 2.4\) with resistive-plate chambers in the barrel, and thin-gap chambers in the endcap regions.

The luminosity is measured mainly by the LUCID--2~\cite{LUCID2} detector that records Cherenkov light produced in the quartz windows of photomultipliers located close to the beampipe.

Events are selected by the first-level trigger system implemented in custom hardware,
followed by selections made by algorithms implemented in software in the high-level trigger~\cite{TRIG-2016-01}.
The first-level trigger accepts events from the \qty{40}{\MHz} bunch crossings at a rate close to \qty{100}{\kHz},
which the high-level trigger further reduces in order to record complete events to disk at about \qty{1.25}{\kHz}.

A software suite~\cite{SOFT-2022-02} is used in data simulation, in the reconstruction
and analysis of real and simulated data, in detector operations, and in the trigger and data acquisition
systems of the experiment.


%
\section{Simulated event samples}
%

%
%
\label{sec:MC}

Monte Carlo (MC) simulation samples are used to model the expected
signal and background yields, with the exception of certain
data-driven background estimates. The MC samples are normalised using
higher-order cross-section predictions in perturbation
theory.

The neutral and charged current Drell-Yan processes are generated at next-to-leading-order (NLO) using
\POWHEGBOX[v1]~\cite{Nason:2004rx,Frixione:2007vw,Alioli:2008gx,Alioli:2010xd}
and the  \CT[10nlo] PDF~\cite{CT10}, with \PYTHIA[8.1]~\cite{pythia8}       %
to model the parton showers (PS) and hadronisation.
The effect of
QED final-state radiation was simulated with \PHOTOSpp[3.52]~\cite{Golonka:2005pn,Davidson:2010ew}.
The \EVTGEN[1.2.0] program~\cite{Lange:2001uf} was used to decay bottom and charm hadrons.
This MC sample is
hereafter referred to as the \textsc{Powheg+Pythia} sample.
Using invariant-mass dependent $k$-factors~\cite{Butterworth:2014efa}
these predictions are corrected to account for
next-to-next-to-leading-order (NNLO) QCD and NLO electroweak (EW) effects. The QCD corrections are computed using
\textsc{Vrap}~v0.9~\cite{Anastasiou:2003ds} and also
weight the predictions to the  \CT[14nnlo] PDF set~\cite{Dulat:2015mca}.
The EW corrections excluding QED final state radiation (FSR)
are computed using \textsc{MCsanc}~\cite{Arbuzov:2015yja}
and vary from about $-3\%$ at low \mtw to about $-10\%$
at \mtw $= 1\,$\TeV.  The additive approach is used to
combine the EW and QCD corrections.

An alternative simulation of the signal process is used for the
assessment of systematic uncertainties arising from different
modelling choices. It is generated
using \SHERPA[2.2.11]~\cite{Sherpa:2019gpd} at NLO in QCD for up to two
additional partons and at LO for up to five partons, using the
\NNPDF[3.0nnlo] PDF set~\cite{Ball:2014uwa}. The sample is
then reweighted to the \CT[18nnlo] PDF set~\cite{Hou:2019efy}.
The prediction is corrected to approximate NNLO accuracy with a $k$-factor of $0.95$ calculated
using \textsc{Matrix} v2~\cite{Grazzini:2017mhc,ATL-PHYS-PUB-2023-026}.
NLO EW corrections are implemented in \SHERPA. The exponentiated approach for
combining EW and QCD corrections is used, which yields corrections
lying between the additive and multiplicative approaches. The
differences between all three corrections is however small and
typically below $1\%$.

The background from \ttbar\ production is the dominant background with
isolated prompt leptons from electroweak boson decays.  It is
estimated at NLO using  \POWHEGBOX[v2]~\cite{Frixione:2007nw} with a top-quark mass of
$m_{\textrm{top}}=172.5$~\GeV and the \NNPDF[3.0nlo] PDF set~\cite{Ball:2014uwa},
with \PYTHIA[8.2]~\cite{Sjostrand:2014zea} for parton showering and hadronisation.
The decays of bottom and charm hadrons
were performed by \EVTGEN[1.6.0].  %
The \ttbar MC samples are reweighted in the \pt of the top quark and in the mass of the \ttbar pair
to inclusive %
cross-section calculations at NNLO accuracy in QCD including
NLO EW corrections~\cite{Czakon:2017wor}.
The single-top-quark background, consisting of $s-$ and
$t-$channel processes and the $tW$ process, is also simulated under the same
conditions.  The dynamic scale diagram subtraction
scheme~\cite{Re:2010bp} is used to account for the interference
between the $tW$ and \ttbar\ production diagrams.

Further important background contributions are due to diboson ($WW$,
$WZ$ and $ZZ$) production decaying into final states with at least
one charged lepton. The diboson processes are generated with
\SHERPA[2.2.1] or \SHERPA[2.2.2] depending on the process, using the
\NNPDF[3.0nnlo] PDF set. They were generated using
matrix elements at NLO accuracy in QCD for up to one additional parton
and at LO accuracy for up to three additional parton
emissions.

All MC samples used in the analysis include the effects of QED FSR, multiple $pp$ interactions per bunch crossing
(``\pileup''), and detector simulation. The QED FSR is simulated using
\PHOTOS~\cite{fsr_ref}, except for samples generated by \Sherpa
which includes a native FSR simulation.
The effect of pile-up was modelled by overlaying
the simulated hard-scattering event with
inelastic $pp$ events generated with \PYTHIA[8.1]     %
using the \NNPDF[2.3lo] PDF set~\cite{Ball:2012cx} and the
A3 set of tuned parameters~\cite{ATL-PHYS-PUB-2016-017}.
The interactions of particles with the detector are
modelled using a full \mbox{ATLAS} detector
simulation~\cite{SOFT-2010-01} based on \GEANT~\cite{geant}.  Finally,
several corrections are applied to the simulated samples, accounting
for differences between data and simulation in the lepton trigger,
reconstruction, identification, and isolation efficiencies as well as
lepton resolution and muon momentum scale.

%
%
%
%
%
%
%
%
%
%
%
%
%
%
%
%
%
%
%
%
%
%


%
\section{Event reconstruction and selection}
%
%
%
%
%

\label{sec:selection}

Events are required to be recorded during stable beam condition periods
and must pass detector and data-quality requirements~\cite{DAPR-2018-01}.
Due to differences in the detector response to electrons and
muons the nominal selection is optimised separately for
each channel and is described in the following.
In addition to the nominal selection two further selections are defined.
Both correspond to less stringent selections needed for either the background estimation (``loose'') or
in order to veto events with a second lepton (``veto'').

\subsection{Object reconstruction and selection}
\label{sec:sel_objects}

\subsubsection{Electrons}
\label{sec:sel_elec}

Electrons are reconstructed from clusters of energy deposited in the
EM calorimeter and matched to tracks reconstructed in the
ID~\cite{EGAM-2018-01}.  An energy scale correction determined from $Z\to
e^+e^-$, and $J/\psi\to e^+e^-$ decays is applied
to data.  Electron candidates are required to have a pseudorapidity within the
inner detector tracking region, $\left|\eta\right|<2.4$, excluding a
region, $1.37<\left|\eta\right|<1.52$, where the transition between
the barrel and endcap electromagnetic calorimeters is not well
modelled in the simulation.

The primary interaction vertex is taken to be the one with the largest sum of squared transverse momenta of all associated tracks~\cite{ATL-PHYS-PUB-2015-026}.
Electrons must originate from the primary vertex and are selected by requiring
$|z_0\sin\theta|<0.5$~mm, where $z_0$ is the coordinate of the track
at the point of closest approach to the beam-line. The significance of
the transverse impact parameter, defined by the distance of closest
approach of the track to the beam-spot in the $r-\phi$
projection $|d_0|$, divided by its estimated uncertainty $\sigma(d_0)$,
is required to satisfy $|d_0|/\sigma(d_0)< 5$.
~

Candidates must satisfy the \textit{LooseAndBLayer} likelihood-based
identification requirements~\cite{EGAM-2018-01} based on EM shower
shapes, track quality, and track–cluster matching.

Leptons produced in the Drell--Yan process are expected to be well
isolated from energy depositions not associated with the lepton.
Therefore the electron candidates also need to fulfil the \textit{FCLoose} isolation working point ~\cite{EGAM-2018-01},
which has two distinct definitions of the degree of isolation: one is based on calorimeter information and one is based on tracks in the ID.
The calorimeter-based isolation is defined
as the scalar sum of transverse energy, $\sum E_{\textrm{T}}$,
contained in a cone of size
$\Delta R=\sqrt{(\Delta\phi)^2+(\Delta\eta)^2}$ around the electron,
omitting the electron transverse energy, $E_{\textrm{T}}$.  It is
given by $\sum E_{\textrm{T}}(\Delta R =0.2)<0.20\cdot
E_{\textrm{T}}$. The track-based criterion is based on the scalar sum
of transverse momenta, $\sum p_{\textrm{T}}$, of additional tracks
around the electron, and is required to be below $15\%$ of the
electron $p_{{\textrm{T}} }$, i.e. $\sum p_{\textrm{T}}(\Delta R)<0.15 \cdot
p_{{\textrm{T}} }$. Here, the $\Delta R$ cone  size shrinks with
increasing transverse momentum of the electron, $\Delta R = \min
(\frac{10~\textrm{GeV}}{\pt} , 0.2 )$.

The criteria above define the ``loose'' and the ``veto'' level as they are identical for electrons.
The nominal selection requires electrons to in addition satisfy the \textit{Tight} likelihood-based identification requirements
and an additional electron  isolation requirement that uses the \textit{FCHighPtCaloOnly}
isolation working point~\cite{EGAM-2018-01}, based on calorimeter
information only via
$\sum E_{\textrm{T}}(\Delta R =0.2)< \text{max}( 0.015\cdot E_{\textrm{T}} , 3.5~\GeV)$.

\subsubsection{Muons}
\label{sec:sel_muon}
Muon candidates are reconstructed from tracks in the muon spectrometer
matched to tracks from the ID and satisfy $|\eta| < 2.4$.
Muons originating from the primary vertex are selected by requiring
$|z_0\sin\theta|<0.5$~mm and $|d_0|/\sigma(d_0)< 3$.

Except for the ``veto'' level the muon identification uses
the \textit{High-$p_{\textrm{T}}$} working point~\cite{MUON-2018-03}
designed to optimise the resolution for muons with $p_{\textrm{T}}$
above $100$~\GeV. This working point requires: agreement in the
charge-to-momentum ratio of the muon as measured in the ID and the MS,
taking into account their uncertainties; at least three precision
space-points in the MS for improved sagitta measurement; and upper
limits on the uncertainty of the charge-to-momentum ratio. These
requirements
cannot be fulfilled in some regions of the MS, for example
in the transition region between barrel and endcap with 1.01 < \abseta < 1.1.
The \textit{Medium} working point~\cite{MUON-2018-03},
which particularly requires only two precision space-points in the MS, is used
when vetoing additional muons.

The nominal selection requires muons to in addition fulfil an isolation
criteria.
It is defined using the scalar sum of transverse
momenta, $\sum p_{\textrm{T}}$, of additional tracks divided by
$p_{\textrm{T}}$, the transverse momentum of the muon.
The selection requires $\sum p_{\textrm{T}}(\Delta R=0.2)<0.06\cdot \pt$,
providing a good discrimination against the background arising from the
semileptonic decays of heavy quarks.

\subsubsection{Jets}
\label{sec:sel_jets}

Jets are reconstructed and used in the determination of the missing
transverse momentum in the event. A particle flow
algorithm~\cite{PERF-2015-09} is used which optimally combines
calorimetric and tracking information.  The anti-$k_t$
algorithm~\cite{Cacciari:2008gp,Fastjet} is used with radius parameter
$R=0.4$. They are calibrated using in situ measurements and
simulation~\cite{JETM-2018-05}. Jets are required to satisfy
transverse momentum
$p_{\textrm{T}}>25$~\GeV\ and pseudorapidity $|\eta|<2.5$. Jets with
$p_{\textrm{T}}<60$~\GeV\ and $|\eta|<2.5$ are also required to pass
the \textit{tight} jet vertex tagger
criteria~\cite{PERF-2014-03}.
Identification of jets containing $b$-hadrons is performed with the DL1r algorithm~\cite{FTAG-2019-07}
with an average tagging efficiency of $b$-jets from simulated dileptonic \ttbar events of $70\%$.

\subsubsection{Overlap removal}
\label{sec:sel_overlap}

Particles identified by the ATLAS detector may be reconstructed as
multiple different objects. Ambiguities for electrons are resolved by applying a
sequential algorithm to select the best choice for each object. If a
pair of electron candidates share a track, the lower $p_{\textrm{T}}$
object is removed. Any jet within $\Delta R=0.2$ of an electron
candidate is removed. Then, any electron candidates found within
$\Delta R=0.4$ of a jet are removed.
Ambiguities for muons are not resolved at this stage as they are beneficial for the
estimation of the amount of non-prompt muons in this measurement (see Section~\ref{sec:bg_muon_mj}).

\subsubsection{\met}
\label{sec:sel_met}
The missing transverse momentum, with magnitude
\met, is determined from the sum of transverse momenta of
reconstructed and calibrated objects in the event.
A so-called soft term accounts for
the \pt of all remaining tracks associated to the primary
vertex~\cite{JETM-2020-03}.

\subsection{Event selection}
\label{sec:sel_signal}
The electron data are collected by a set of triggers which use
calorimetric information to identify a compact electromagnetic energy
deposition~\cite{TRIG-2018-05}. Identification algorithms use calorimeter shower shape
information to find candidate electrons with a minimum transverse
energy of $60$ or $120$~\GeV\ in 2015. \textit{Medium} identification
criteria~\cite{EGAM-2018-01} are applied at the lower threshold and
\textit{loose} criteria for the higher threshold. In the remaining
years the higher transverse energy threshold was changed to
$140$~\GeV. To validate backgrounds two additional low-threshold triggers
were used with trigger thresholds $24$~\GeV
and \textit{medium} identification criteria in 2015, and a $26$~\GeV
threshold with \textit{tight} identification in subsequent years.
In the muon channel data are collected using a single muon trigger
with a transverse momentum above $50$~\GeV~\cite{TRIG-2018-01}.

Events are required to have exactly one reconstructed electron with $\et>65$~\GeV\
or exactly one reconstructed muon with $p_{\textrm{T}}>65$~\GeV.  The electron or
muon has to satisfy the nominal selection criteria.
In both cases the lepton is required
to match a corresponding trigger.
Events containing additional muons or electrons with $p_{\textrm{T}}>20$~\GeV\ fulfilling the ``veto'' level criteria
are rejected.  The \met\ is required to be greater than $85$~\GeV\
and the reconstructed \mtw\ must exceed $200$~\GeV. It is determined from the
relationship given in Section~\ref{sec:intro} using the
reconstructed \met, lepton \pt, and the azimuthal angle between them.

%
%
%
%
%
%
%
%
%
%
%
%
%
%
%
%
%
%


%
\section{Background estimation}
%

The background from processes with one prompt isolated final-state
lepton is estimated from MC simulation and includes: $Z\to\ell\ell$ where
one lepton is not reconstructed due to detector acceptance or
efficiency effects; $W\to\tau\nu$ and $Z\to\tau\tau$ where a $\tau$-lepton
decays leptonically; diboson production of $ZZ$, $WZ$ and $WW$ in
which one boson decays leptonically; and \ttbar and single top production
(hereafter termed the top-quark background). The MC predictions for
some of the major contributions are validated in dedicated kinematic
regions.

A further background contribution arises from light- and heavy-flavour
multijet production in which non-prompt leptons are produced or from the
misidentification of jets as fake electrons. In the following the
multijet background is defined as the sum of both contributions.  This
background is not well modelled in simulation and is determined using
data-driven techniques described in detail below.

\subsection{Top-quark background contribution}
\begin{figure} [t]
\centering
\subfloat[\label{subfig:1jetno}]{\includegraphics[width=0.48\textwidth]{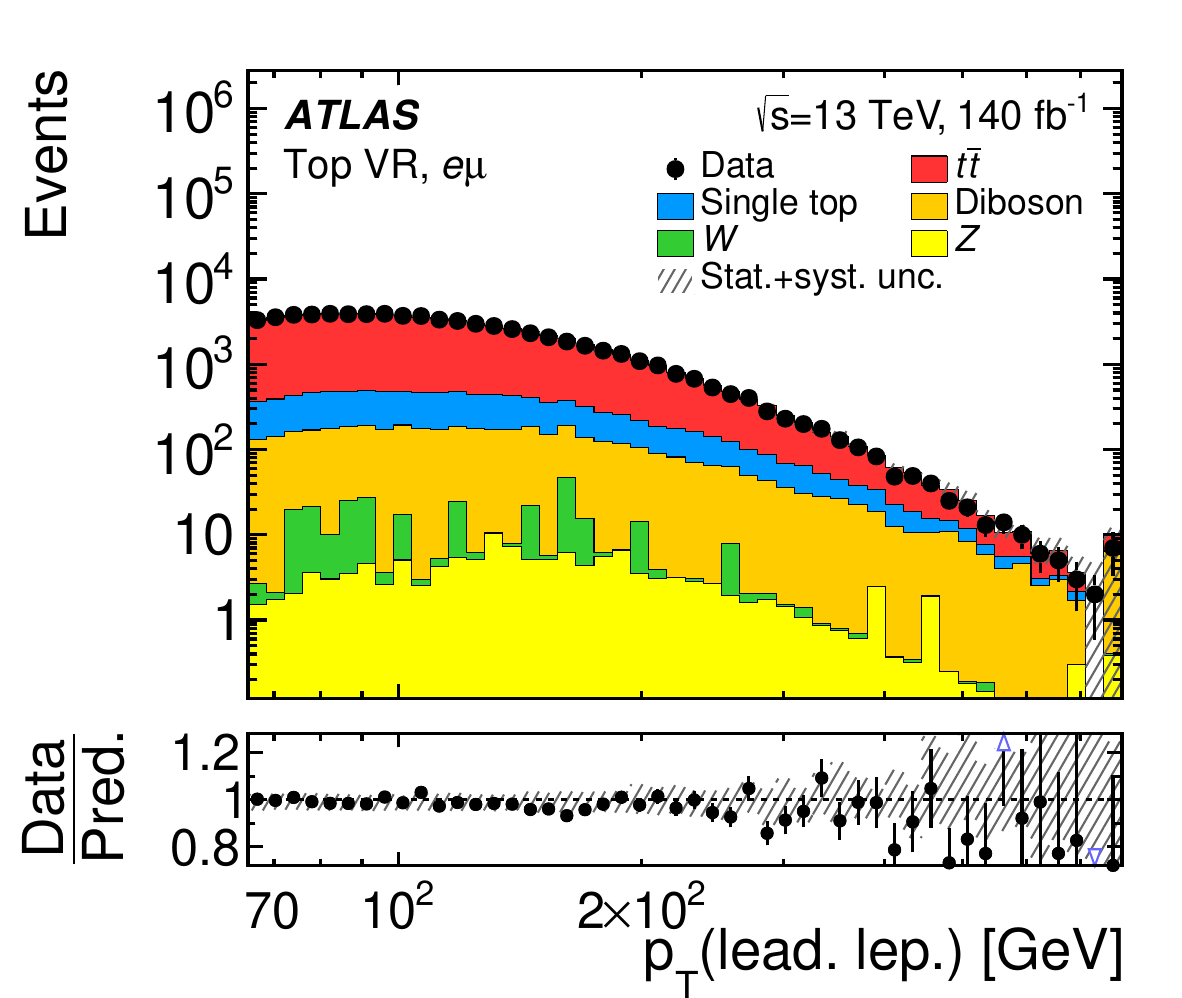}}
\subfloat[\label{subfig:1mtw}]{ \includegraphics[width=0.48\textwidth]{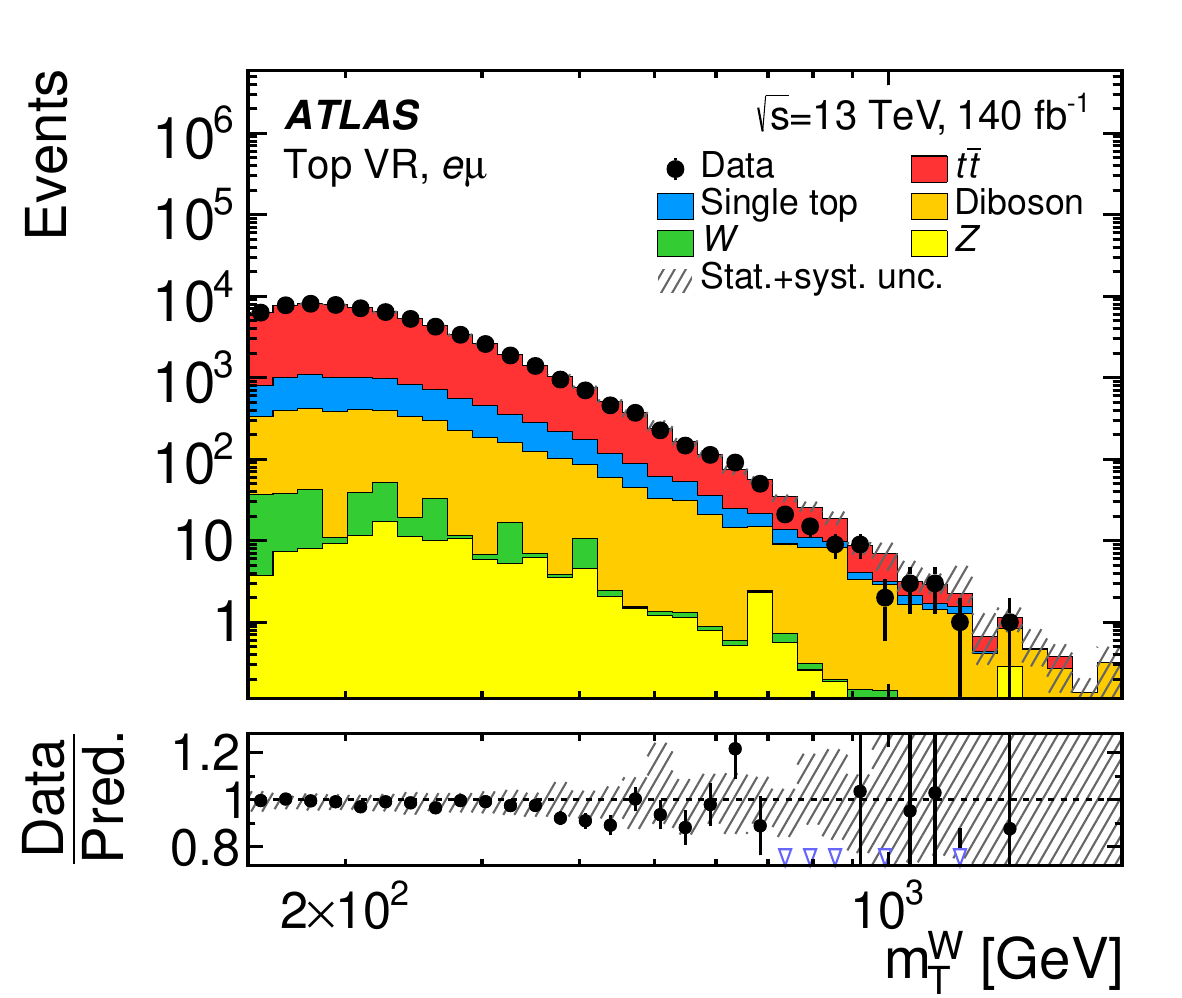}}
\caption{
Comparisons of the data and prediction in the top validation region
for the (a) \pt of the leading lepton and the (b) transverse mass of the W boson.
Statistical and systematic uncertainties of the prediction are included in the uncertainty band.
A blue triangle in the bottom panel indicates data points outside the vertical range shown.
}
\label{fig:topVR}
\end{figure}

The background contribution from top-quark production is
estimated using the MC simulation, and is expected to be large in the
phase space of this measurement. It is therefore important to check
the validity of the MC simulation using data.
As the dominant contribution stems from dileptonic decays of the \ttbar pairs,
a top-quark-enriched
sample of events is selected by requiring two ``tight'' leptons -- one
electron and one muon both with $p_{\textrm{T}}>30$~\GeV. The leading
lepton is further required to have $p_{\textrm{T}}>65$~\GeV\ matching
the signal region selection, and this lepton is used with \met in the
calculation of  the \mtw observable. The \met\ is required to be
above $85$~\GeV, and the \mtw is required to exceed
$150$~\GeV. This validation region closely matches the kinematic region
of the measurement and is found to be highly enriched in \ttbar\ and single-top
production with fractions of $85\%$ and $9\%$ respectively, where the
single-top production is dominated by the $tW$ process.
The signal and multijet contributions are negligible in this sample. All kinematic observables
are found to be well described in shape and in normalisation by the
sum of MC expectations. Examples are shown in Figure~\ref{fig:topVR}
for the leading lepton \pt and the \mtw observable.

The top-quark background is further cross checked in two variations of the validation
region where the \met requirement is removed or where an additional b-jet
is required. The former validation region allows a check of the region of small \met\
used in the multijet estimate, while the latter is a pure (> 99\%) top-quark control region.
The simulation is again found to describe the data well in both regions.

\subsection{Multijet background estimate via the matrix method}
\label{sec:bg_matrix_method}

The matrix method~\cite{STDM-2014-06} is used to estimate the multijet
background in the electron and muon channels.        %
The method makes use of the different sets of selection criteria
described in Sections~\ref{sec:sel_elec} and ~\ref{sec:sel_muon}.
It relates the number of observed leptons satisfying      %
the nominal (``tight'') requirement, $N_T$, or only the less stringent (``loose'') requirements, $N_L$, to
the number of prompt (``real'') leptons, $N_R$,  and to the number of non-prompt leptons or jets (``fake''), $N_F$.
They are connected by a pair of linear equations,
using MC estimates
of the real efficiency, $\epsilon_r$, and data-driven estimates of the
fake efficiency, $\epsilon_f$.
The real efficiency is defined to be the
fraction of prompt leptons satisfying at least the loose selection which also pass
the tight selection,
$\epsilon_r = N^T_R/(N^L_R+N^T_R)$.  %
The fake efficiency is analogously defined as the fraction of non-prompt
leptons and jets satisfying at least the loose selection which also pass the tight
selection $\epsilon_f = N^T_F/(N^L_F+N^T_F)$.   %
The relationship %
is given
as:

\begin{equation*}
\label{ff_matrix_full}
\begin{pmatrix} N_{T}  \\ N_{L} \end{pmatrix}
=
\begin{pmatrix}
\epsilon_r            &       \epsilon_f              \\
1-\epsilon_r       &    1-\epsilon_f             \\
\end{pmatrix}
\begin{pmatrix} N_R \\ N_F \end{pmatrix}.
\end{equation*}

The equation can be inverted to derive the multijet background
as
$$N^{\textrm{multijet}} = \epsilon_f N_F = \frac{\epsilon_f}{\epsilon_r
- \epsilon_f}  \left[ \epsilon_r (N_L+N_T) - N_T   \right] .$$

This general method
is used to derive the distribution of multijet events as a function of any variable of interest for both the electron and the
muon channels.

\subsubsection{Multijet background estimate in the electron channel}
\label{sec:bg_elec_mj}
The real efficiency is determined using the signal MC sample in which
the reconstructed prompt electron is matched to the corresponding
Born-level particle.
The \met and \mtw criteria described in Section~\ref{sec:sel_signal}
are not applied.  The $\epsilon_r$ are determined in 132
two-dimensional bins of electron $p_{\textrm{T}}$ and $|\eta|$. The
resulting factors are found to vary between $90-95\%$ except at the
largest $|\eta|$ and close to the transition region between the barrel and endcap calorimeter.

\begin{figure}
\centering
\subfloat[]{\includegraphics[width=.48\textwidth]{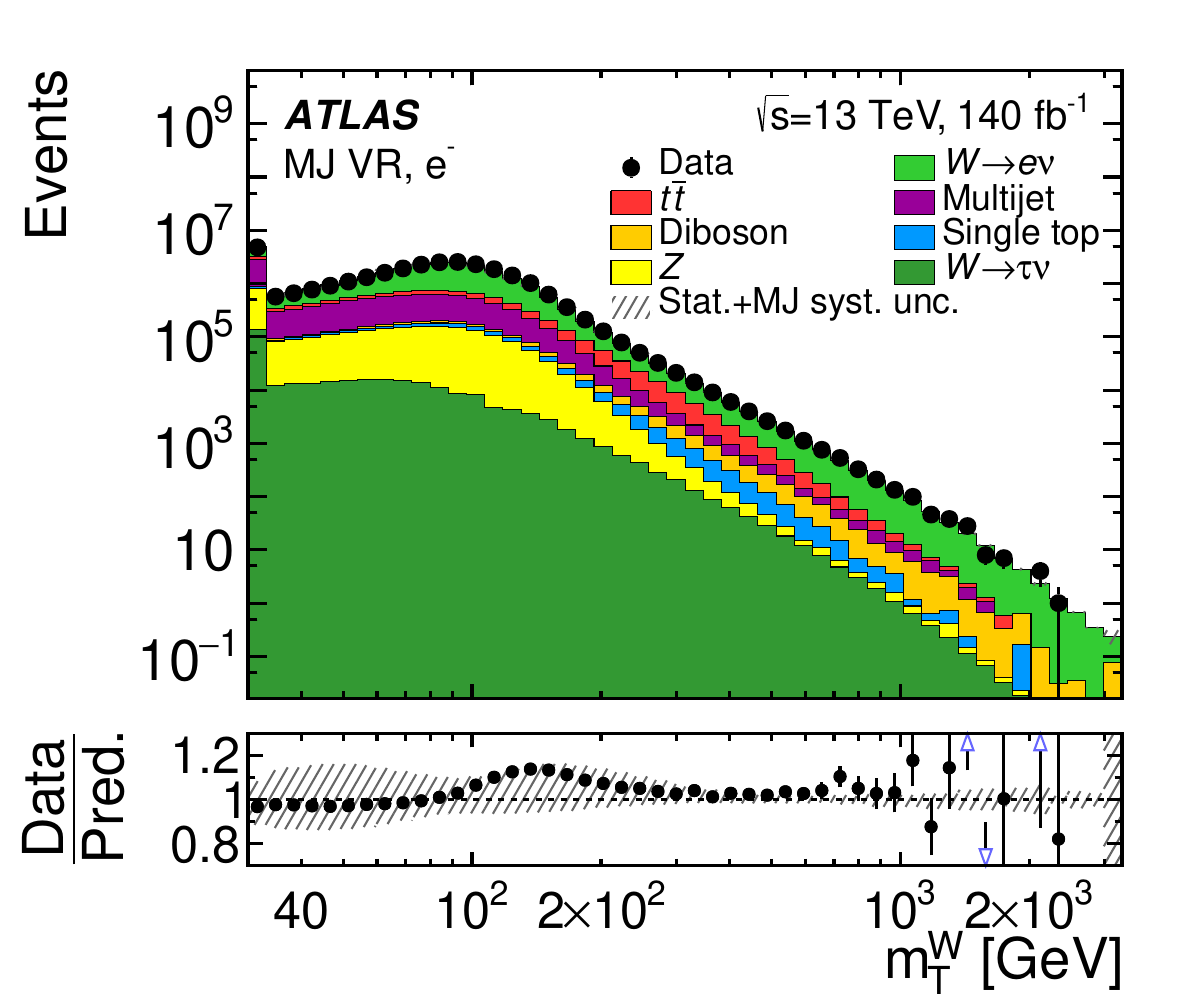}}
\subfloat[]{\includegraphics[width=.48\textwidth]{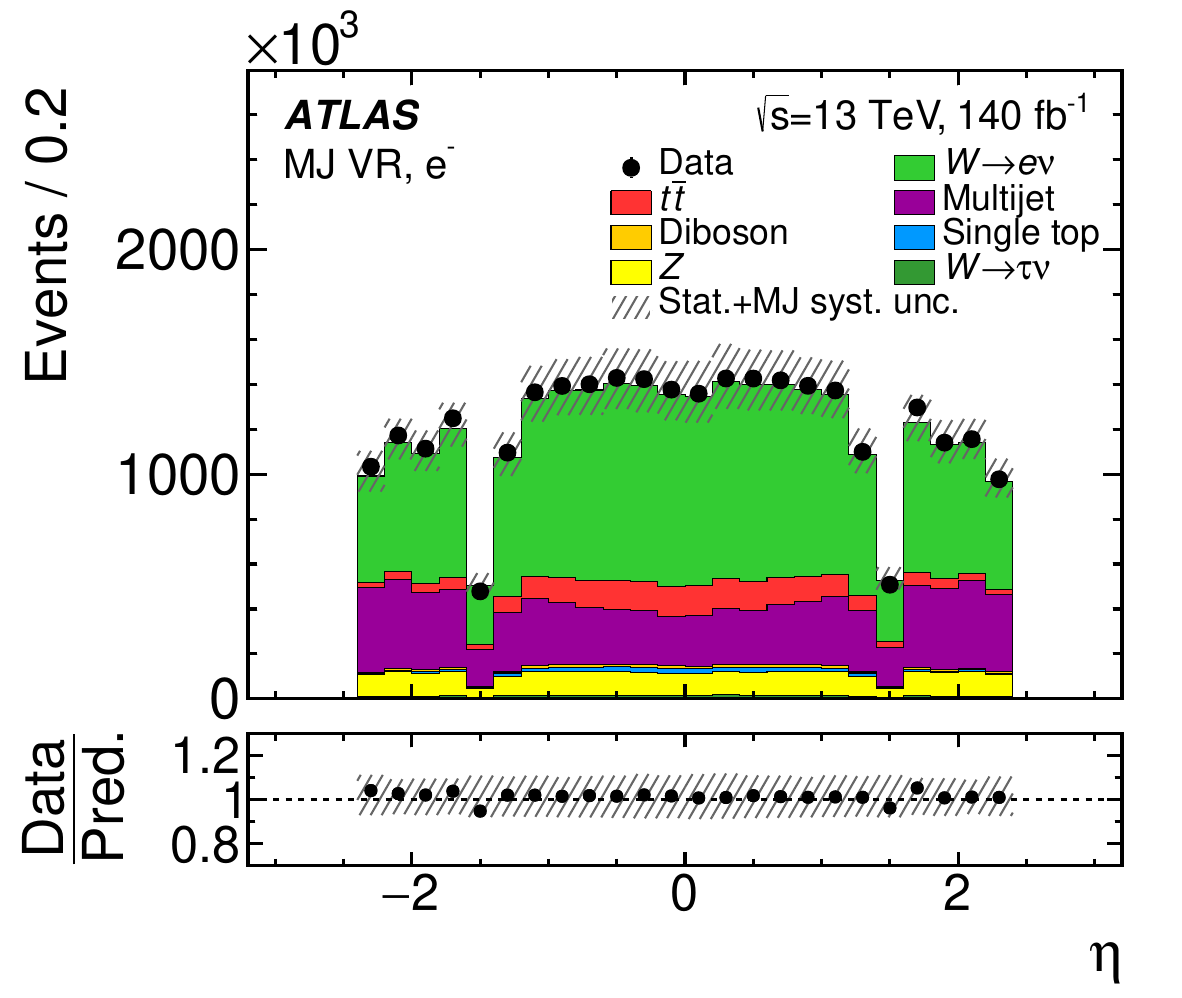}}
\caption{
Comparisons of the data and prediction in the multijet validation region
for the (a) transverse mass of the W boson and the (b) pseudorapidity of the electron  for the $e^-$ final state.
The grey shaded band represents the statistical uncertainty of the prediction and the systematic
uncertainties on only the multijet estimate. A blue triangle in the bottom panel indicates data points outside the vertical range shown.
}  \label{fig:MJ_elecminus}
\end{figure}

The fake efficiency is measured using a data sample enriched in fake
electrons by removing the \mtw criterion and requiring the \met
to be below $65$~\GeV, which also ensures orthogonality between the
signal and fake enriched selections.
The residual real electron
contribution is subtracted using the MC predictions.
The fake
efficiency has the same binning in $p_{\textrm{T}}$ and $|\eta|$
as the real efficiency, but depends simultaneously also on the difference
in azimuthal angle between the electron and the \met, $\Delta\phi(e,\met)$,
leading to 660 three-dimensional bins.
The fake efficiencies are found
to be typically below $50\%$ with a marked reduction at larger
$p_{\textrm{T}}$.
The resulting efficiencies are validated by
comparing the simulation and predicted multijet background
contribution to the data in a wider fake enriched region in which
the \met condition is removed completely.
Figure~\ref{fig:MJ_elecminus} shows the data and the predicted event
yields for the \mtw and $\eta$ spectra in this region for
the $e^-$ measurement channel.
The data are described by the prediction given the
multijet and statistical uncertainties shown in the figure. The
dominant uncertainty, particularly at low \met, arises from the
choice of calibration applied to loose objects in the \met
calculation as discussed in Section~\ref{sec:sysElec}.

\subsubsection{Multijet background estimate in the muon channel}
\label{sec:bg_muon_mj}
The multijet background estimation in the muon channel largely follows
that of the electron channel described above.
The real efficiency is
determined from the signal Drell--Yan MC simulation from events in which the
reconstructed muon is matched to the Born-level particle.
The \met\ and \mtw\ selection criteria are dropped.
The real and fake efficiencies, $\epsilon_r$ and $\epsilon_f$, are determined
in only 18 two-dimensional bins of reconstructed muon $p_{\textrm{T}}$ and $|\eta|$ due
to the limited sample size available to the fake efficiency estimate when compared to the case of electrons.
The real efficiencies are found to exceed $99\%$ in all bins.

The fake enriched region in the muon channel is created by first
removing the \mtw\ requirement and requiring $\met<65$~\GeV, as described
in Section~\ref{sec:bg_elec_mj}. Since the dominant source of fake muons
arises from heavy flavour $b$- and $c$-hadron decays within jets,
further selections are used to enhance the yield of this class of
events. Dijet topologies are targeted in which an additional jet is
required (see Section~\ref{sec:sel_objects}) which is back-to-back in the
azimuthal angle with respect to the
muon, $\Delta\phi(j,\mu)>\frac{5}{6}\pi$. The azimuthal angular separation
between the
muon
and the \met is required to be less than
$\frac{\pi}{6}$. This criterion efficiently suppresses the signal
contribution. Finally, the transverse impact parameter significance
$|d_0|/\sigma(d_0)$ is required to be greater than $1.5$. The
remaining contributions from EW processes and the signal process are
subtracted using the predictions from MC simulation.

An algorithm similar to the one used to resolve particle object ambiguities
in the electron channel within an event as described in Section~\ref{sec:sel_objects}
is not applied in the muon channel as non-prompt muon arise predominantly
from within jets. Therefore,  the contribution from non-prompt muons is increased in the
fake enriched region while these muons are efficiently suppressed by the isolation requirement
for muons which is part of the tight selection.

The efficiency for selecting fake muons is found to be about $10\%$ at low
$p_{\textrm{T}}$, rising to between $30\%$ for central
pseudorapidity and $65\%$ at the highest pseudorapidity.
It is validated by estimating the multijet background in an
enlarged kinematic region with respect to the region used to estimate
the fake efficiency. This is achieved by first removing the \met\
selection criterion.
\begin{figure}[t!]
\centering
\subfloat[]{\includegraphics[width=0.46\textwidth]{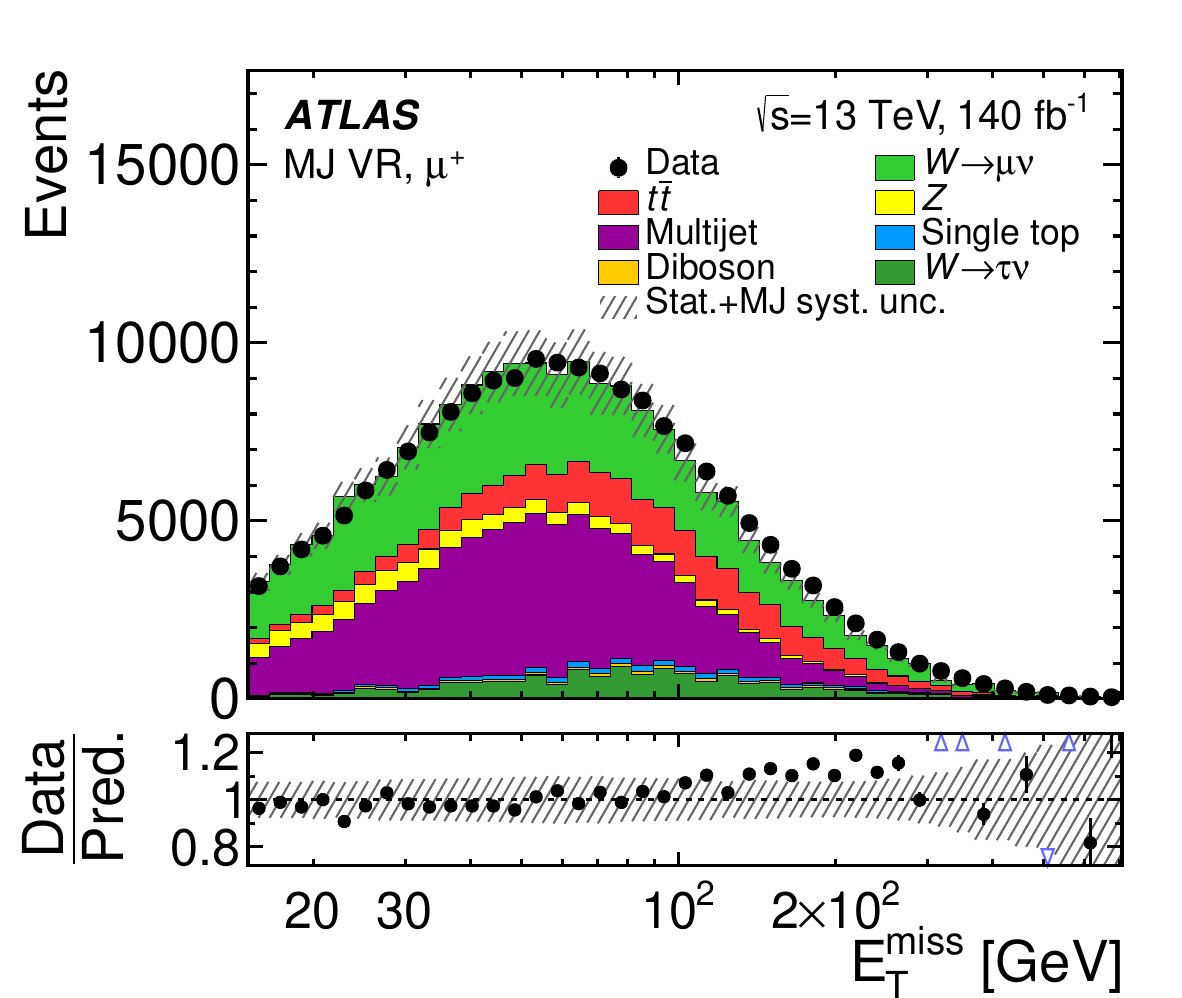}}
\subfloat[]{\includegraphics[width=0.46\textwidth]{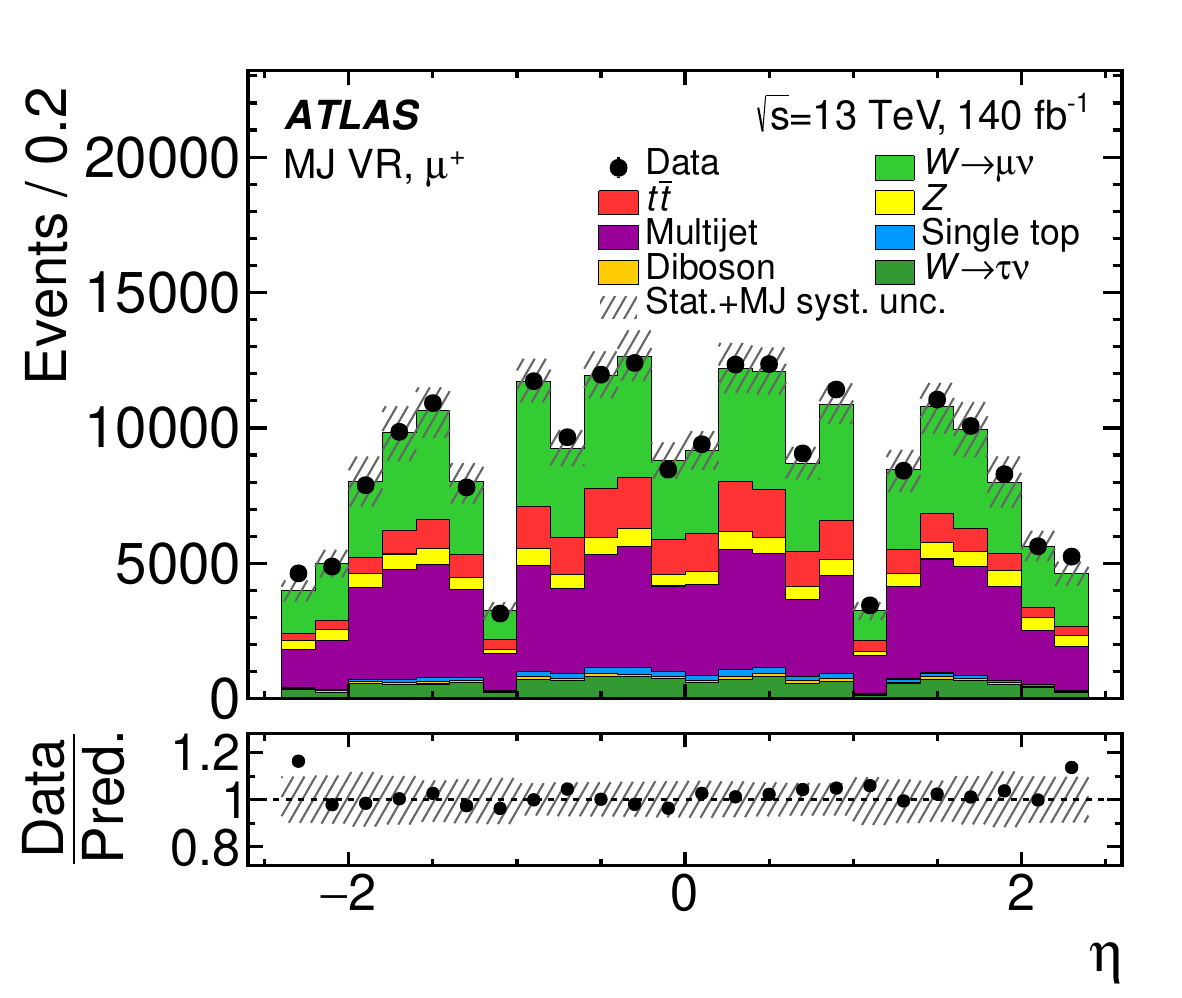}}
\caption{
Comparisons of the data and prediction in the multijet validation region
for the (a) missing transverse energy and the (b) pseudorapidity of the muon  for the $\mu^+$ final state.
The grey shaded band represents the statistical uncertainty of the prediction and the systematic
uncertainties on only the multijet estimate. A blue triangle in the bottom panel indicates data points outside the vertical range shown.	}
\label{fig:MJ_muonplus}
\end{figure}
Figure~\ref{fig:MJ_muonplus} shows the data and
the predicted event yields for the extrapolation in \met\ and for
$\eta$
for the $\mu^+$ measurement channel.
Up to $\met \approx 150$~\GeV where the multijet contribution is large,
the data are well described by the predictions within the systematic
uncertainty of only the multijet estimate and the statistical
uncertainty of the prediction.  However, the region of high \mtw\
cannot be validated as the requirement on $\Delta\phi(\mu,\met)$
suppresses the event yield. Therefore a second validation region is
defined by removing the requirement on $\Delta\phi(\mu,\met)$ and requiring
at least three jets instead.
The resulting phase space is found to be composed of approximately equal contributions from multijet events,
signal, and top quark processes. The data in this phase space are well described by the prediction.

\subsection{Data and prediction comparisons}
\label{sec:control_plots}

The number of expected events is calculated as the sum of the
data-driven and simulated background estimates, and the expected event
yield predicted by the signal Drell--Yan MC simulations.

\begin{figure}[ht!]
\centering
\subfloat[\label{fig:SR_ele_pt_plus}]{\includegraphics[width=0.46\textwidth]{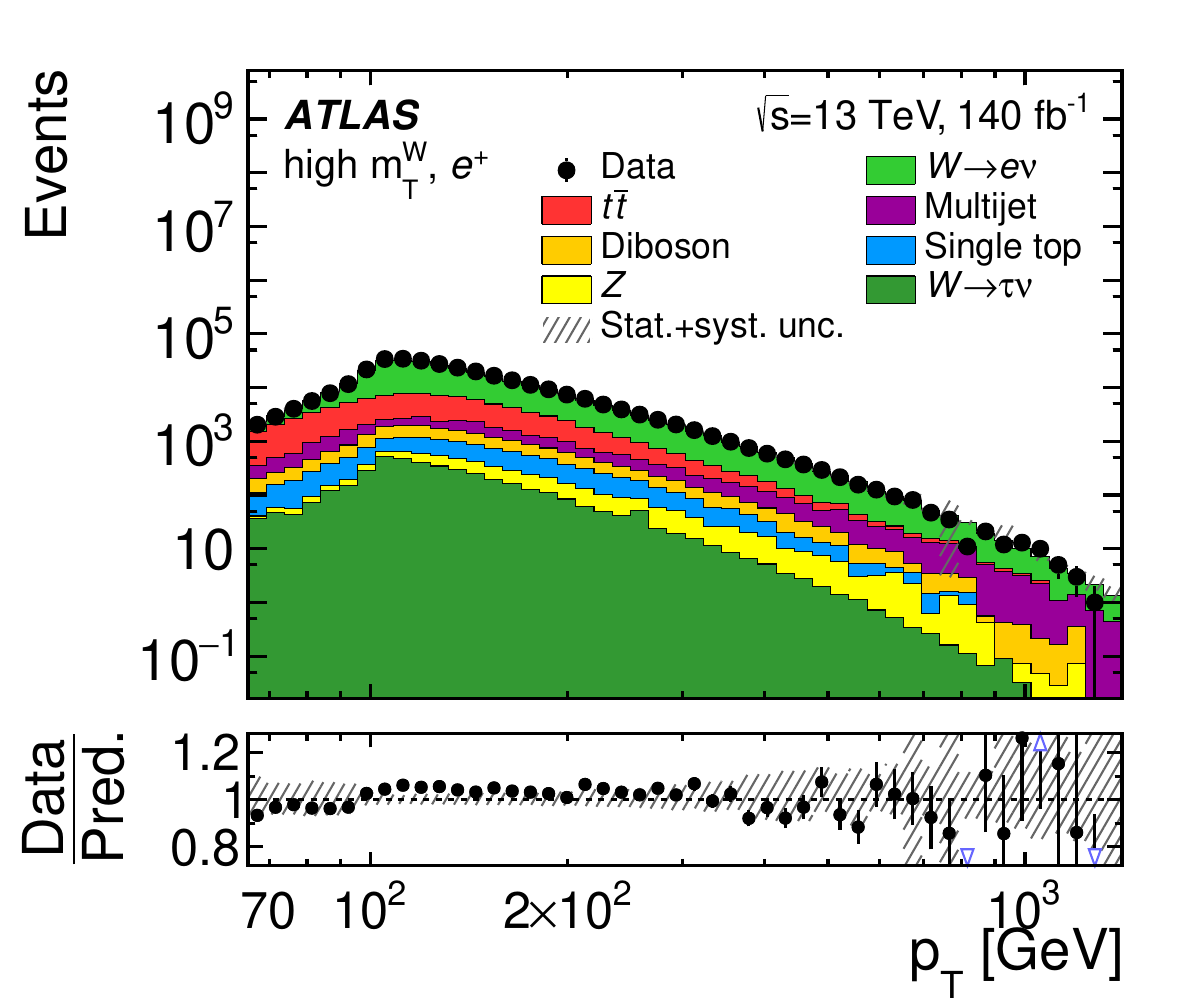}}
\subfloat[\label{fig:SR_ele_pt_minus}]{\includegraphics[width=0.46\textwidth]{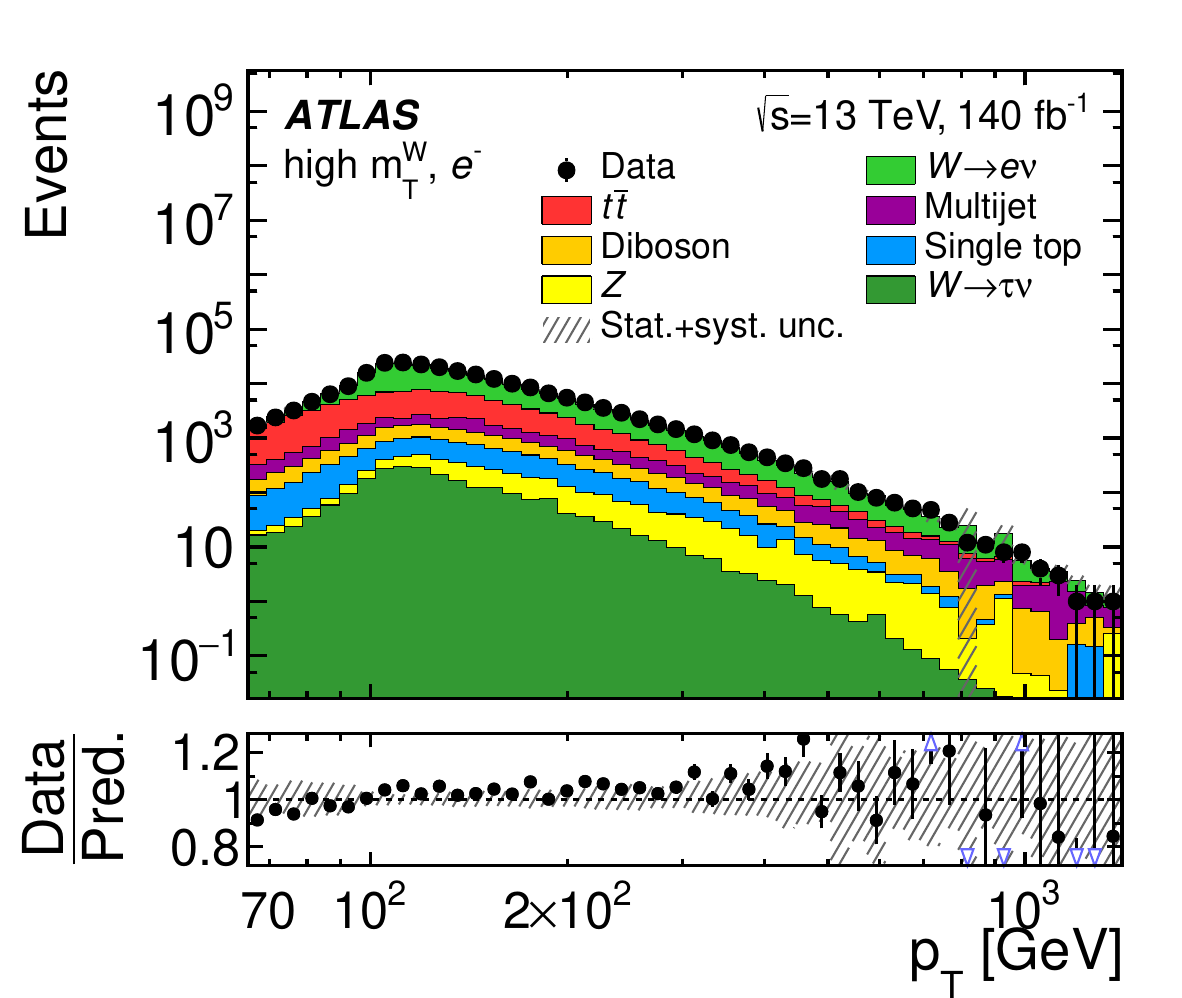}}\\[-2ex]
\subfloat[\label{fig:SR_ele_met_plus}]{\includegraphics[width=0.46\textwidth]{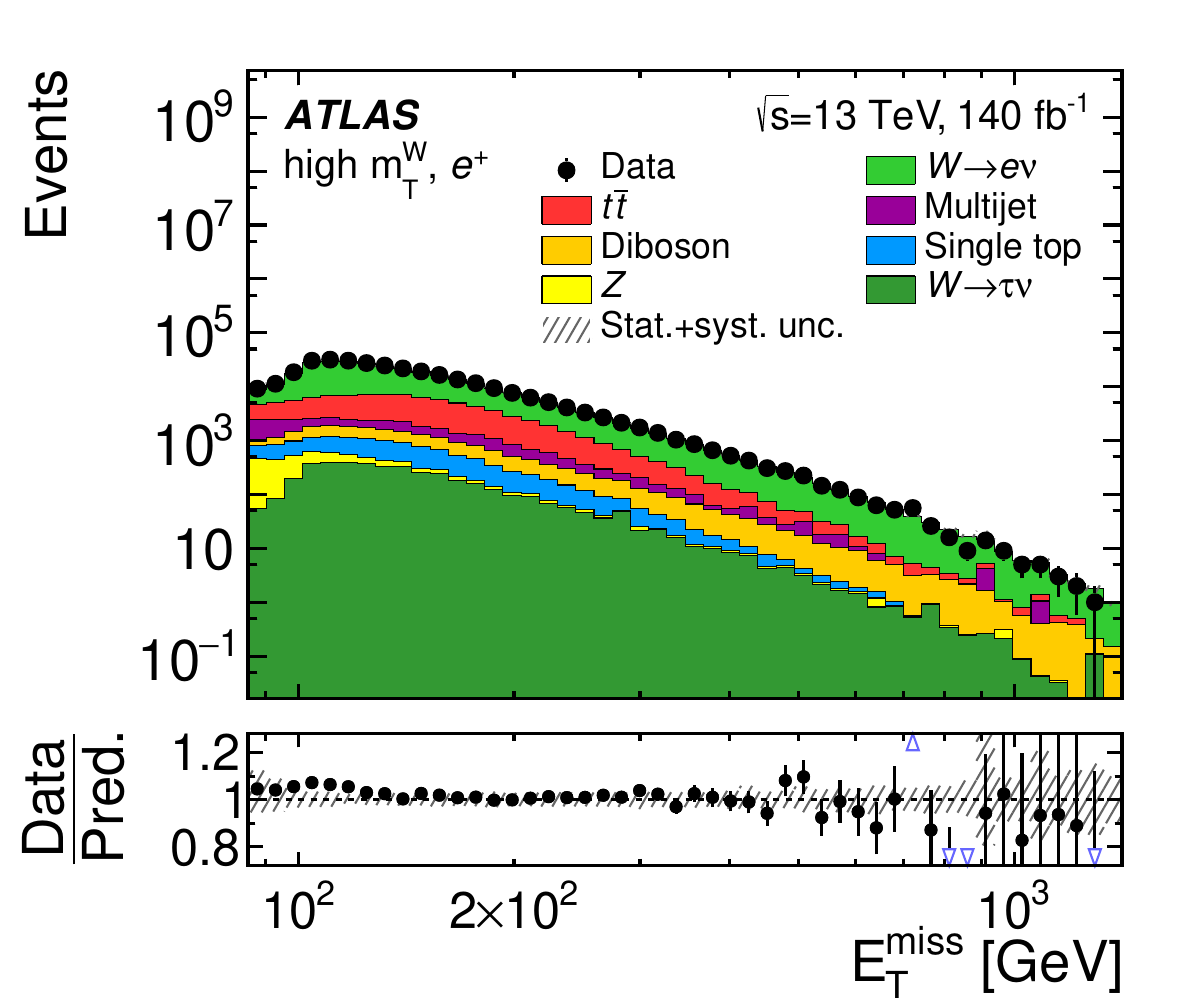}}
\subfloat[\label{fig:SR_ele_met_minus}]{\includegraphics[width=0.46\textwidth]{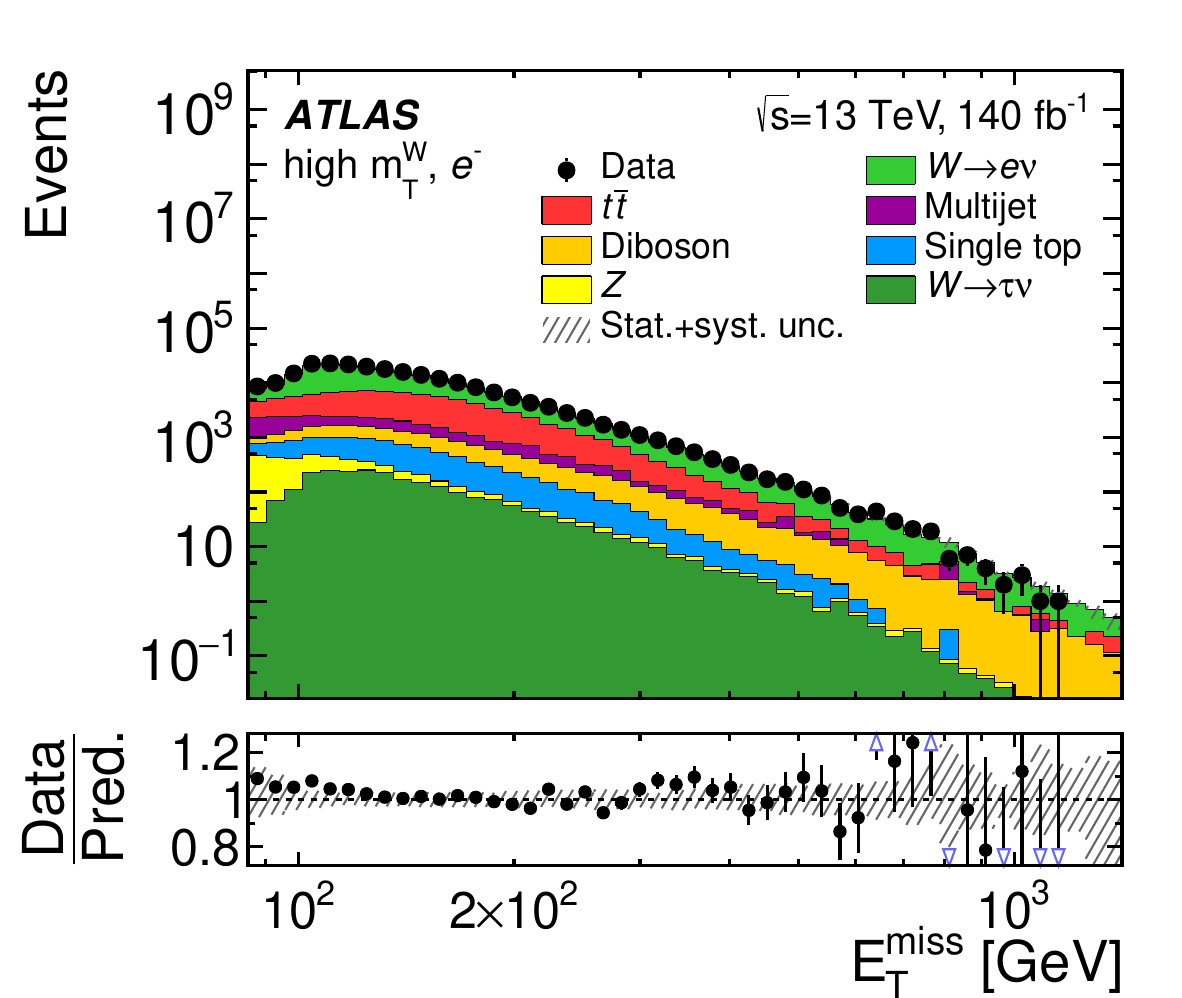}} \\[-2ex]
\subfloat[\label{fig:SR_ele_mtw_plus}]{\includegraphics[width=0.46\textwidth]{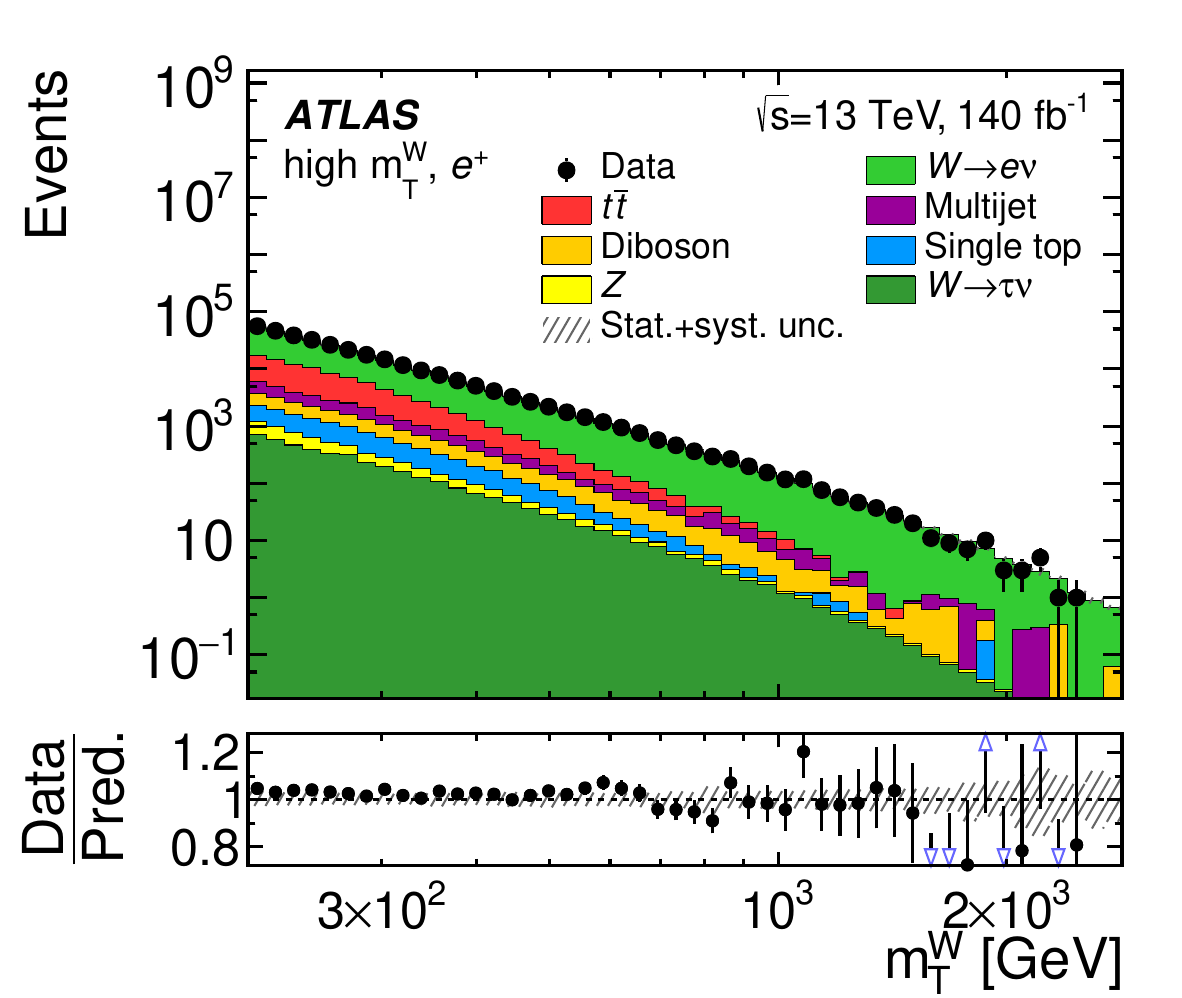}}
\subfloat[\label{fig:SR_ele_mtw_minus}]{\includegraphics[width=0.46\textwidth]{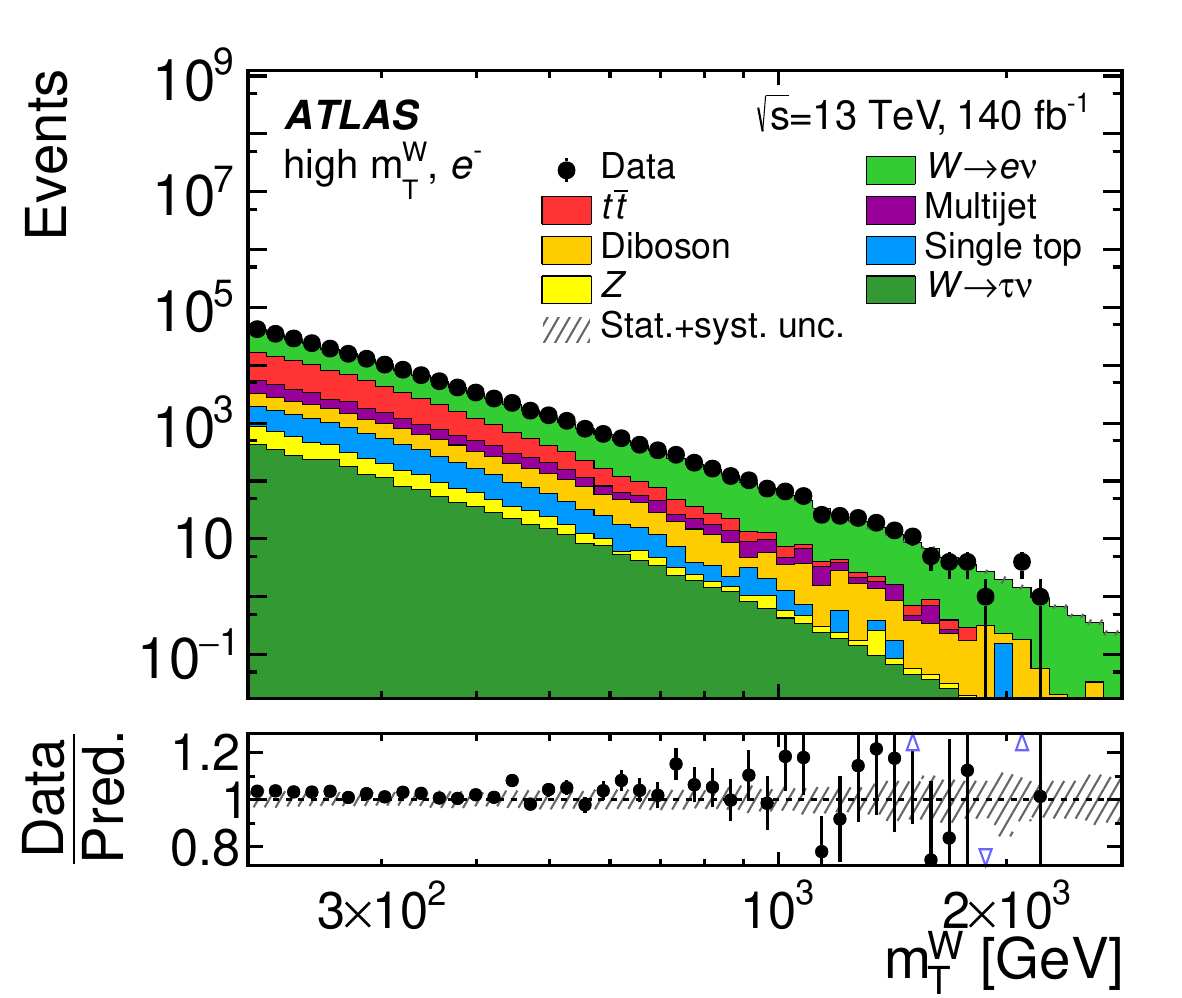}}\\
\caption{
Comparisons of the data and prediction in the signal
region for the (a, b) transverse
momentum of the lepton, (c, d) missing transverse momentum, and
(e, f) transverse mass \mtw\ in the (a, c, e) $e^+$ and (b, d, f) $e^-$  final
states.
Statistical and systematic uncertainties of the prediction (except for theoretical uncertainties on the signal)
are included in the uncertainty band.        A blue triangle in the bottom panel indicates data points outside the vertical range shown.
}
\label{fig:SR_elec_plusminus}
\end{figure}

\begin{figure}[bt!]
\centering
\subfloat[\label{subfig:SR_pt_plus}]{\includegraphics[width=0.46\textwidth]{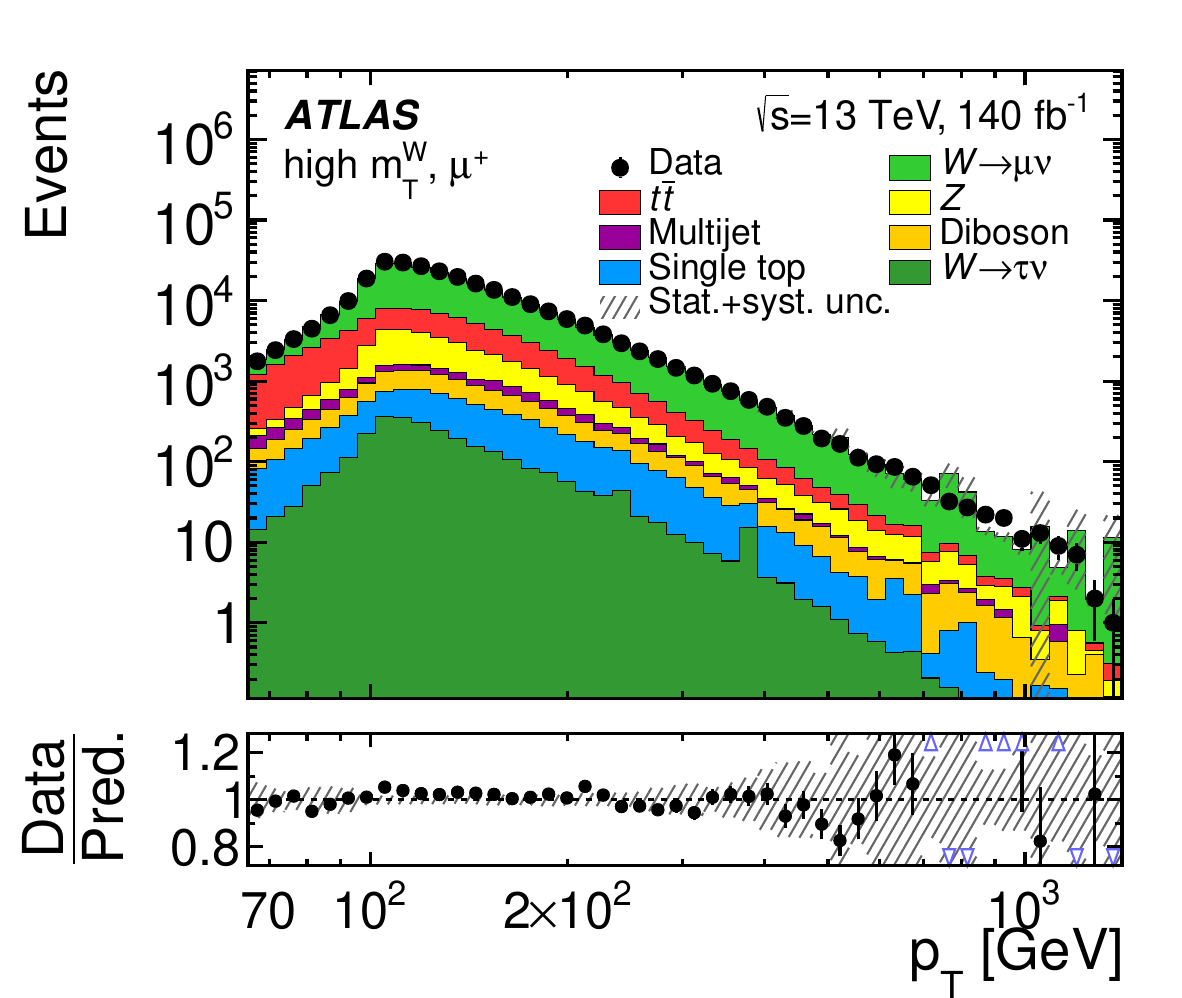}}
\subfloat[\label{subfig:SR_pt_minus}]{\includegraphics[width=0.46\textwidth]{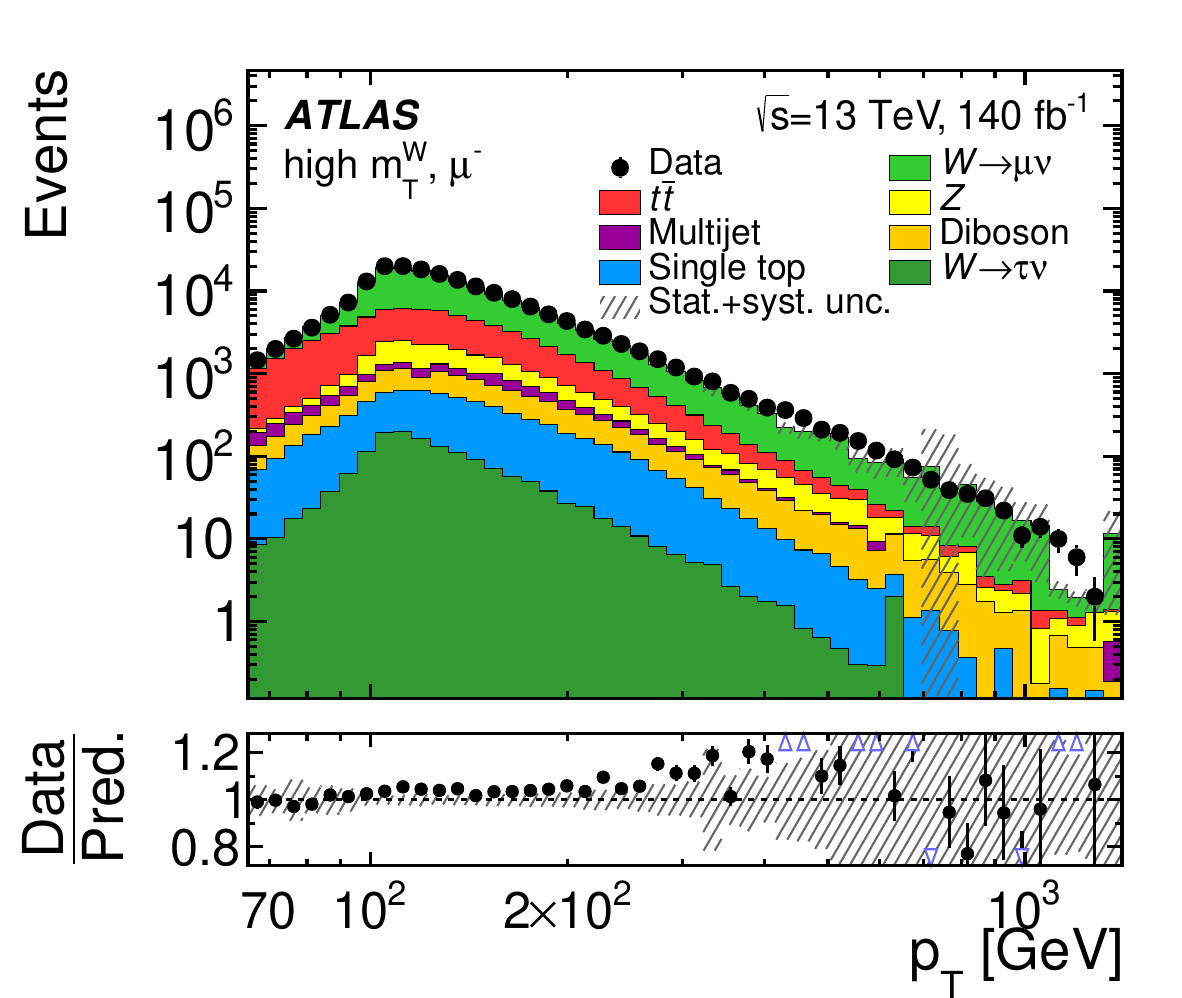}}\\[-2ex]
\subfloat[\label{subfig:SR_met_plus}]{\includegraphics[width=0.46\textwidth]{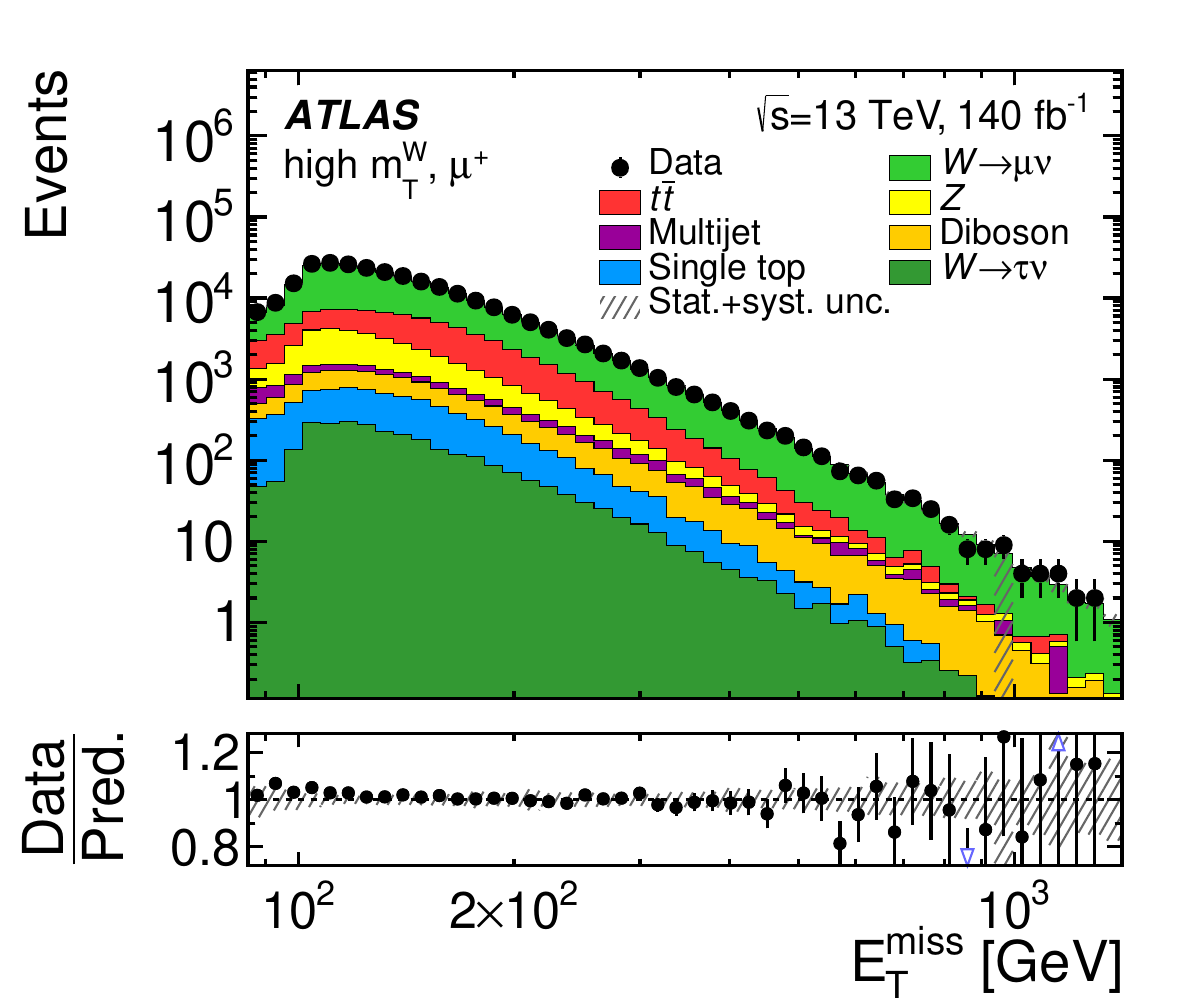}}
\subfloat[\label{subfig:SR_met_minus}]{\includegraphics[width=0.46\textwidth]{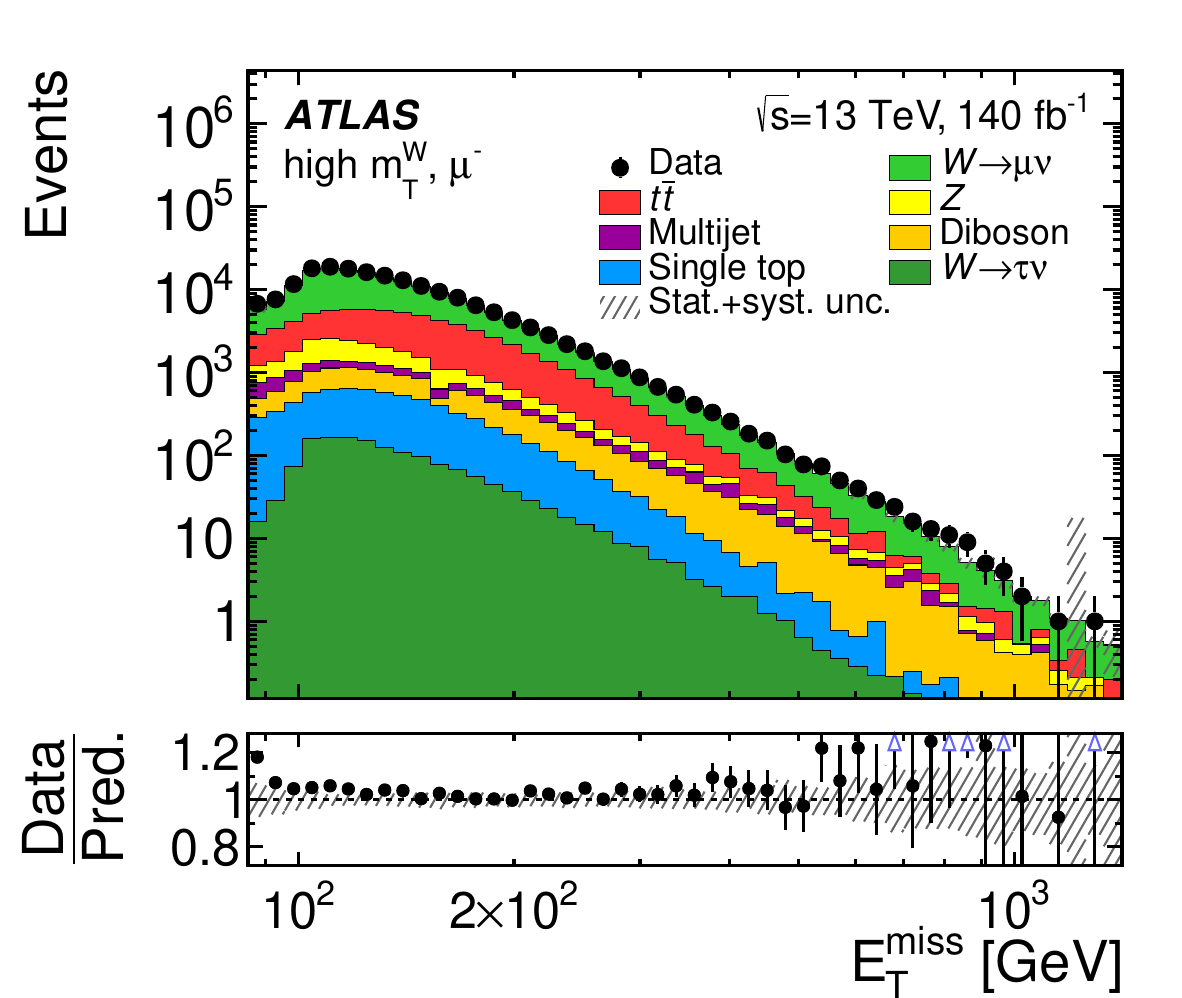}} \\[-2ex]
\subfloat[\label{subfig:SR__mymtw_plus}]{\includegraphics[width=0.46\textwidth]{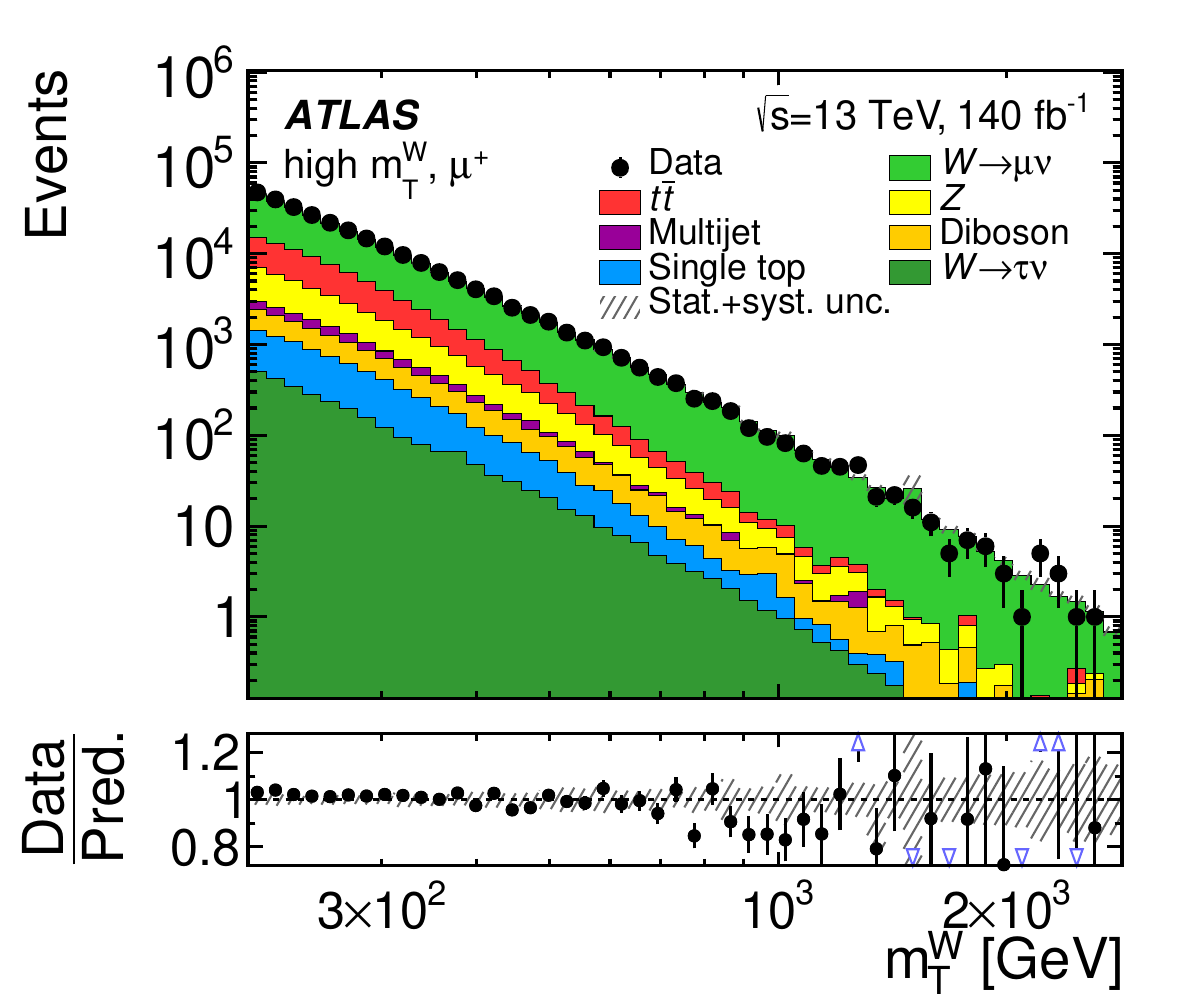}}
\subfloat[\label{subfig:SR__mymtw_minus}]{\includegraphics[width=0.46\textwidth]{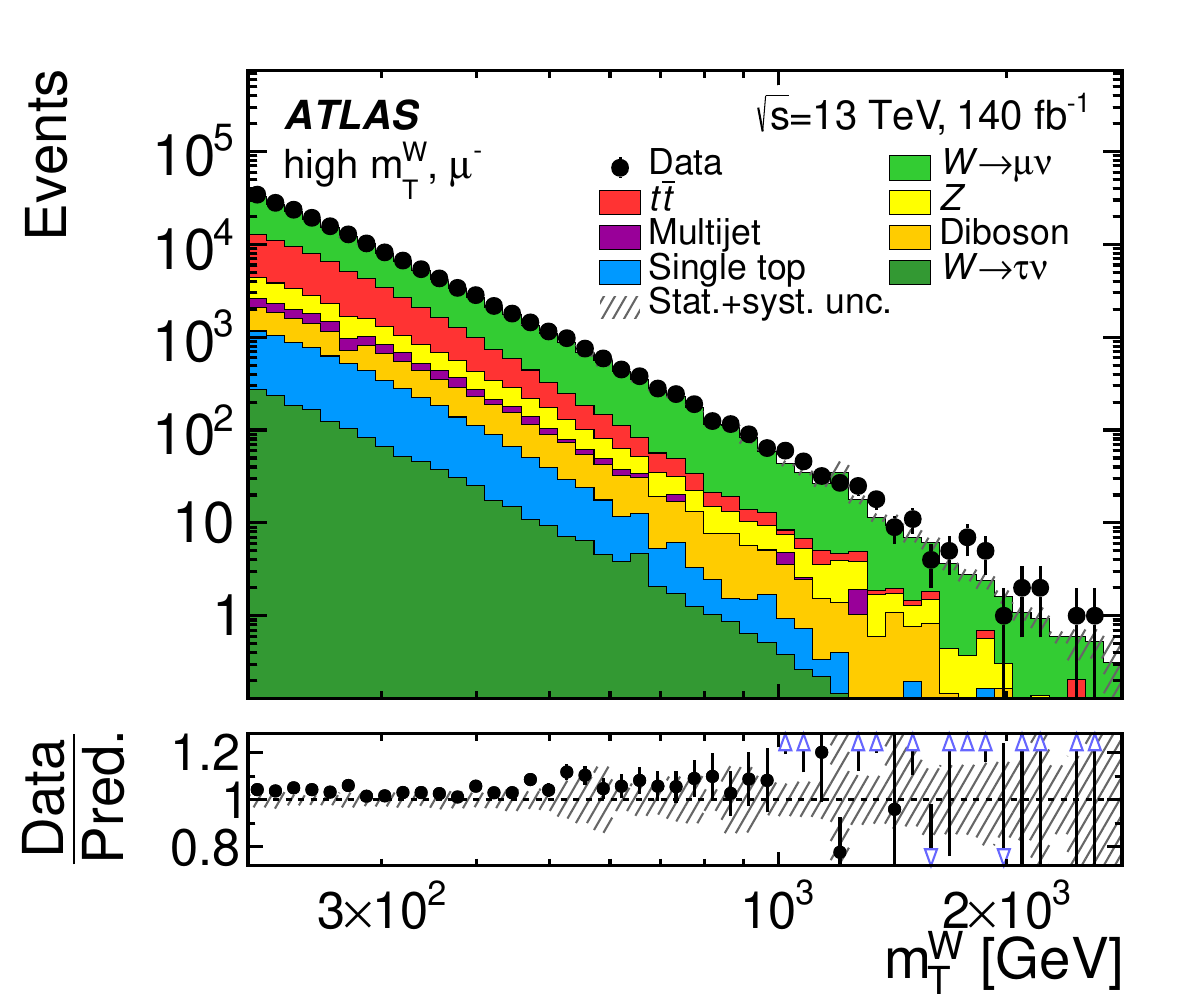}}
\caption{Comparisons of the data and prediction in the signal
region for the (a, b) transverse
momentum of the lepton, (c, d) missing transverse momentum, and
(e, f) transverse mass \mtw\ in the  (a, c, e) $\mu^+$ (right) and (b, d, f) $\mu^-$ final
states. Statistical and systematic uncertainties of the prediction (except for theoretical uncertainties on the signal) are
included in the uncertainty band.        A blue triangle in the bottom panel indicates data points outside the vertical range shown. }
\label{fig:SR_muon_plusminus}
\end{figure}

Figures~\ref{fig:SR_elec_plusminus}
and~\ref{fig:SR_muon_plusminus} compare data and expectation for the
reconstructed lepton $p_{\textrm{T}}$, \met, and \mtw\ distributions. In the phase space of the measurement the
three largest background contributions in the electron channel are the
top-quark (\ttbar\ and single top), multijet and diboson contributions which are found to be
approximately $23\%$, $4\%$, and $3\%$ respectively in the $W\to
e^+\nu$ channel.  The corresponding contributions in the
$W\to e^- \nu$ channel are $33\%$, $5\%$, and $4\%$.

In the muon channel the three largest background contributions are
from top-quark (\ttbar\ and single top), $Z$+jets, and diboson production. They
constitute about $20\%$ , $8\%$, and $2\%$ of the total expectation in
the $W\to \mu^+\nu$ channel. In the $W\to \mu^-\nu$ measurement the
contributions are found to be approximately $31\%$ , $6\%$, and $3\%$
of the expectation event yield. The multijet contribution in the muon
channels is below $2$\%.

The data event yields and spectra are found to be well described by
the expectation in both the
$W^\pm \rightarrow e^\pm \nu$ and the
$W^\pm \rightarrow \mu^\pm \nu$
channels.

\FloatBarrier


%
\section{Cross-section measurement}
%

\label{sec:cross-section}

The charged current Drell--Yan cross-sections are measured
for each charge and lepton flavour separately. The differential cross-sections are measured in 12 \mtw\ bins in the range
$200 \leq$ \mtw\  $ \leq 5000$~\GeV. Double differential cross-sections in five \mtw\ bins and up to $12$ $|\eta|$ bins
are also presented covering the region $0 \leq |\eta| \leq 2.4$, and $200 \leq$
\mtw\  $ \leq 2000$~\GeV, as listed in Table~\ref{tab:bins}.

The binning in \mtw\ is chosen such that the bin widths are larger than the detector resolution
to minimise migrations between bins and that the bins at the highest values of \mtw\ have moderate statistical uncertainties.
In particular, the muon channel is expected to have worse
resolution at high \mtw\ than the electron channel since the muon is
fully reconstructed from track-based measurements only. The muon
$p_{\textrm{T}}$ dominates the resolution for \mtw$>400$~\GeV ,
but below this region it is the \met\ which gives the
leading contribution. In the fiducial region the single differential bin purity, defined
as the yield ratio of particle-level signal MC events to reconstructed
signal MC events, is found to be typically $60\%$. In the muon channel
this decreases in the $2-5$~\TeV \mtw bin, whereas in the electron
channel it increases, reaching unity at the highest \mtw.

\begin{table}[t]
\caption{Kinematic bin edges for the two-dimensional cross-section measurements.}
\label{tab:bins}
\centering
\small
\begin{tabular}{c | l l l }
\hline
\hline
\mtw /~\GeV     & $200$ ,  $300$ ,  $425$                   \\
$|\eta|$        & $0.0$ , $0.2$,  $0.4$,   $0.6$,  $0.8$,   $1.0$, $1.2$,  $1.4$,   $1.6$, $1.8$, $2.0$ , $2.2$ ,   $2.4$       \\
\hline
\mtw /~\GeV     & $425$ , $600$ , $900$               \\
$|\eta|$        & $0.0$ , $0.4$,   $0.8$,   $1.2$,  $1.6$ ,  $2.0$ , $2.4$ \\
\hline
\mtw /~\GeV     & $900$ , $2000$               \\
$|\eta|$        &  $0.0$ , $0.6$,   $1.2$,  $1.8$ ,  $2.4$ \\
\hline
\hline
\end{tabular}
\end{table}

\subsection{Fiducial definition}
The measurements are unfolded to a common fiducial region that closely
matches the kinematic selection described in
Section~\ref{sec:selection}. It is defined at particle level as $|\eta| < 2.4$,
$p^{\ell}_{\textrm{T}}>65$~\GeV, the transverse momentum of the
(anti-)neutrino $p^{\nu}_{\textrm{T}}>85$~\GeV, and $200 \leq$ \mtw\
$ \leq 5000$~\GeV for the single differential measurements, or
$200 \leq$ \mtw\ $ \leq 2000$~\GeV for the double differential
measurements. The lepton charge is included in the definition of the
fiducial region.

The fiducial region is defined at the Born particle level prior to any
QED radiation from the final-state charged lepton.  The measurements
can also be corrected to the ``dressed'' particle level by multiplying
the reported cross-sections by the multiplicative factor
$C_{\textrm{fsr}}$ defined as the ratio of the dressed-level cross-section to the Born level, and is provided in the data tables (see
Section~\ref{sec:results}). The dressed particle level is obtained from MC
simulation by merging the Born-level leptons  with any prompt photons within a cone of $\Delta
R < 0.1$ around the charged lepton. The $C_{\textrm{fsr}}$ are close
to unity and found to vary weakly with \mtw.

\subsection{Unfolding procedure}
After the signal region selection has been applied and each of the
background contributions has been subtracted from the data, the observed distributions
are corrected for detector effects in an unfolding procedure~\cite{Adye:2011gm}.
The differential one-dimensional cross-section ${\textrm{d}
\sigma^j}/{\textrm{d} \mtw} $ is determined using
\begin{equation*}
\label{eq:diff_xsec}
\frac{\textrm{d} \sigma^j}{\textrm{d} \mtw} = \frac{1}{\Delta \mtw \cdot  \mathcal{L}_{\textrm{int}} \cdot  \epsilon^j} \sum_{i} R_{ij}^{-1} \cdot f^i_{\textrm{in}} \cdot (N_{d}^i-N_{b}^i),
\end{equation*}
where $ \Delta \mtw$ is the bin width,
$\mathcal{L}_{\textrm{int}} $ the integrated luminosity,
and
$N_{d}^i$ and $N_{b}^i$ are the numbers of observed data and
estimated background events in the $i$-th bin, respectively.
The factor  $f^i_{\textrm{in}}$ corrects for signal events that pass the
detector-level selection but not the fiducial selection, i.e. for
events which migrate into the measurement region, also called
``in-smearing''.
The factor $\epsilon^j$ corrects for signal events that pass the fiducial
selection but not the detector-level selection, accounting for selection
efficiency and acceptance.
The matrix $R_{ij}$ is the detector response matrix
containing the number of events in the $i$-th reconstructed
measurement bin and the corresponding prediction in the $j$-th
particle-level bin. It gives the probability of a selected event
reconstructed in a given bin $i$ to have originated at particle level from
the fiducial bin $j$.
The signal MC sample is used to obtain $f^i_{\textrm{in}}$, $\epsilon^j$ and $R_{ij}$.
The equation is trivially extended for the
two-dimensional case.

Iterative Bayesian unfolding~\cite{DAgostini:1995} is used to approximate the inversion of the matrix $R_{ij}$.
The procedure is regularised through the number of iterations in which
the particle-level prediction is used as a biased prior. The prior is
replaced by the unfolded output of the previous iteration, thereby
reducing the bias, but potentially increasing the variance. The
measurement is optimised using two iterations.

\subsection{Event migration studies}
The reconstructed event selection efficiency including acceptance
losses, is found to be $\epsilon^j \approx 70\%$ -- $90\%$ in the electron channel and $\epsilon^j \approx 50\%$ -- $75\%$
in the muon channel.
The factors $f_{\textrm{in}}^i$ are used to correct the data for
in-smearing in particular from the %
$W$-boson resonance production region into the kinematic phase space
of this analysis. As the correction is done without iteration
it is better to maximise its value, reducing the impact of
measurement biases. This was studied by reducing the \mtw selection to
$150$~\GeV and introducing an additional bin in the iterative
unfolding procedure covering the region $150 - 200$~\GeV. This
modification is found to increase the value of $f^i_{\textrm{in}}$ in
the measurement bin $200<\mtw < 250$~\GeV from $70\%$ to $85\%$, but
has no impact at larger \mtw.

The measurement is not extended down to \mtw$=150$~\GeV because
$f^i_{\textrm{in}}$ is only $40\%$ meaning $60\%$ of reconstructed events in this extended bin are
predicted to originate from mainly the $W$-boson resonance.  For
this reason the unfolding is performed with (a) a selection cut
of \mtw$>150$~\GeV, and (b) the introduction of a so-called ``shadow
bin'' with $150 \leq \mtw \leq 200$~\GeV which serves to stabilise the
unfolding. The final cross-section measurements are only reported
for \mtw$\geq 200$~\GeV.

\begin{figure}[tb!]
\subfloat[\label{subfig:2D_ele_fiin}]{\includegraphics[width=.48\textwidth]{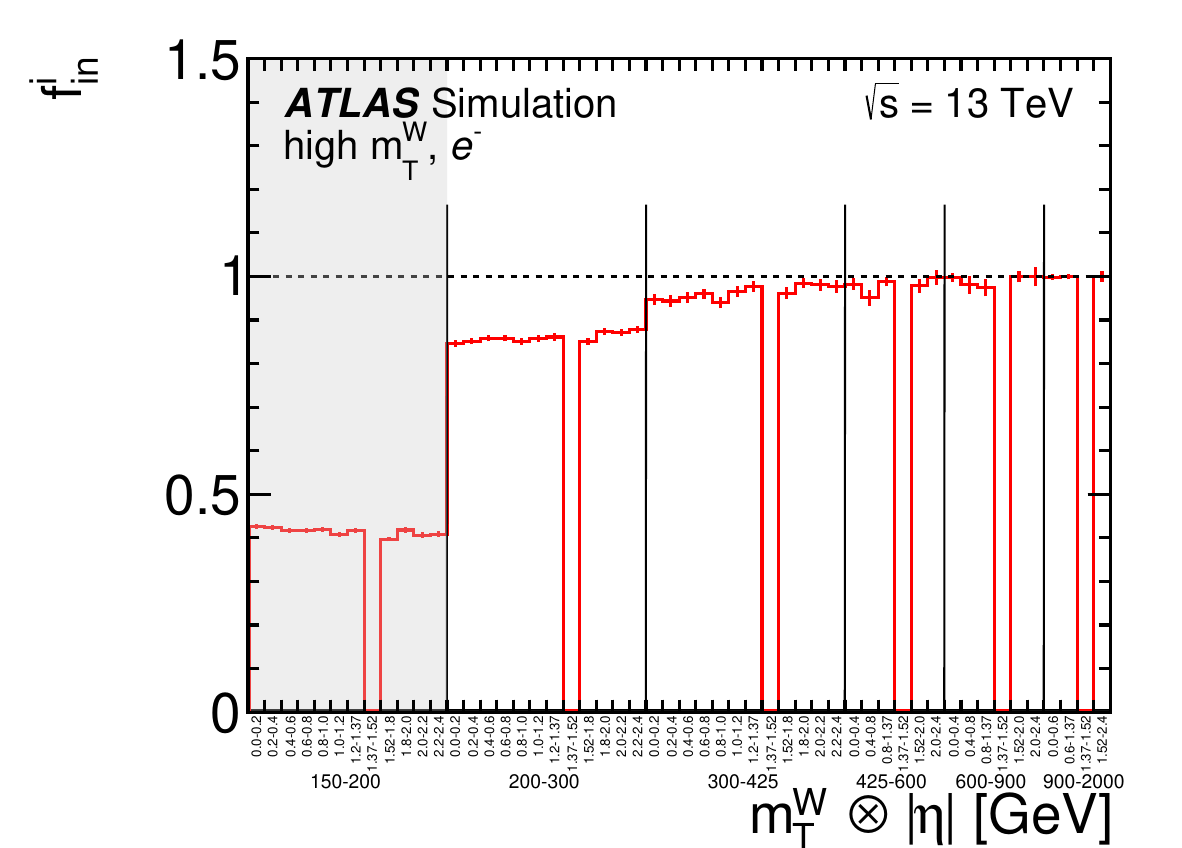}}
\subfloat[\label{subfig:2D_fiin}]{\includegraphics[width=.48\textwidth]{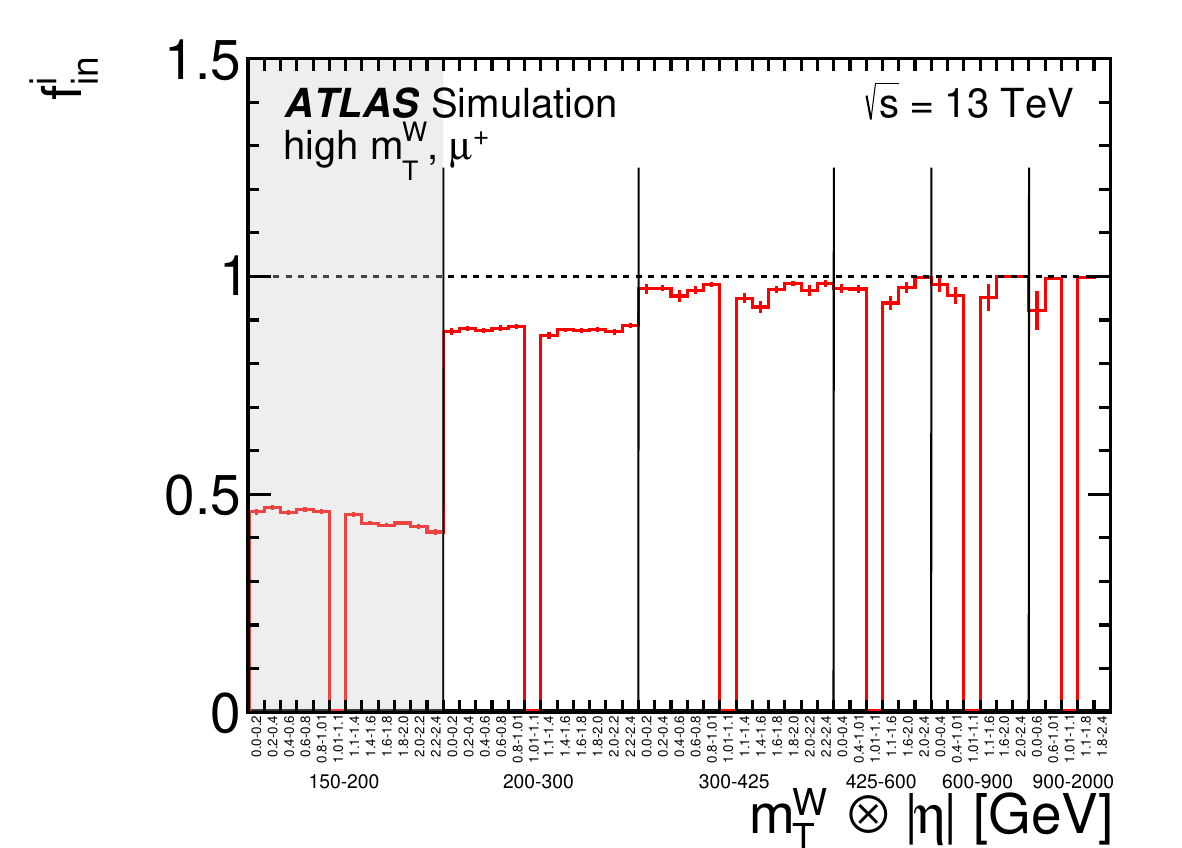}}
\caption{ The correction factor $f^i_{in}$ that corrects for events
not generated in the fiducial measurement range are
shown for a (a) negatively charged  electron and (b) positively charged muon.
On the horizontal-axis the double-differential measurement binning in the transverse mass of the W boson and the absolute pseudorapidity of the lepton
is shown.}
\label{fig:2D_correction_factors}
\end{figure}

Figure~\ref{fig:2D_correction_factors} shows the two-dimensional
correction factor $f^i_{\textrm{in}}$ for the negatively charged
electron and positively charged muon analyses.  The shadow bin (shaded region) is also
shown for comparison.  The in-smearing factors are found to be very
low in the shadow bin, but reach $85\%$ -- $90\%$ in the first measurement
bin. They are observed not to depend strongly on \abseta, as expected.

\subsection{Optimisation of number of iterations}
The robustness of the unfolding procedure is tested in a number of
studies using the signal MC. These include using the \Sherpa
MC as pseudo-data unfolded using the \textsc{Powheg+Pythia} MC sample;
unfolding the $W^+$ MC pseudo-data with the $W^-$ MC sample and vice versa; and
applying an arbitrary linear \mtw dependent event weight to the signal MC
which changes the predicted cross-section by $\pm 20\%$ and $\pm 50\%$
at low \mtw. In all cases the resulting biases and variances were
studied for between one and five iterations. The optimum number of
iterations is chosen to be two, mainly based on a scan of global correlation coefficients~\cite{Schmitt:2016orm}
but also on the linear \mtw dependent reweighting discussed above.

\subsection{Observed and predicted event yields}
\begin{figure}[tbh!]
\centering
\subfloat[\label{subfig:mtw_mybins_ele_plus}]{\includegraphics[width=.48\textwidth]{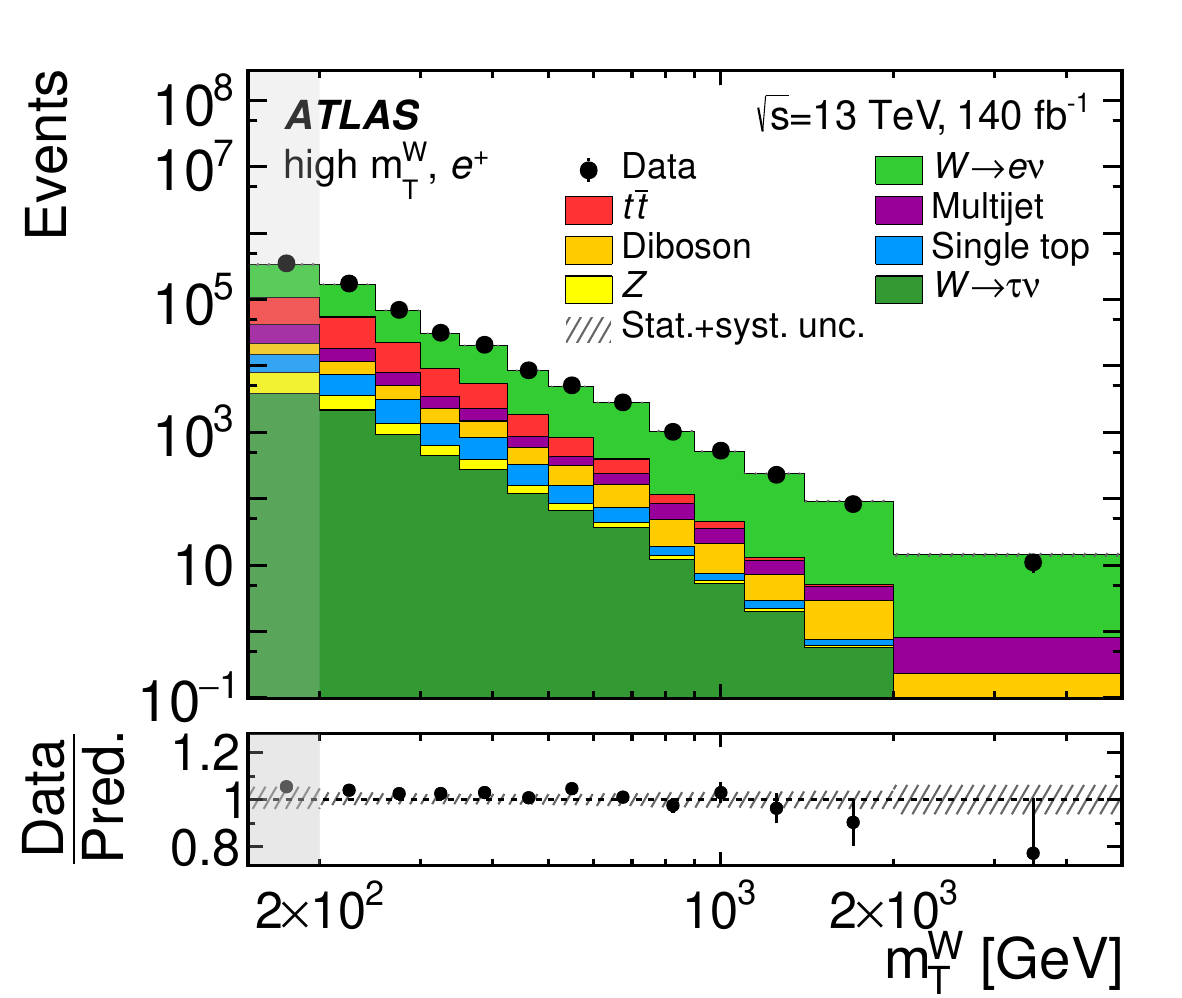}}
\subfloat[\label{subfig:mtw_mybins_ele_minus}]{\includegraphics[width=.48\textwidth]{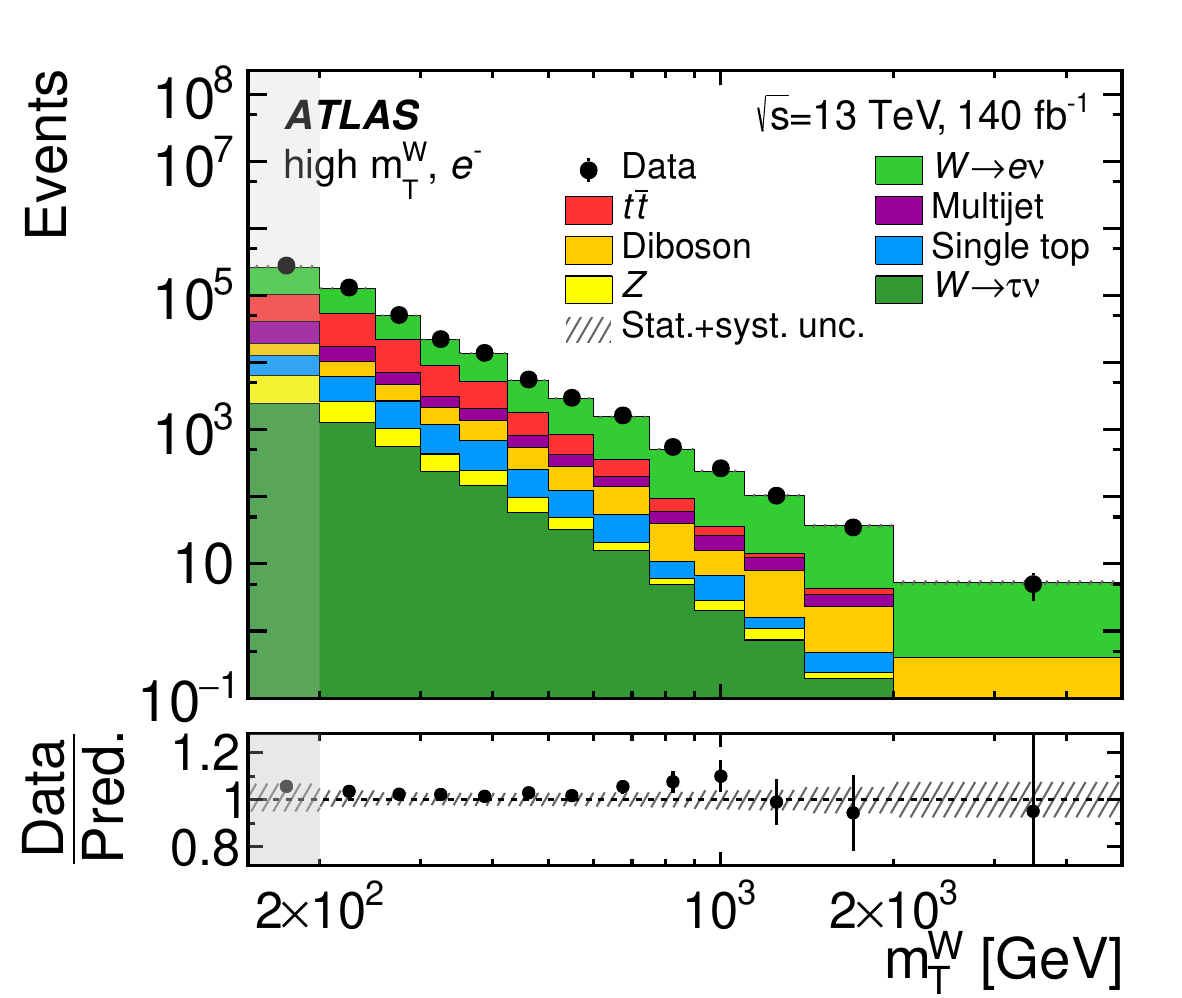}}  \\
\subfloat[\label{subfig:mtw_mybins_plus}]{\includegraphics[width=.48\textwidth]{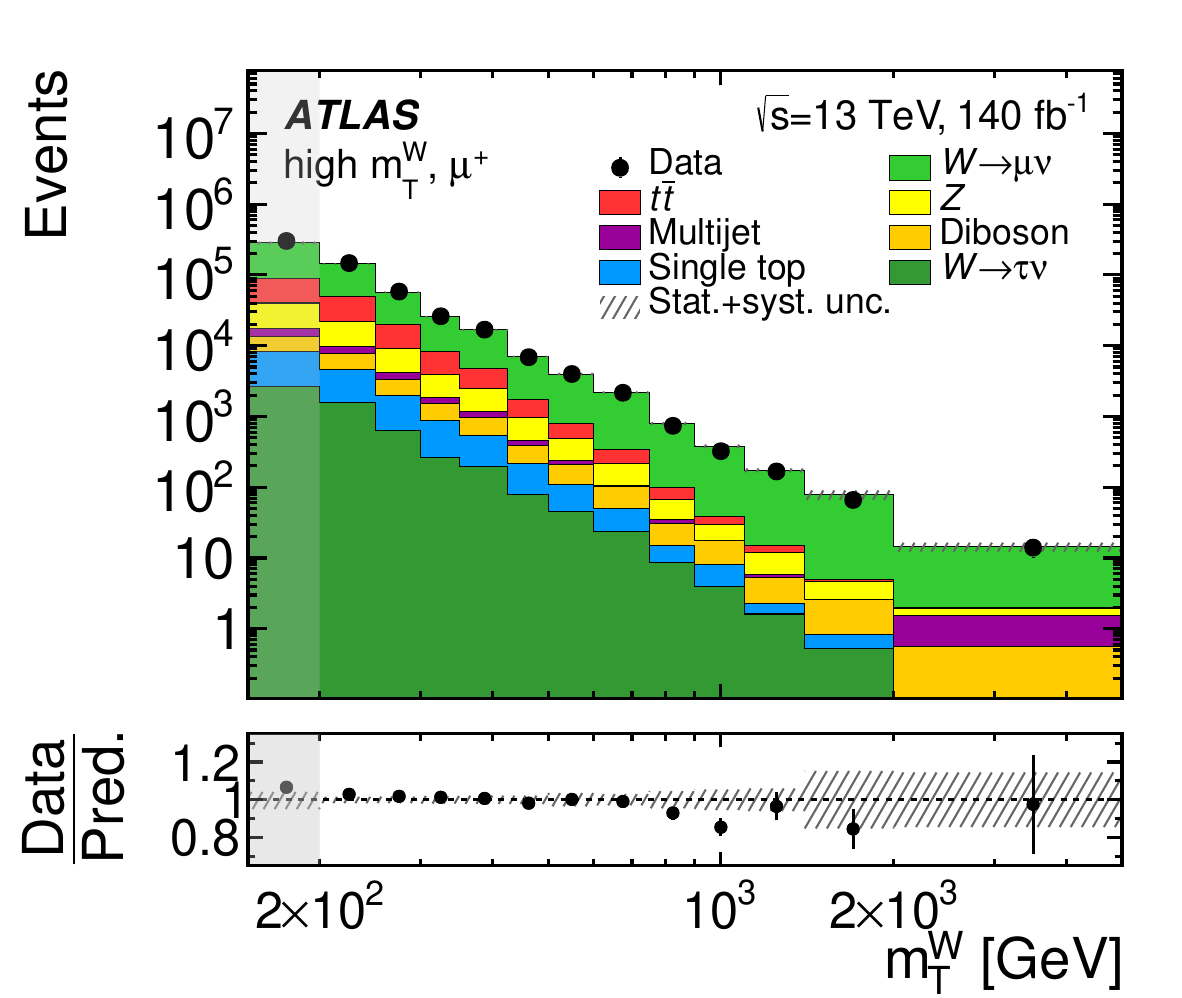}}
\subfloat[\label{subfig:mtw_mybins_minus}]{\includegraphics[width=.48\textwidth]{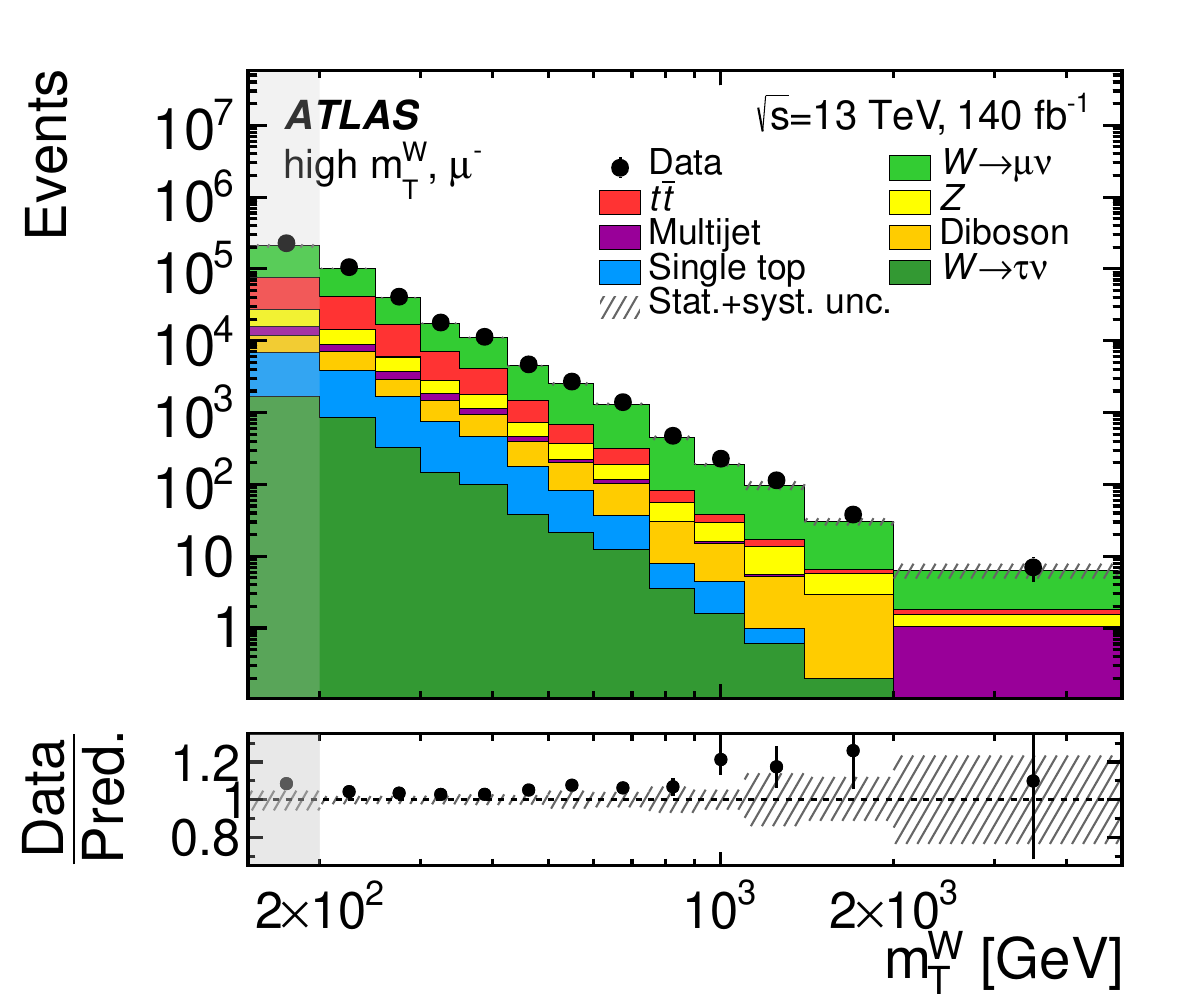}}
\caption{Comparisons of the data and the predictions in the signal region
shown in the measurement binning of
the transverse mass of the $W$ boson,
for the (a) $e^+$, (b) $e^-$, (c) $\mu^+$ and (d) $\mu^-$ final states.
Statistical and systematic uncertainties of the prediction (except for theoretical uncertainties on the signal)
are included in the uncertainty band.  }
\label{fig:mtw_1d}
\end{figure}

In Figure~\ref{fig:mtw_1d} the predicted event yields are compared to
data in the one-dimensional measurement binning separated by charge
and lepton flavour.
The uncertainty band represents the combined
statistical and systematic uncertainties.
The overall agreement of data and prediction, taking into
account the uncertainty band, is very good.

\begin{figure}[ht!]
\centering
\subfloat[]{\includegraphics[width=.48\textwidth]{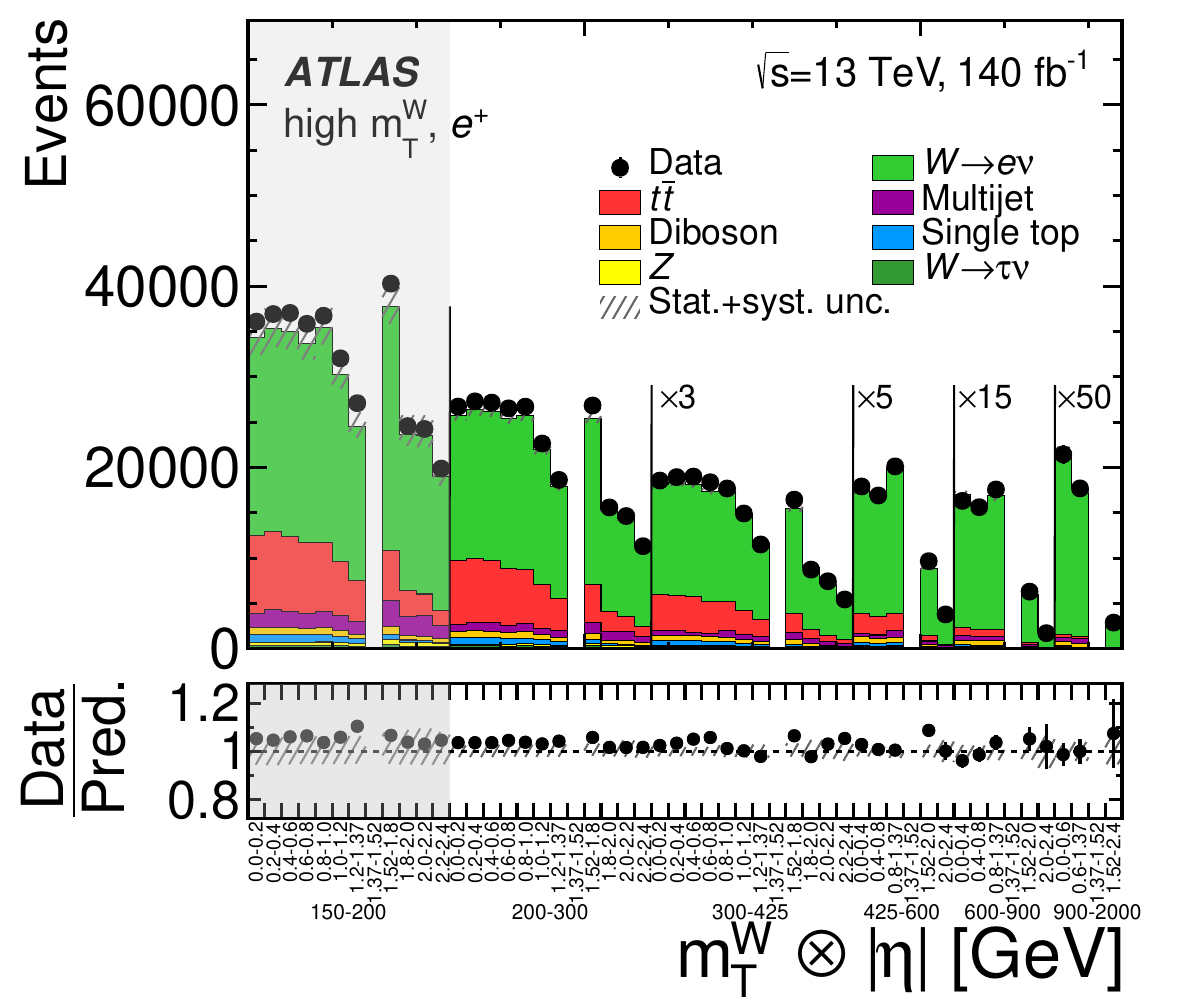}}
\subfloat[]{\includegraphics[width=.48\textwidth]{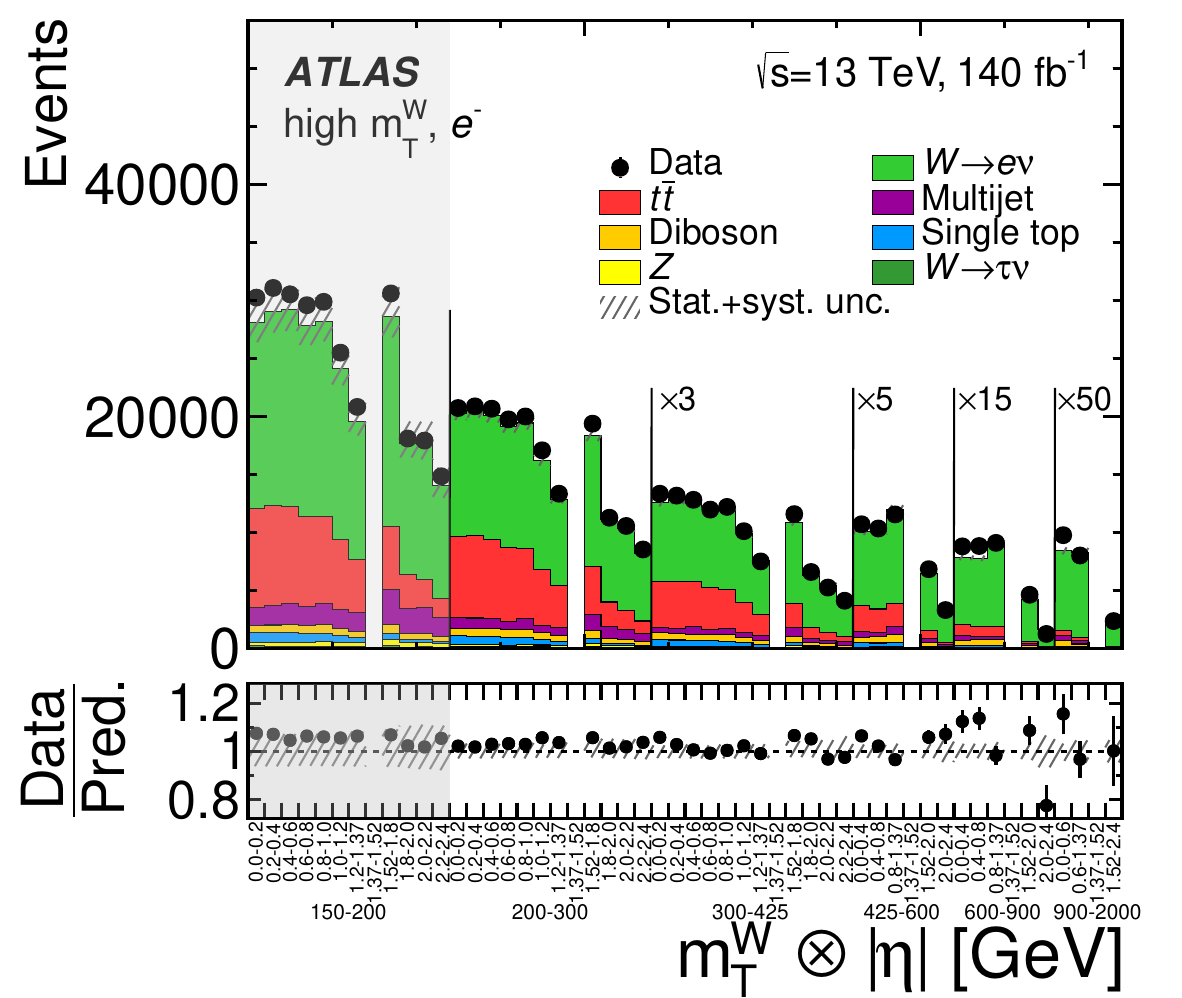}}\\
\subfloat[]{\includegraphics[width=.48\textwidth]{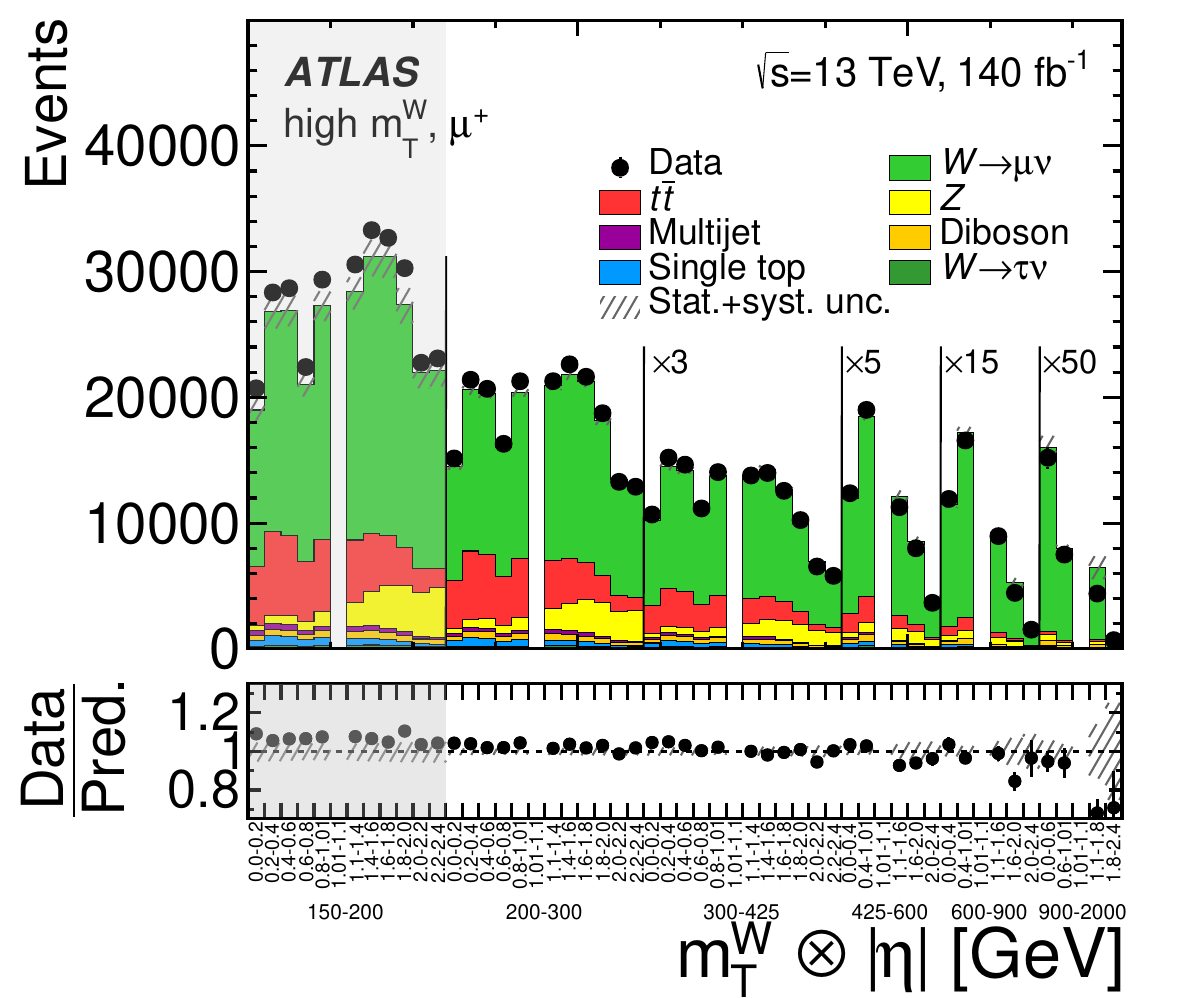}}
\subfloat[]{\includegraphics[width=.48\textwidth]{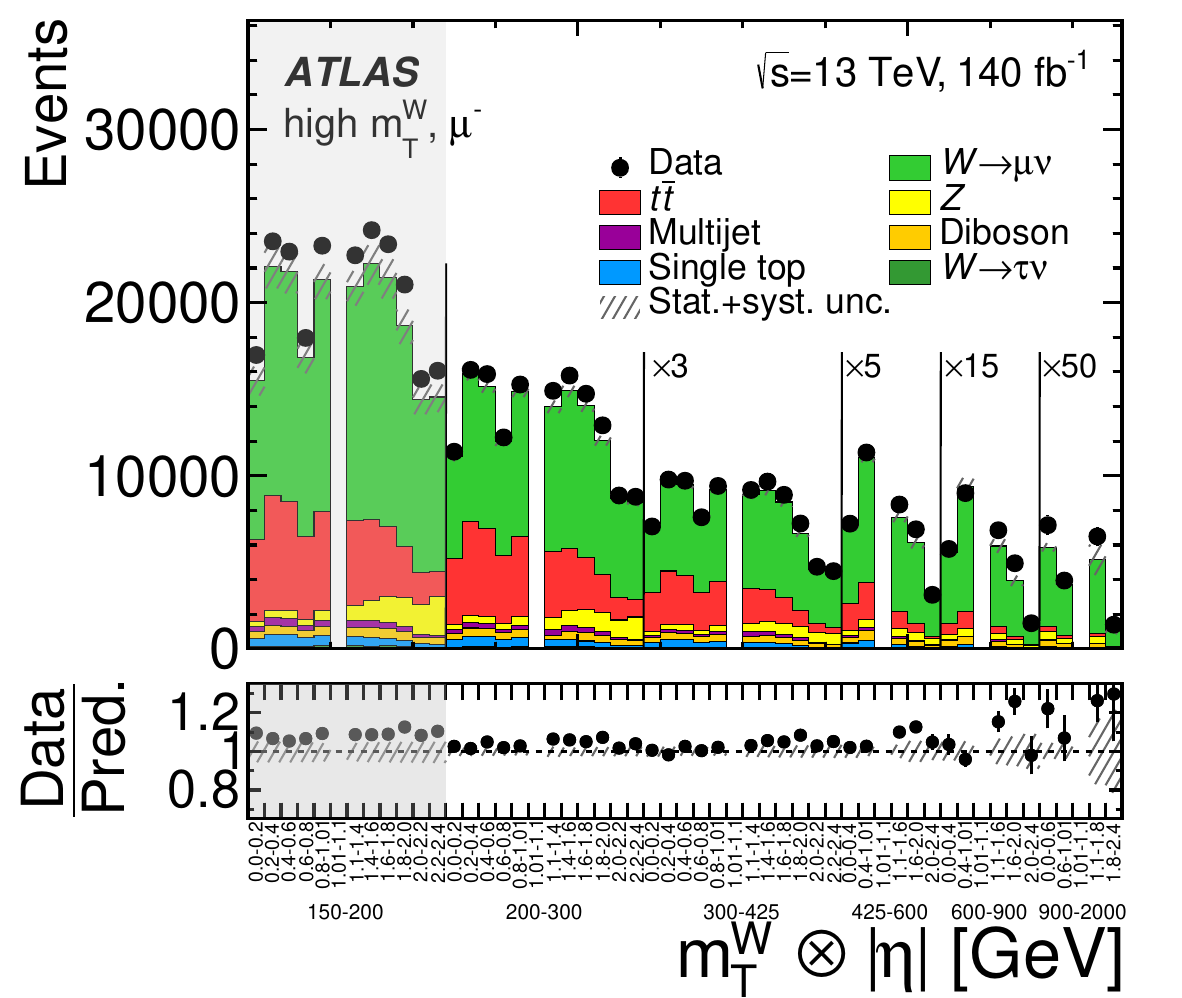}}\\
\caption{Comparisons of the data and the predictions in the signal region
shown in the double-differential measurement binning in
the transverse mass of the $W$ boson and the absolute pseudorapidity of the lepton for the (a) $e^+$, (b) $e^-$, (c) $\mu^+$ and (d) $\mu^-$ final states.
Statistical and systematic uncertainties of the prediction (except for theoretical uncertainties on the signal)
are included in the uncertainty band.
For better readability, the numbers of predicted and data events are scaled up by factors of 3, 5, 15 and 50, respectively,  in the last four \mtw bins.
}
\label{fig:mtw_eta_2d}
\end{figure}

In Figure~\ref{fig:mtw_eta_2d} the predicted two-dimensional event
yields are compared to data separated by charge and lepton flavour.
The distributions are shown in the final binning used to report the
cross-section measurements. The distribution in the shadow bin is also
shown for comparison.

\FloatBarrier


%
\section{Systematic uncertainties}
%

\label{sec:systematics}

The systematic uncertainties on the measurements are discussed
separately for those sources which arise only in the electron channel,
those which arise only in the muon channel, and those which are common
to both measurements.
Each source is classified as being correlated or uncorrelated
between measurement bins in a single channel. All bin-to-bin
correlated sources are also correlated between measurements of the
same lepton flavour but opposite lepton charge. In addition, all common
bin-to-bin correlated sources are considered to be correlated
between the two lepton flavour channel
measurements.

The correlated uncertainty contributions are propagated by the offset method in
which the values from each source are coherently shifted upwards and
downwards by one standard deviation and the magnitude of the change in
the measurement is computed.
Only the systematic uncertainties with an impact of at least
$0.5\%$ in at least one bin of any of the measurement distributions are propagated.
By default the computed systematic uncertainties are symmetrised unless
they are one-sided or strongly asymmetric.

\begin{figure}[htb!]
\centering
\subfloat[]{\includegraphics[width=.48\textwidth]{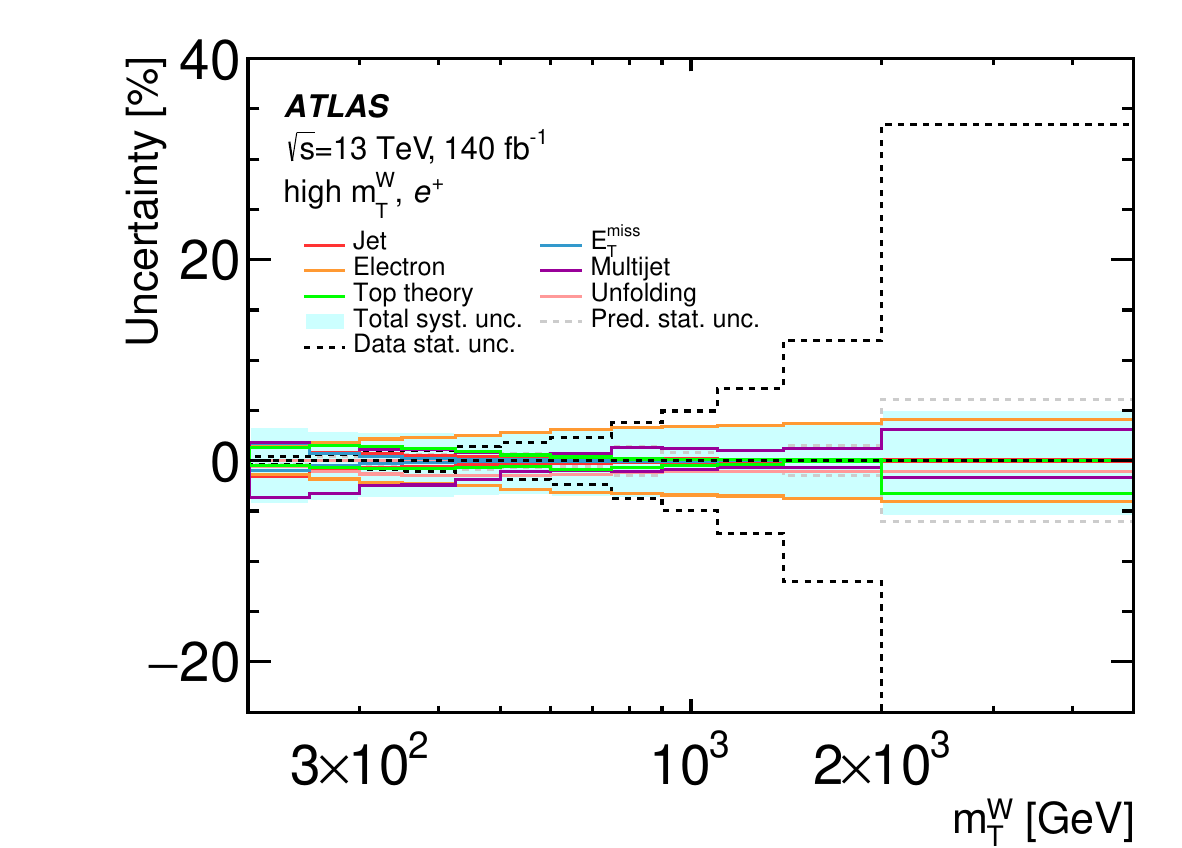}}
\subfloat[]{\includegraphics[width=.48\textwidth]{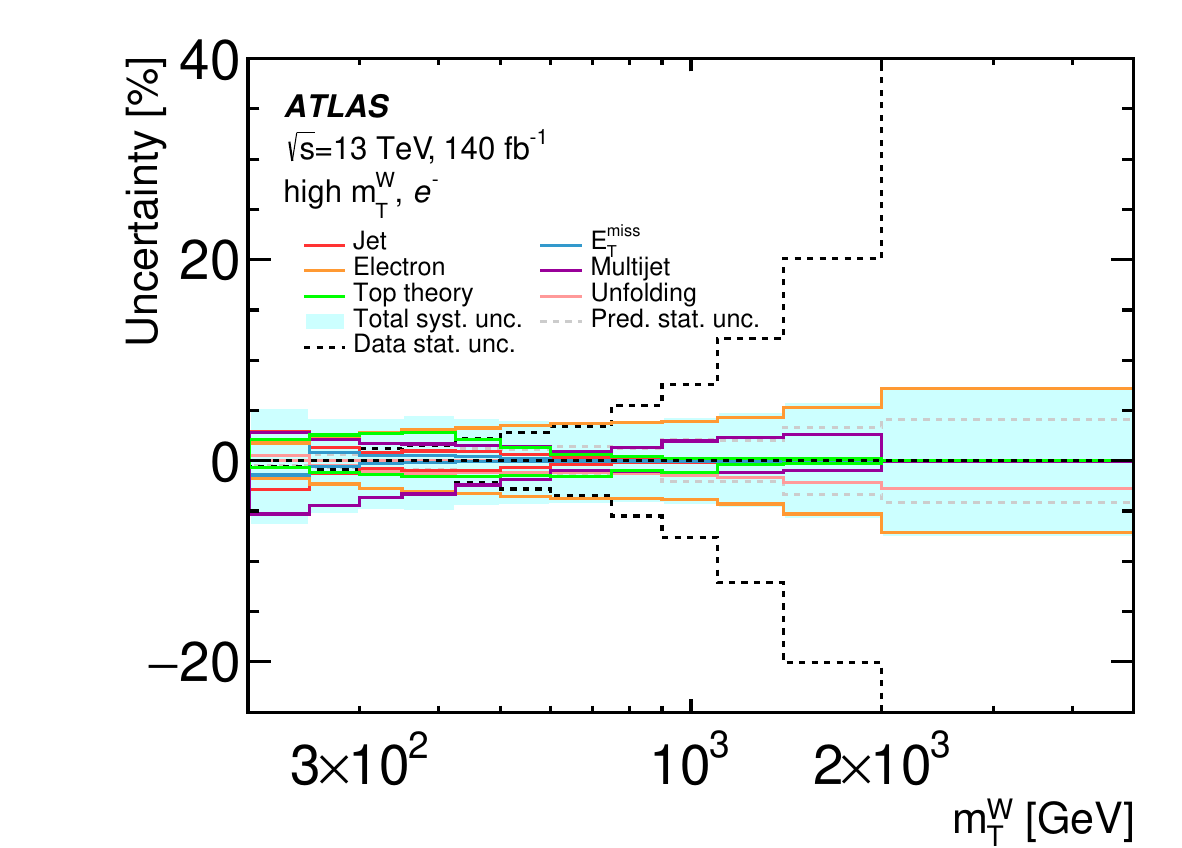}} \\
\subfloat[]{\includegraphics[width=.48\textwidth]{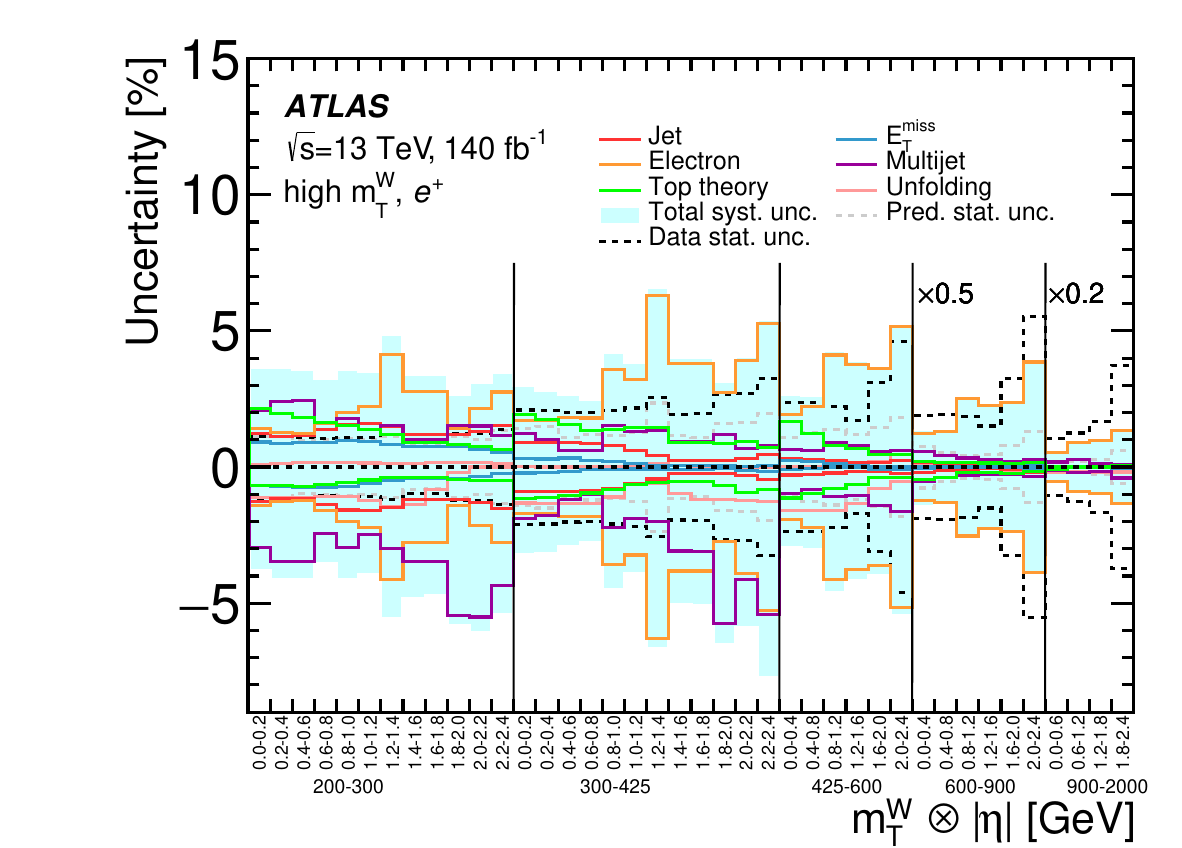}}
\subfloat[]{\includegraphics[width=.48\textwidth]{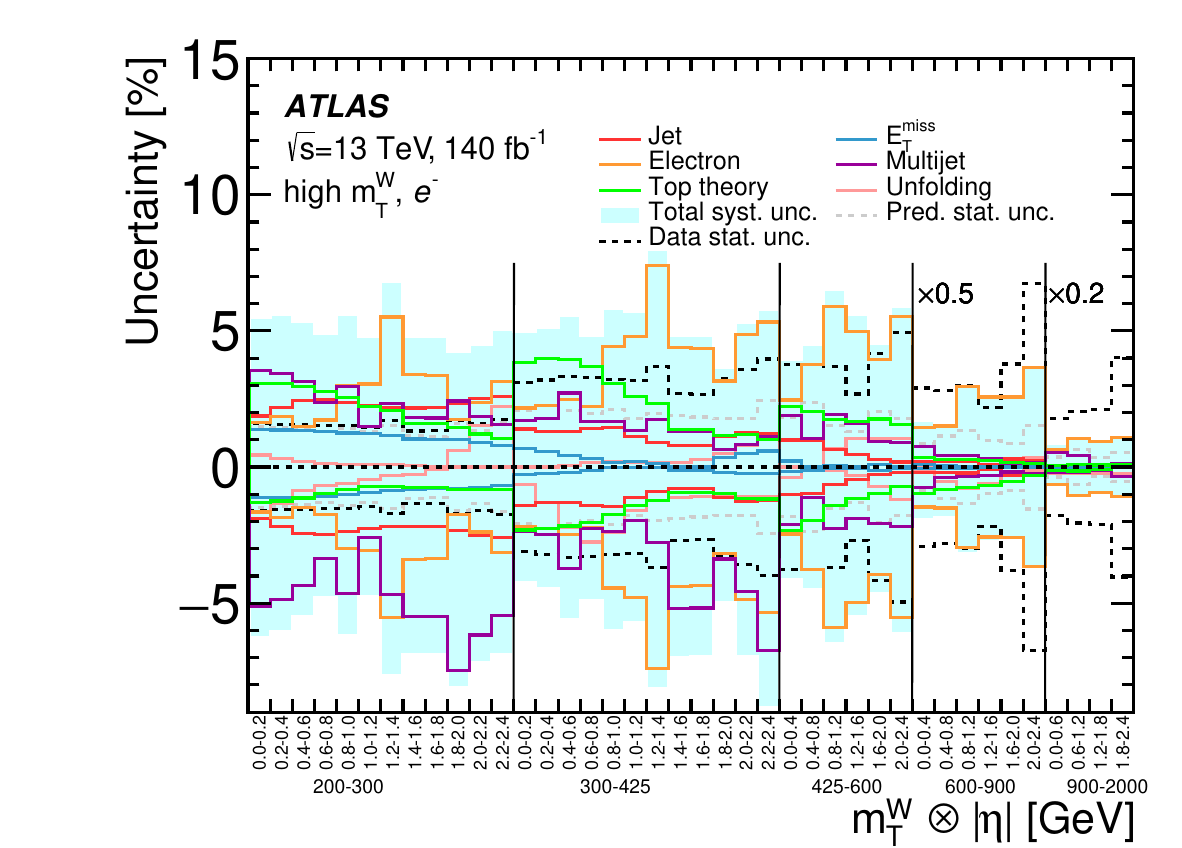}}
\caption{Summary of the uncertainties in the unfolded
(a, b) single-differential and (c, d) double-differential
cross-sections for the (a, c) $e^+$ and (b, d) $e^-$ final state.
Similar sources of uncertainties are combined
via their quadratic sum. The light blue band indicates the
total systematic uncertainty. In addition the
statistical uncertainties in the prediction and in the data are shown.}
\label{fig:syst_summary_ele}
\end{figure}

\begin{figure}[htb!]
\centering
\subfloat[]{\includegraphics[width=.48\textwidth]{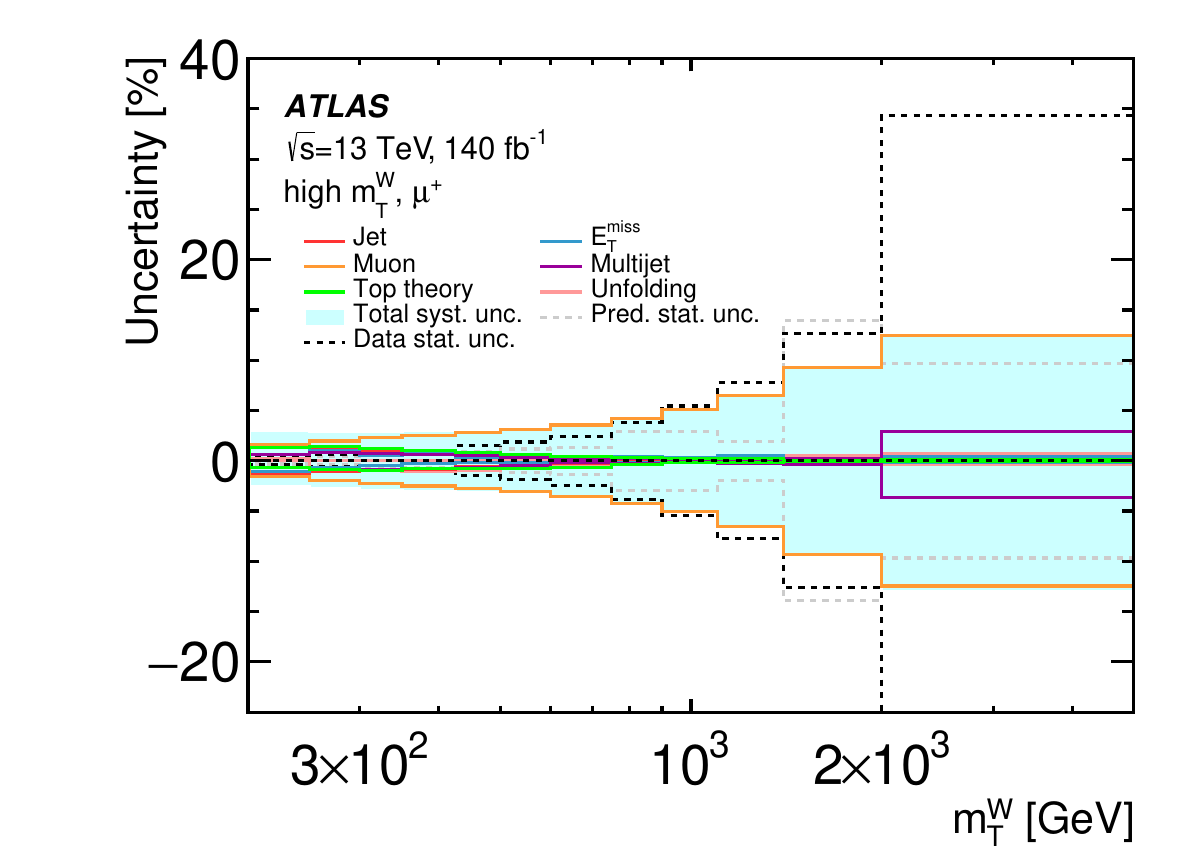}}
\subfloat[]{\includegraphics[width=.48\textwidth]{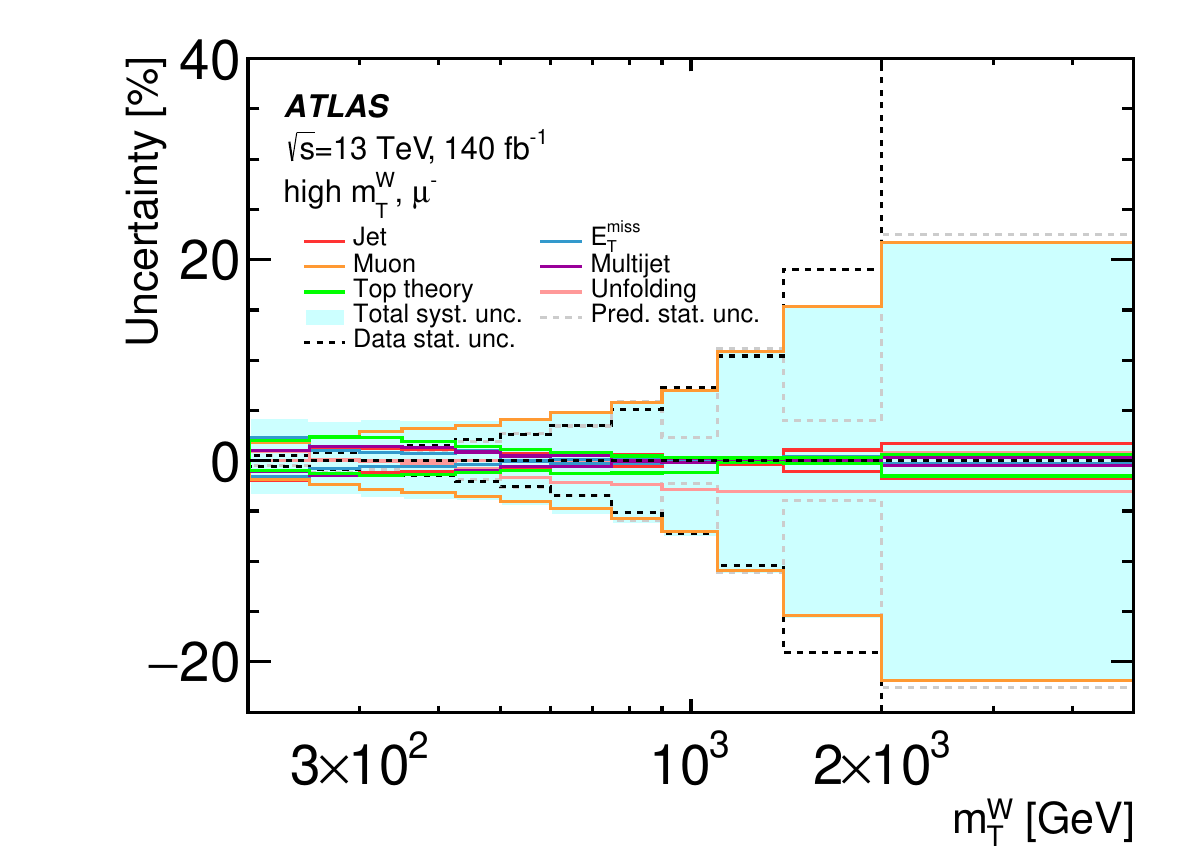}} \\
\subfloat[]{\includegraphics[width=.48\textwidth]{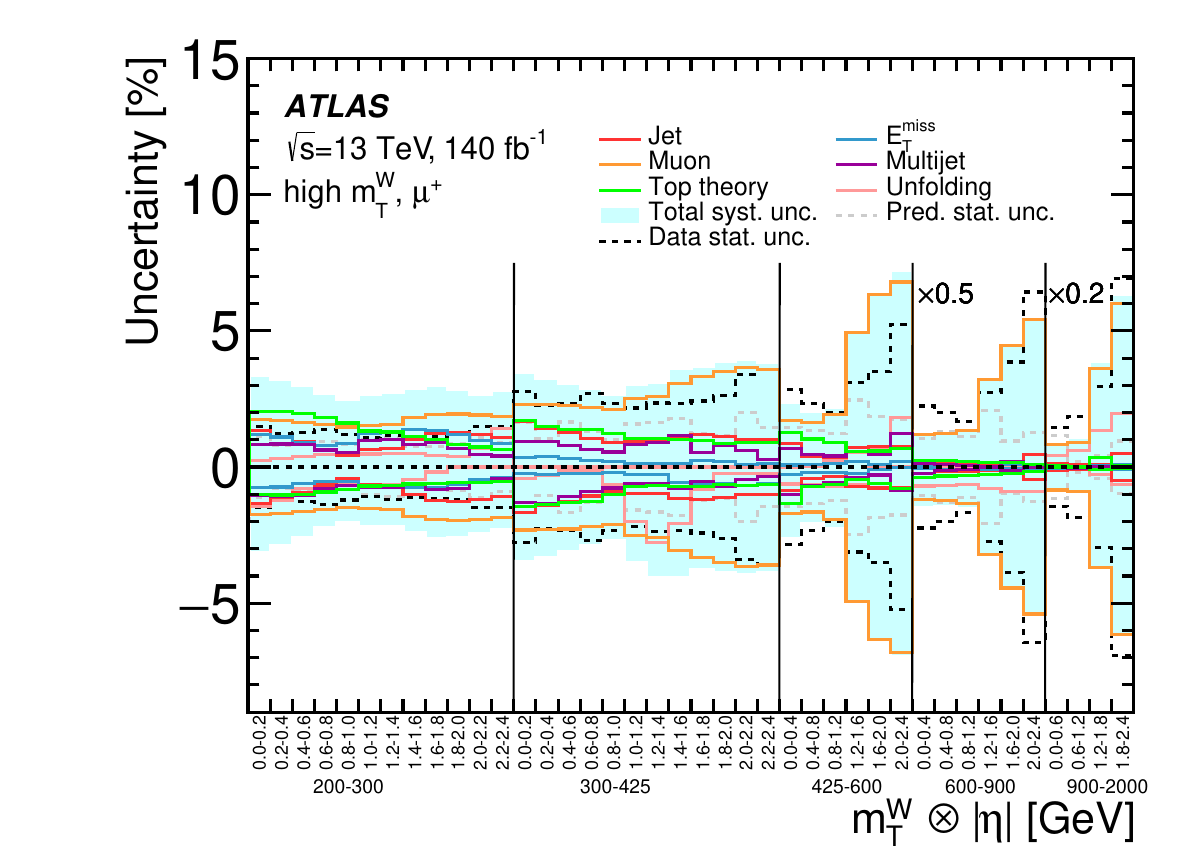}}
\subfloat[]{\includegraphics[width=.48\textwidth]{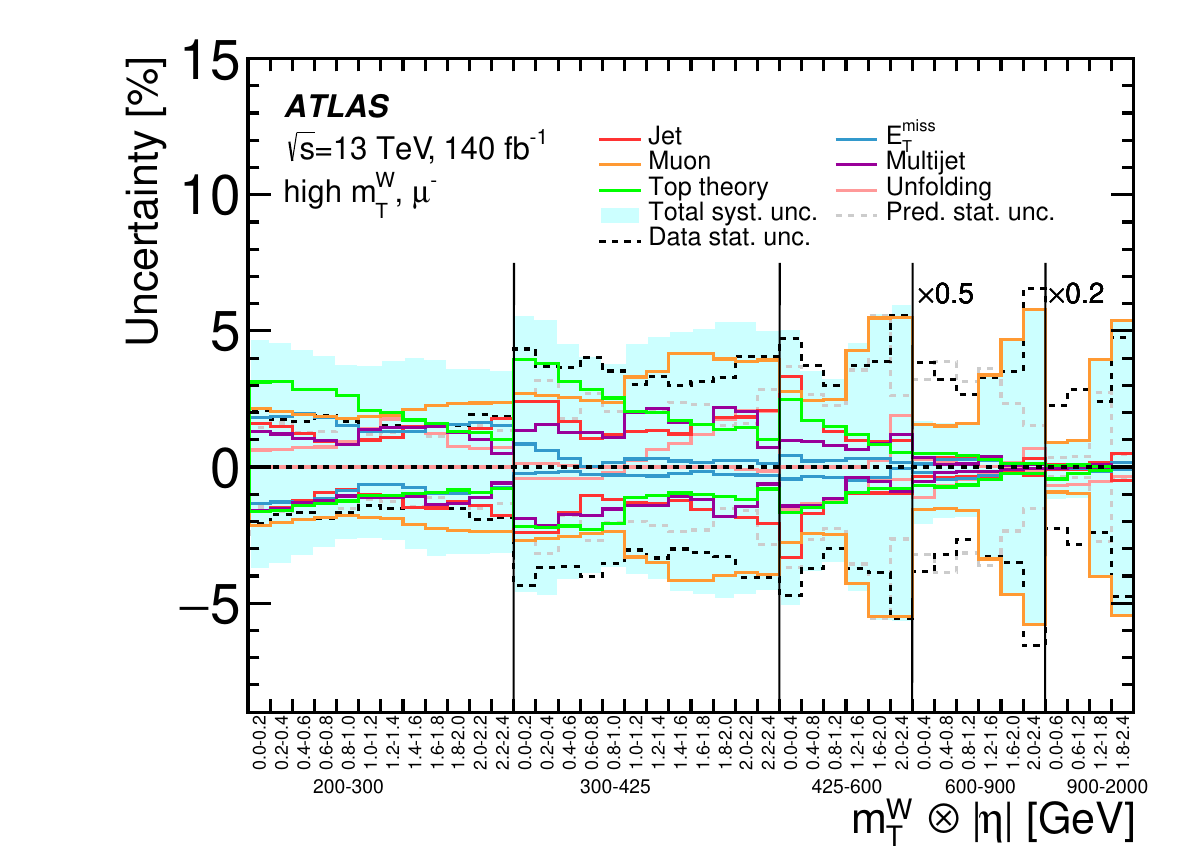}}
\caption{Summary of the uncertainties in the unfolded
(a, b) single-differential and (c, d) double-differential
cross-sections for the (a, c) $\mu^+$ and (b, d) $\mu^-$ final state.
Similar sources of uncertainties are combined
via their quadratic sum. The light blue band indicates the
total systematic uncertainty.  In addition the
statistical uncertainties in the prediction and in the data are shown.   }
\label{fig:syst_summary}
\end{figure}

A summary of the statistical and systematic uncertainties in the single- and double-differential cross-sections
are shown in Figure~\ref{fig:syst_summary_ele} for the electron channel and Figure~\ref{fig:syst_summary} for the muon channel.
The systematic uncertainties in the cross-section in the
electron channel are dominated primarily by the uncertainties in the
determination of the multijet background described in Section~\ref{sec:bg_elec_mj} and
by the electron energy scale.
In the muon channel the charge-dependent impact in the muon momentum scale calibration (sagitta bias) is the dominant
uncertainty at larger values of \mtw and $|\eta|$, while various sources contribute on a similar level
at smaller values of \mtw.

\subsection{Electron channel}
\label{sec:sysElec}

\paragraph{Multijet background}
The uncertainty in the multijet background is driven by the
statistical and systematic uncertainties in the fake efficiency estimate,
as the real efficiency is taken from the MC simulation directly.
The fake efficiency depends on the selections used to define the
multijet enriched regions and the modelling of the
backgrounds with real electrons in these regions.
Biases from the selection are estimated by varying the \met range from
\met $<65$~\GeV to \met $<30$~\GeV and 30 < \met $<65$~\GeV
and by asking for an additional jet
well separated from the electron.
The modelling of the real electron contributions is varied by using the alternative Sherpa signal sample
normalised to the \textsc{Powheg+Pythia} prediction and by scaling the cross-sections of the MC samples by $\pm10\%$.
The value of 10\% used here is larger than the uncertainties used in the main analysis due to the unusual phase space
of the fake enriched region, see Section~\ref{sec:bg_elec_mj}.
Finally, as \met\ itself depends on the sum of transverse momenta
of all reconstructed and calibrated objects in the event an ambiguity can arise depending on which
calibration is applied to a certain object. Applying e.g. a jet calibration or an electron calibration to
a certain reconstructed object typically results in a factor of two difference in transverse momentum.
While not a problem for well identified (``tight'') objects this is different
for objects that only satisfy the less stringent (“loose”) requirements utilised
in the matrix method, for which the true nature (electron or jet) is unknown.
A systematic uncertainty is assigned by attempting
to estimate the actual contribution of true electrons and jets as a function of the event kinematics.

While asking for an additional jet leads to negligible differences, all other uncertainties
have a $1\%$-$3\%$ impact on the measurement at low \mtw, which is decreasing for higher values of \mtw.
They constitute the dominant source of uncertainty in the measurement at low \mtw.

\paragraph{Electron trigger, reconstruction, identification and isolation efficiencies}
The efficiencies of the trigger, electron reconstruction and
identification, and of the isolation criteria are estimated with $Z\to
e^+e^-$ data using a tag-and-probe method~\cite{EGAM-2021-01,TRIG-2018-05}. The
uncertainties are given in a scheme using a single nuisance parameter
for each of the isolation, reconstruction and trigger efficiency
contributions and two nuisance parameters for the charge misidentification.
The remaining identification uncertainties are given by
16 sources correlated between measurement bins. An additional 24 nuisance parameters describe the impact of sources uncorrelated
in \abseta and \pt, but fully correlated between electron charges.

Among those, the largest uncertainties are due to the isolation and identification efficiencies.
The former is largest for \mtw$\sim 1$~\TeV where it reaches $2\%$.
The latter is dominated by one bin-to-bin correlated source in
particular which reaches $1\%$ in size for \mtw above $0.5$~\TeV, while the uncorrelated
sources contribute typically at the level of 1\% in their respective \abseta bin.

\paragraph{Electron energy scale and resolution}
The determination of the electron energy scale and resolution is
presented in Ref.~\cite{EGAM-2018-01}. The related uncertainties are
described by 25 nuisance parameters for the energy scale and a
further nine for the resolution. The energy scale is the larger
contribution to the measurement uncertainty rising from $0.5\%$ at low
\mtw to $3\%$ in the highest \mtw bin.

\subsection{Muon channel}
\label{sec:sysMuon}

\paragraph{Multijet background}
As for the electron channel, the multijet background estimation in the
muon channel depends on the selections used to define the fake
enriched region and the modelling of the real electron contribution in this
region.  %
Biases from these contributions are estimated by varying the \met
range used in the same way as in the electron channel and, additionally,
varying the jet multiplicity, the impact parameter selection, or  the
$\Delta\phi(\mu,\met)$ selection individually.
The modelling of the real muon contribution is varied in the same way as described
for the electron channel in Section~\ref{sec:sysElec}.
All of these uncertainty estimates are found to contribute
less than $1\%$ to the final cross-section measurement.

\paragraph{Muon trigger, identification, vertex association and isolation efficiency}
The muon efficiency corrections are obtained for the full Run~2 data
using $Z\to\mu^+\mu^-$ data and simulation with a tag-and-probe method
described in Refs.~\cite{MUON-2018-03,TRIG-2018-01}.
The vertex association efficiency corrects for losses
arising from the $d_0$ and $z_0$ impact parameter selection
criteria. Each of these contributions is associated with two nuisance
parameters for a correlated and uncorrelated statistical component of
the uncertainty. The largest impact on the measurement is from the
identification efficiency which rises from $1\%$ at low \mtw to $5\%$
in the highest bin.

\paragraph{Muon momentum scale and resolution}
The uncertainties in the determination of the muon momentum and
resolution are discussed in Ref.~\cite{MUON-2022-01}, and are determined
using similar samples as for the muon efficiency corrections above.
In total six nuisance parameters are used to describe the impacts of these uncertainties.
They relate to separate resolution effects in the ID and muon spectrometer,
biases in the momentum scale,
and three contributions related to the sagitta bias, which are
anti-correlated between the muon charges. The three contributions are
a global residual bias, a specific component for the high-\pt extrapolation of muons with $|\eta|>1.1$ and $p_{\textrm{T}} > 450$~\GeV,
and a contribution from local biases in $|\eta|$ and $\phi$ of the muon. The
first two of these three contributions have the largest impact and
dominate the overall uncertainty for
\mtw$>600$~\GeV, reaching up to 10\% at the highest values of \mtw for the former and up to 20\% at the highest values of \mtw and \abseta
for the latter.

\subsection{Common uncertainties}
\label{sec:sysCommon}

\paragraph{Top-quark background}
The \ttbar background uncertainties are divided into a number of
contributions. %
The influence of hadronisation and fragmentation models is determined by
re-showering the nominal top-quark generator-level events with
\HERWIG[7]~\cite{Bellm:2015jjp}.
The sensitivity to initial (ISR) and final (FSR) state QCD
radiation is determined by varying a
parameter of the A14 tune~\cite{ATL-PHYS-PUB-2014-021} corresponding to a variation of $\alpha_S$ for the ISR,
and by variations of the renormalisation scale in the FSR part of the PS sample.
The \POWHEG \hdamp parameter is a resummation damping factor
that effectively regulates the high-\pT radiation against which the \ttbar system recoils.
It is changed from $1.5$ to $3$ times the top-quark mass.
The $\pt$-hard parameter controlling the matching of \textsc{Powheg} matrix elements to the PS
is also varied.
The PDF uncertainty is evaluated by using the RMS of one hundred replicas of the nominal \NNPDF[3.0nlo] PDF set.

The relative size of the systematic uncertainties discussed above are evaluated before
the reweighting of the \ttbar MC sample to the NNLO QCD calculations including NLO EW corrections described in Section~\ref{sec:MC}.
The sequence (top \pt, \ttbar mass) in which the reweighting is performed is another source of uncertainty.
Potential differences in the modelling of additional radiation between
the NNLO calculation and the PS are also taken into account,
separately for the \pT of the top quark and the \ttbar mass.  The
choice of PDF has also been varied in order to assess the impact on
the EW corrections.  Finally, separate variations of the factorisation
and renormalisation scales $\mu_R$ and $\mu_F$, in the NNLO
calculation are taken into account by varying the scales independently
by factors of two.

The interference between the $tW$ and $t\bar{t}$ production processes
is accounted for in the simulation through the diagram subtraction
scheme~\cite{ATL-PHYS-PUB-2023-026} as discussed above, and the
difference to the dynamic scale diagram removal
scheme~\cite{Frixione:2008yi} is used as uncertainty.
The \POWHEG \hdamp parameter is varied for the $tW$ samples
in the same way as for the $t\bar{t}$ samples.
In addition, a normalization uncertainty of 4\% is assigned, arising from the uncertainty in
the theoretical prediction at NNLO~\cite{Kidonakis:2021vob}. This is
also assigned to all single-top channels which is $90\%$ dominated by
$tW$ production.

The largest uncertainties are due to the hadronisation and fragmentation models, the interference
between $tW$ and $t\bar{t}$, and the differences in the modelling of additional radiation between
the NNLO calculation and the PS. At small \abseta and for the negative lepton charges, where the contribution
from top background is largest, these uncertainties
extend up to 2-3\% each.

\paragraph{EW background}
The remaining EW background processes lead to small contributions in
the signal region. A correlated normalisation uncertainty is
applied for each process separately, taken to be $5\%$ for
$Z\to\ell\ell$ and $W\to\tau\nu$~\cite{STDM-2015-03}, and $6\%$ for diboson processes.

\paragraph{Jet uncertainties}
The uncertainties in the jet energy scale and resolution directly
influence the \met reconstruction and are described in detail in
Ref.~\cite{JETM-2018-05}. The Jet Energy Scale (JES) uncertainties are
described using the 29 nuisance parameter scheme, which considers
uncertainties arising from the pile-up correction to jets, calibration
biases, the jet flavour response, the modelling of jets and of the
detector, as well as punch through and the response of high
$p_{\textrm{T}}$ jets.
The Jet Energy Resolution (JER) uncertainties use an eight-parameter scheme.
They are larger than the JES contributions but still remain below $1.5\%$
throughout the measurement range.

\paragraph{Uncertainties in the \met soft term}
The \met soft term uncertainties arise from the momentum scale of this
contribution to the \met, and from the separate parallel and
perpendicular resolution modelling of the tracks comprising the soft
term with respect to the mean \pt  of the
tracks~\cite{JETM-2020-03}. These contributions are largest
at low \mtw, and dominate the uncertainty in the lowest \mtw bin
reaching $1.5\%$. In the shadow bin this rises to $4\%$.

\paragraph{Pile-up modelling}

To account for differences between simulation and data in the pile-up
distribution, the pile-up profile in the simulation is corrected to
match the one in data. The uncertainty in the correction factor is
$\pm 4\%$ and is
applied in the measurement as a variation of the event weight.

\paragraph{MC statistics}
The size of the MC samples used in the analysis leads to uncertainties
that are considered uncorrelated between measurements bins, charges
and flavours.

\paragraph{Luminosity}
The uncertainty of the ATLAS luminosity measurement is $0.83\%$ and is
discussed in detail in Ref.~\cite{DAPR-2021-01}.

\paragraph{Unfolding uncertainties}
The uncertainty in the unfolding procedure is estimated for two sources.   %
The first contribution assesses the bias arising from differences
between data and MC simulation in the measured observables \mtw and
\abseta. This is estimated by reweighting the MC simulation at the particle level
such that its resulting distribution
at reconstruction level matches the distributions observed in the data.
The uncertainty is derived by unfolding the resulting distribution on reconstruction level with the nominal
unfolding procedure and comparing the result to the reweighted particle level MC spectra.
The uncertainty is found to be below $1\%$ throughout the range of the measurement.

A second contribution to the unfolding uncertainty
quantifies the impact of poor modelling in unmeasured observables for
example (but not only) the transverse momentum of the $W$ boson which
is modelled differently by the \textsc{Powheg+Pythia} and \Sherpa
signal MC samples. The uncertainty is estimated by reweighting the
alternative \Sherpa signal MC sample to the \textsc{Powheg+Pythia} spectra at
the particle level in \mtw and \abseta and then unfolding the resulting distribution
on reconstruction level with the nominal
unfolding procedure. The difference between the unfolded distribution
and the reweighted one at particle level
is taken as the uncertainty, and has an impact on the measurement of typically $1\%$.


%
\section{Results }
%

\label{sec:results}

\subsection{Separate differential cross-sections in the electron and muon channels}

\begin{figure}[ht!]
\centering \subfloat[\label{subfig:unf_1D_ele_plus}]{\includegraphics[width=0.48\textwidth]{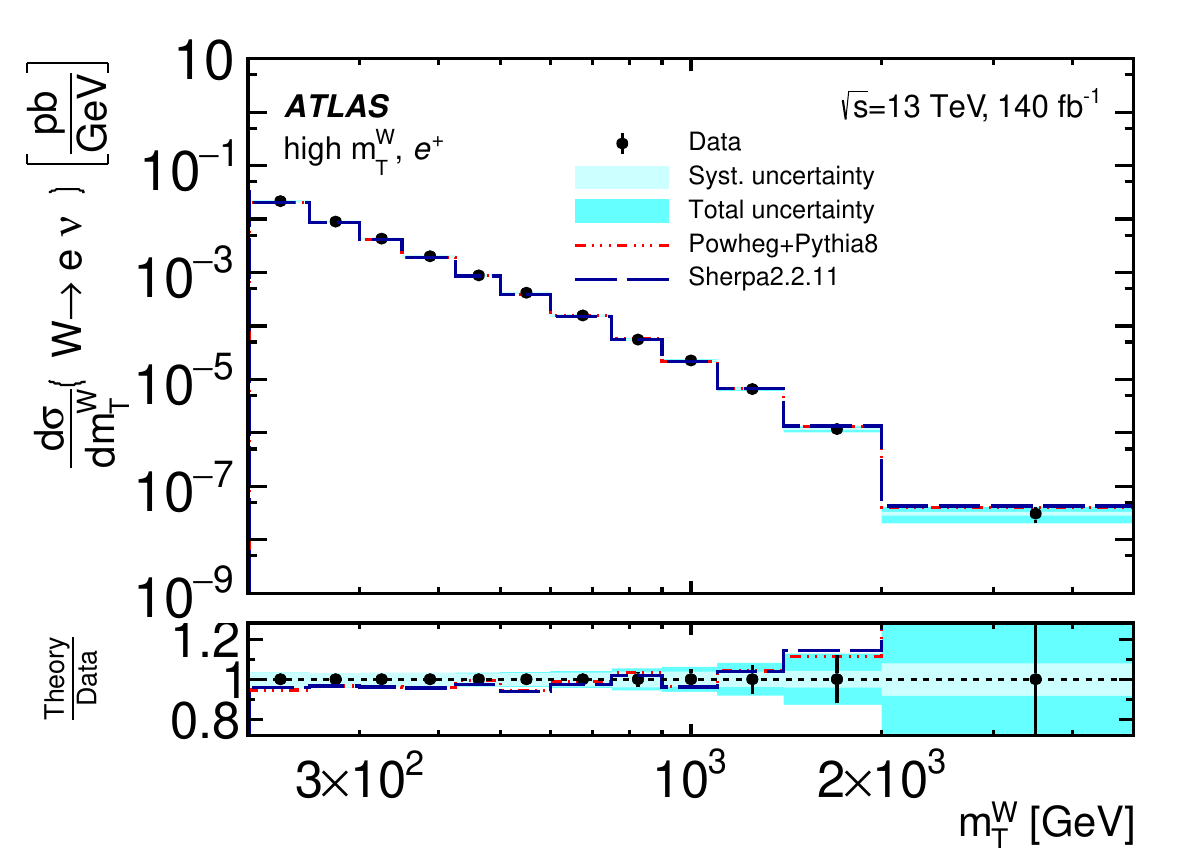}} \subfloat[\label{subfig:unf_1D_ele_minus}]{\includegraphics[width=0.48\textwidth]{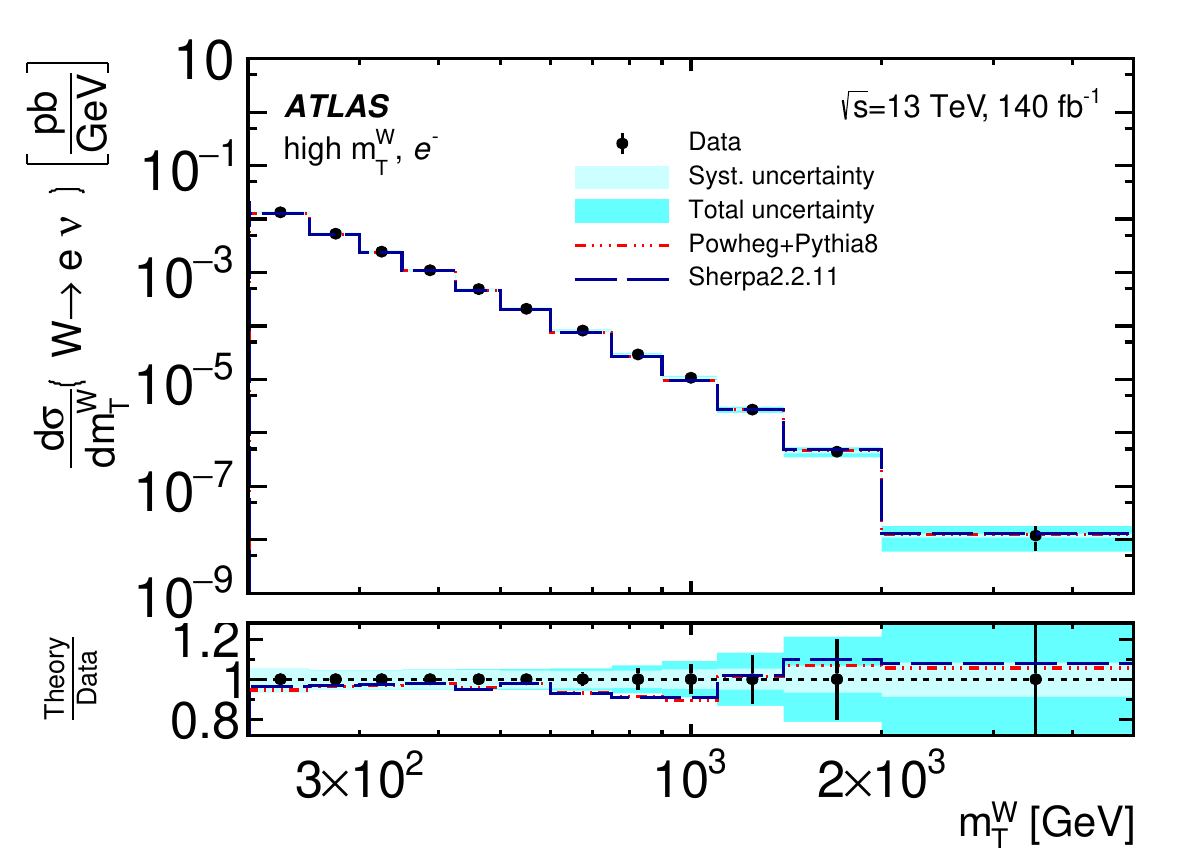}}\\ \subfloat[\label{subfig:unf_1D_muo_plus}]{\includegraphics[width=0.48\textwidth]{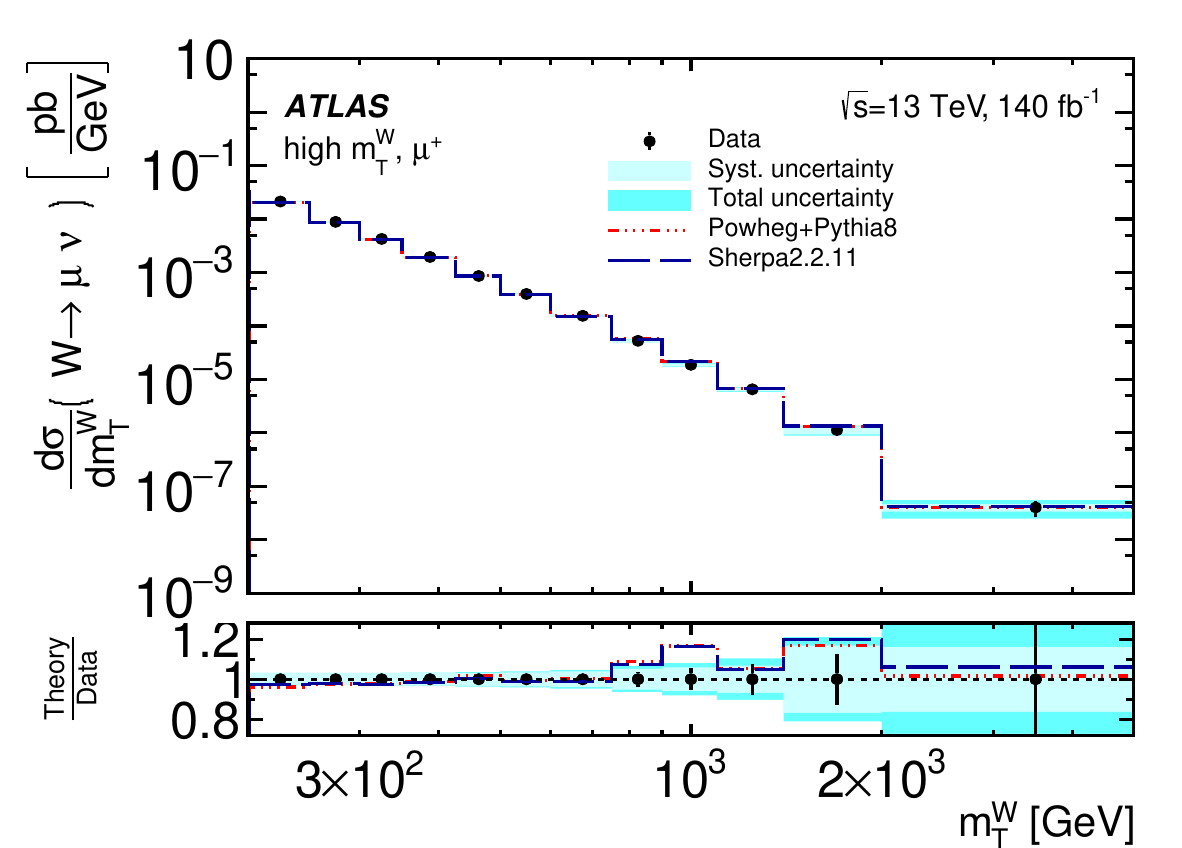}} \subfloat[\label{subfig:unf_1D_muo_minus}]{\includegraphics[width=0.48\textwidth]{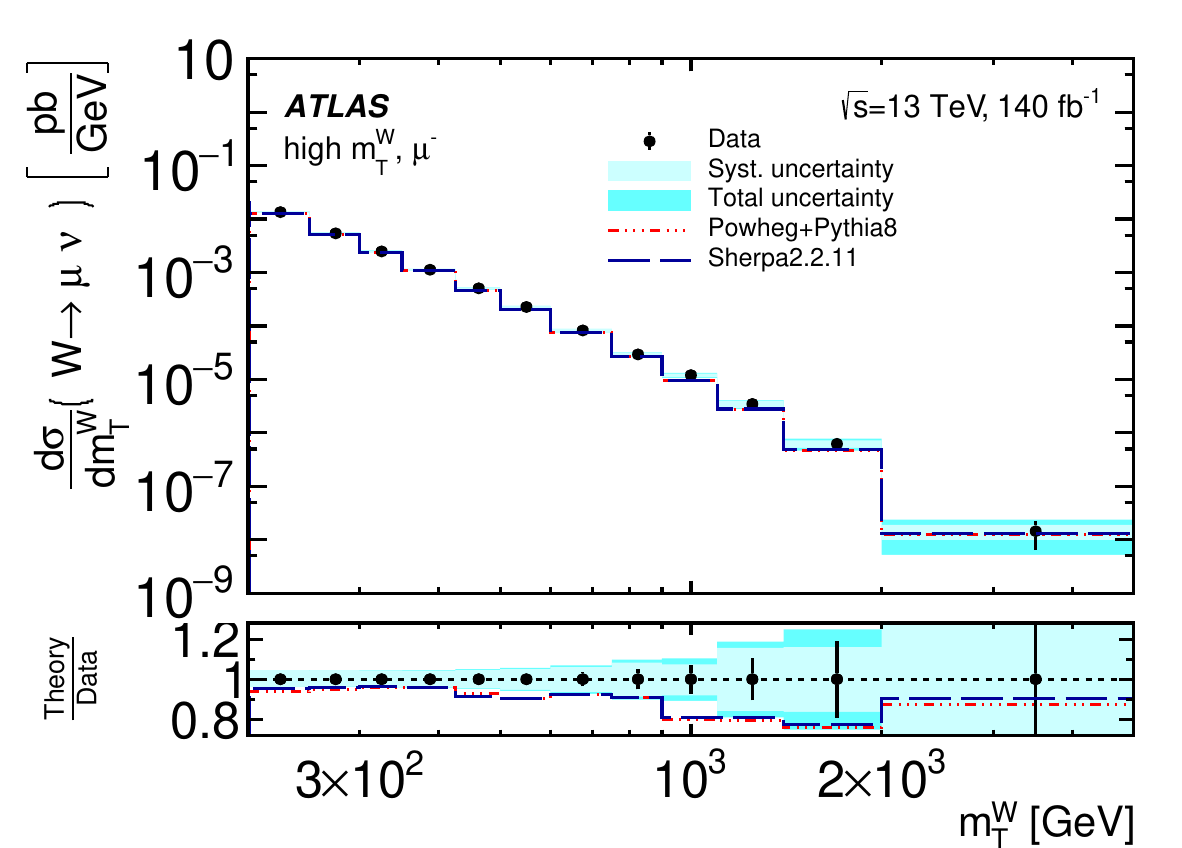}}
\caption{Unfolded
differential cross-section binned in \mtw region for the
(a) $e^+$, (b) $e^-$, (c) $\mu^+$ and (d) $\mu^-$ final states.
The error bars represent the statistical uncertainty. The inner shaded band represents
the systematic uncertainty in the cross-sections, and
the outer shaded band represents the total measurement
uncertainty (excluding the luminosity uncertainty).  The data
are compared to predictions from
\textsc{Sherpa} and \textsc{Powheg+Pythia}.
}\label{fig:xsec1d_eplus_emin_muplus_mumin}
\end{figure}

The unfolded Born-level cross-section measurements $\textrm{d}\sigma/
\textrm{d}\mtw$ including their statistical, systematic and total uncertainties are presented in
Figure~\ref{fig:xsec1d_eplus_emin_muplus_mumin}
separately for each
lepton charge and flavour.
Note that here and in the following the luminosity uncertainty of $0.83\%$ is not shown and not included in the overall systematic and total uncertainty bands.
Detailed tables for all results including the systematic uncertainties separately for each source can be found in \textsc{Hepdata}~\cite{Hepdata,Maguire:2017ypu}.

The data are compared to predictions from \Sherpa,  using the \CT[18nnlo] PDF set, and \textsc{Powheg+Pythia} using the  \CT[14nnlo] PDF set.
The lower panels in
each of the figures show the ratio of the two predictions to the data
measurement and the uncertainty contributions.

The cross-sections are observed to fall over seven orders of
magnitude as the \mtw increases from $200 - 5000$~\GeV.  The $e^+$ and
$\mu^+$ cross-sections are everywhere larger than the corresponding
$e^-$ and $\mu^-$ cross-sections as expected from the difference
between the $u$- and $d$-quark PDFs in the proton.  The predictions
describe the data well within the measurement uncertainties.

The measurements have a data statistical uncertainty ranging from about $0.5\%$ at
low \mtw to about $33\%$ in the $e^+$ and $\mu^+$ channels
and about $50\%$ in the $e^-$ and $\mu^-$ channels at the highest values of \mtw.

In the $e^+$ channel the total systematic uncertainty varies from
$3\%-5\%$ over the \mtw range, compared to $5\%-8\%$ for the $e^-$
measurement.
At low \mtw, the largest contribution to the systematic uncertainty in both cases
are from the \met, JER, and multijet background
estimations.
At higher \mtw the largest contributions to
the experimental uncertainty are from the multijet estimation, the
electron calibration, and the top-quark background, specifically the
choice of the single-top $tW$ subtraction scheme.

In comparison, the muon channel measurements achieve a better systematic
uncertainty at low \mtw, $3\%$ and $4\%$ for $\mu^+$ and $\mu^-$
respectively, due to a variety of smaller contributions.
However, the systematic uncertainty is worse at large values of \mtw, reaching $13\%$ for $\mu^+$
and $22\%$ for $\mu^-$, driven by two muon sagitta uncertainty sources.

\begin{figure}[ht!]
\centering \subfloat[]{\includegraphics[width=0.48\textwidth]{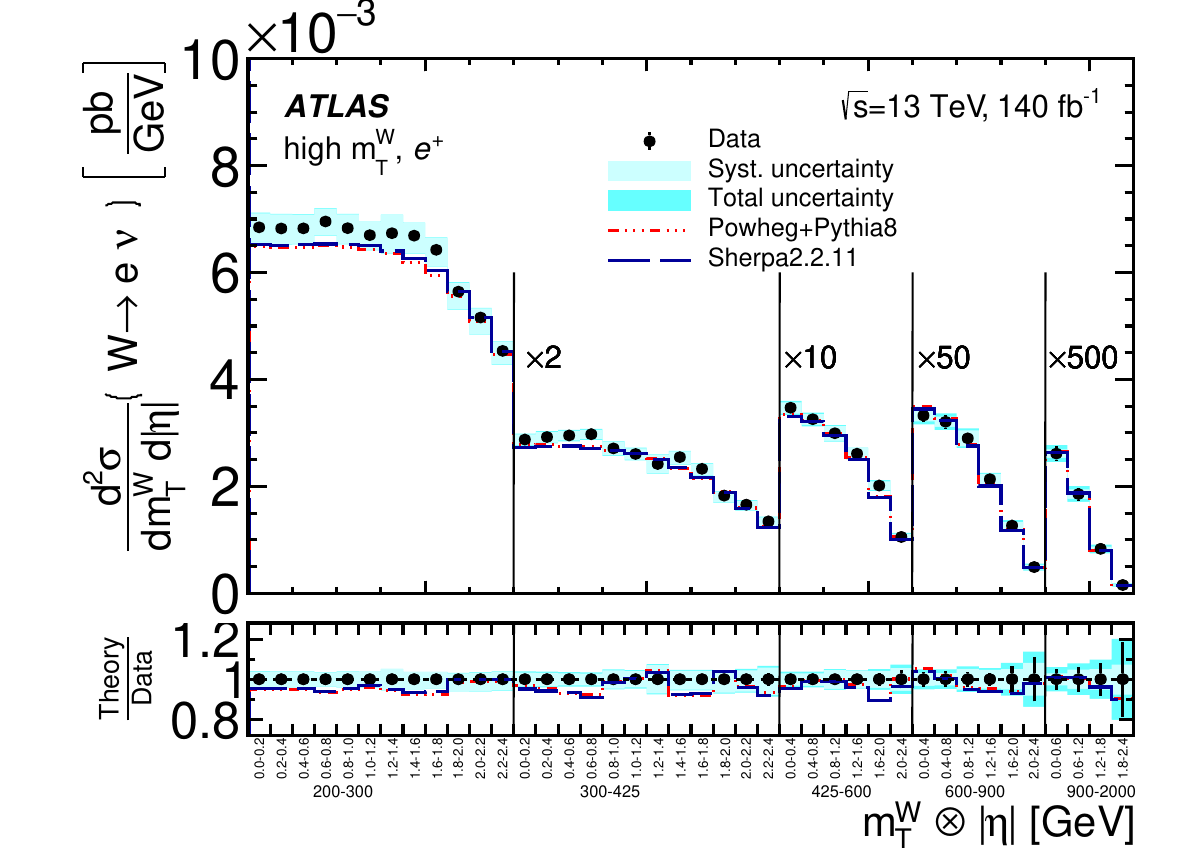}} \subfloat[]{\includegraphics[width=0.48\textwidth]{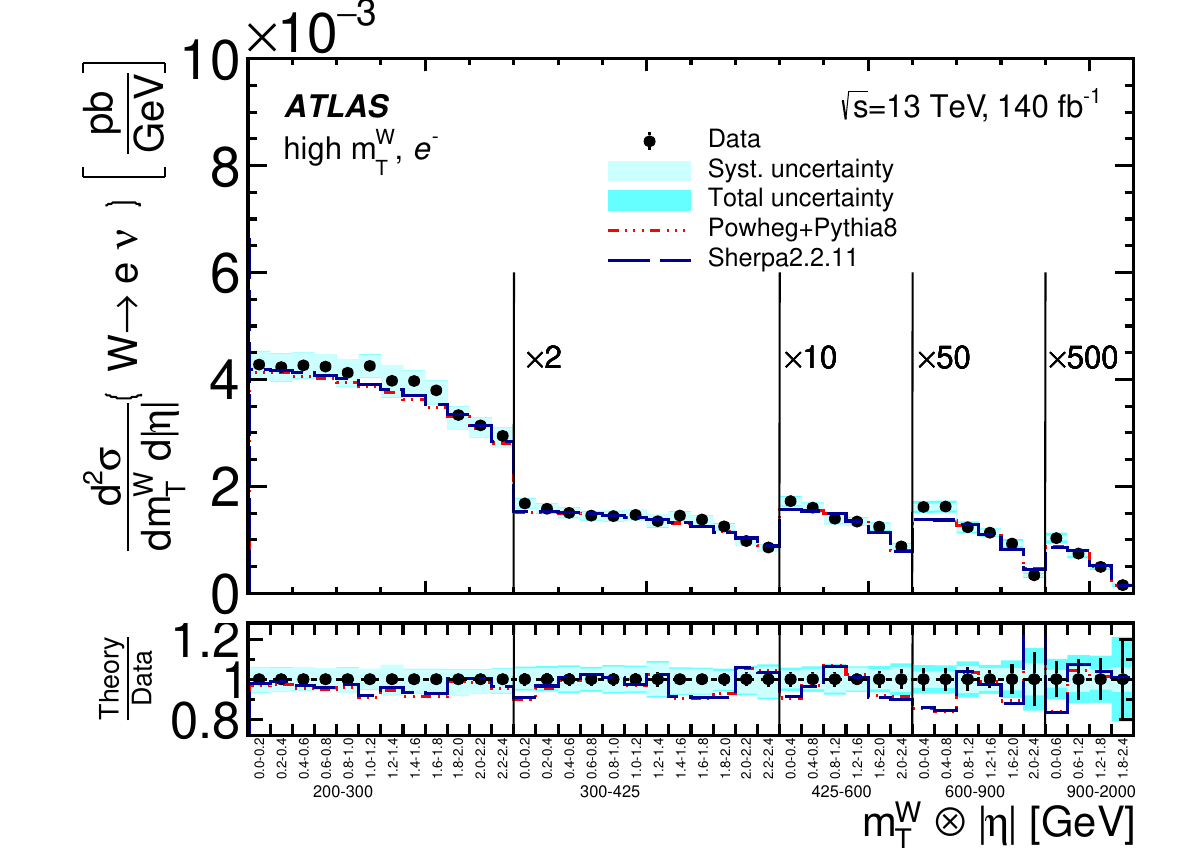}}\\ \subfloat[]{\includegraphics[width=0.48\textwidth]{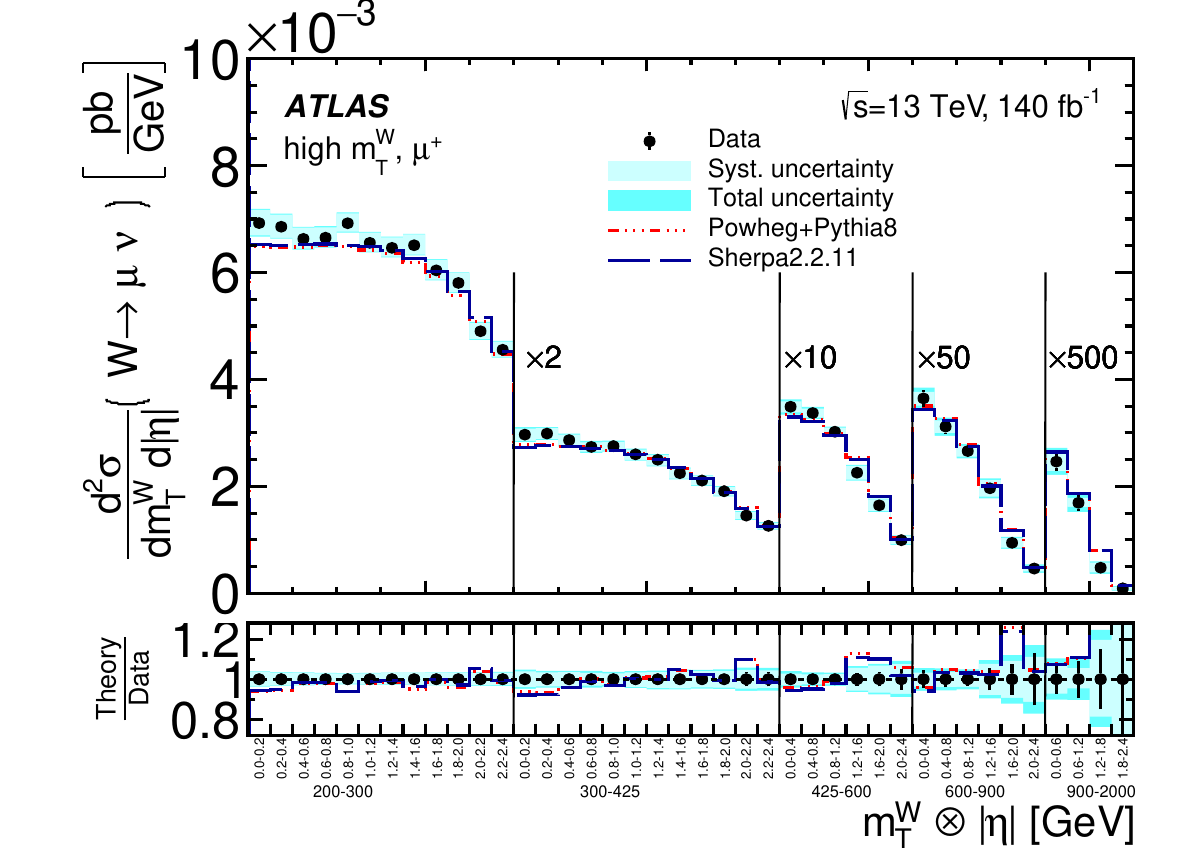}} \subfloat[]{\includegraphics[width=0.48\textwidth]{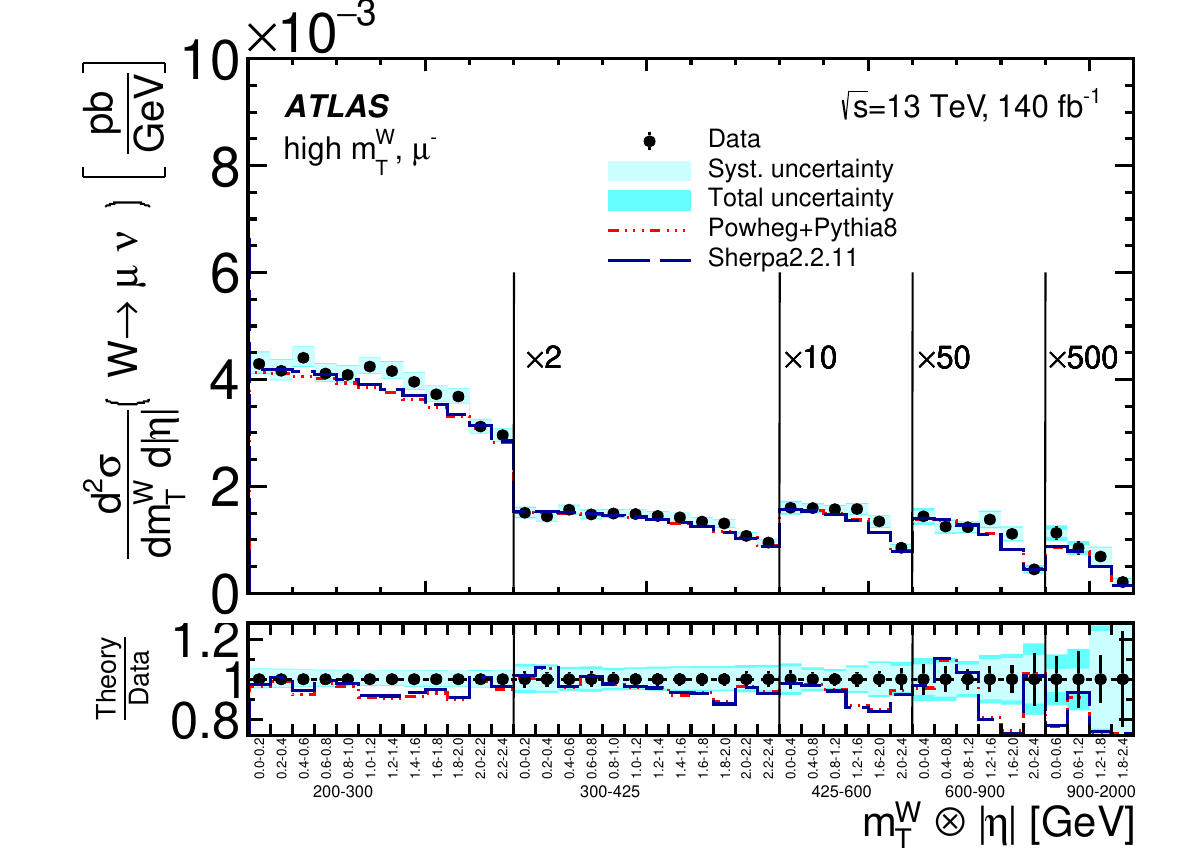}} \caption{Two-dimensional
unfolded cross-section binned in \abseta and \mtw for the
(a) $e^+$, (b) $e^-$, (c) $\mu^+$ and (d) $\mu^-$ final states.
The error bars represent the
statistical uncertainty. The inner shaded band represents the
systematic uncertainty in the cross-sections, and the
outer shaded band represents the total measurement uncertainty
(excluding the luminosity uncertainty).  The data are compared
to predictions from
\textsc{Sherpa} and \textsc{Powheg+Pythia}.
For better readability, the cross-sections in the last four \mtw bins are scaled up by factors of 2, 10, 50 and 500, respectively.
}\label{fig:xsec2d_eplus_emin_muplus_mumin}
\end{figure}

The Born-level two-dimensional cross-sections
$\textrm{d}^2\sigma/ \textrm{d}\mtw\textrm{d}|\eta|$ are shown in
Figure~\ref{fig:xsec2d_eplus_emin_muplus_mumin} separately for each
lepton charge and flavour.
The data are presented in each \mtw bin separated by the vertical lines, and the
measurements for $\mtw \geq 300$~\GeV are scaled by the factors shown
for presentation purposes only. The
measurements are
compared to the
\Sherpa and \textsc{Powheg+Pythia} predictions which
describe the data well.

The cross-sections at low \mtw show a plateau-like behavior for \abseta up
to $1.4$, and then decrease at larger
pseudorapidity. With increasing \mtw the plateau region narrows.

\subsection{Electron-muon ratios of charge-integrated cross-sections}

Charge-integrated single- and double-differential cross-sections have also been extracted
separately for the electron ($e^\pm$) and muon ($\mu^\pm$) channel and can be found in \textsc{Hepdata}.  %

\begin{figure}
\centering
\subfloat[\label{fig:1Demurat}]{\includegraphics[width=.48\textwidth]{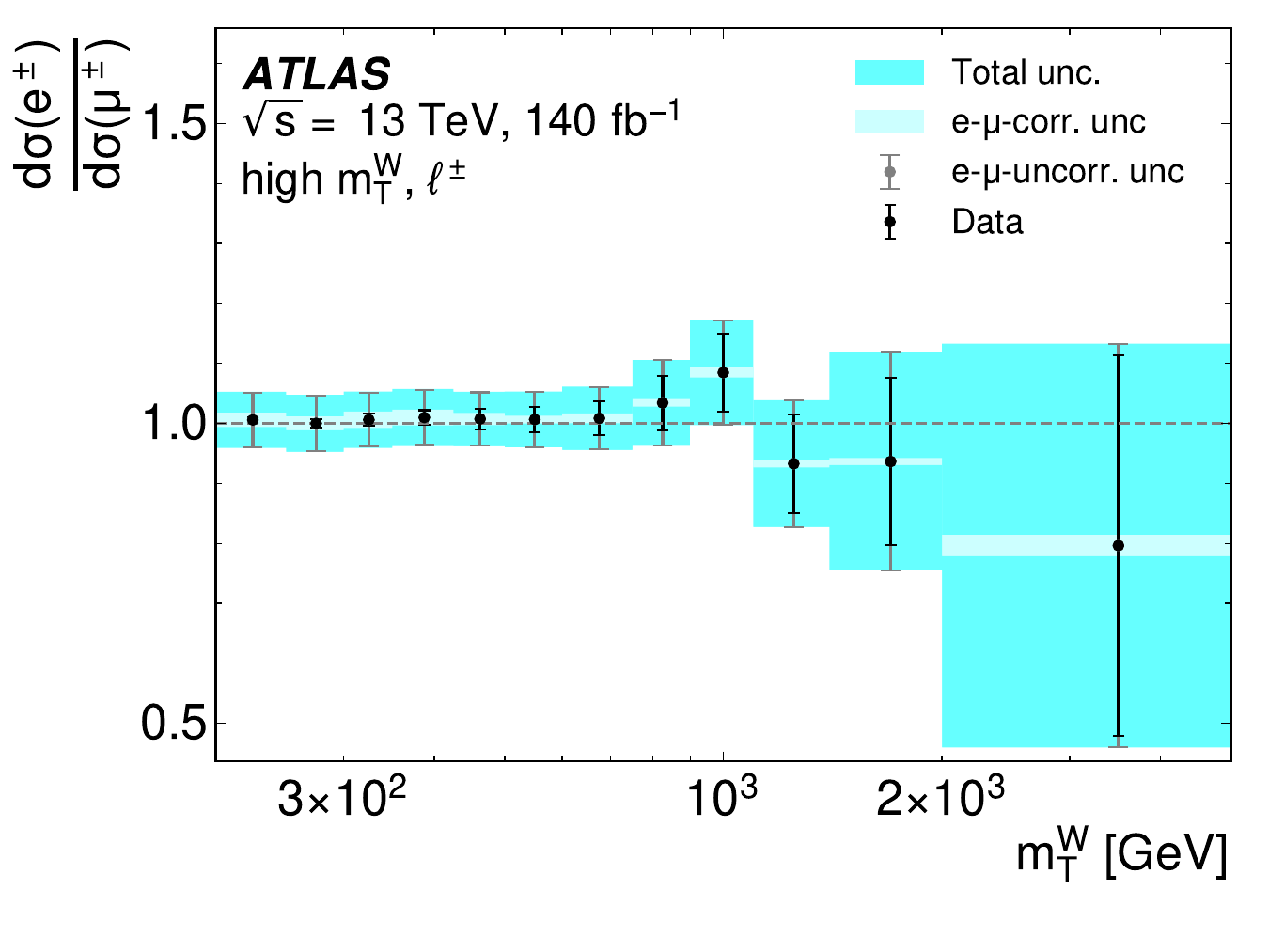}}
\subfloat[\label{fig:2Demurat}]{\includegraphics[width=.48\textwidth]{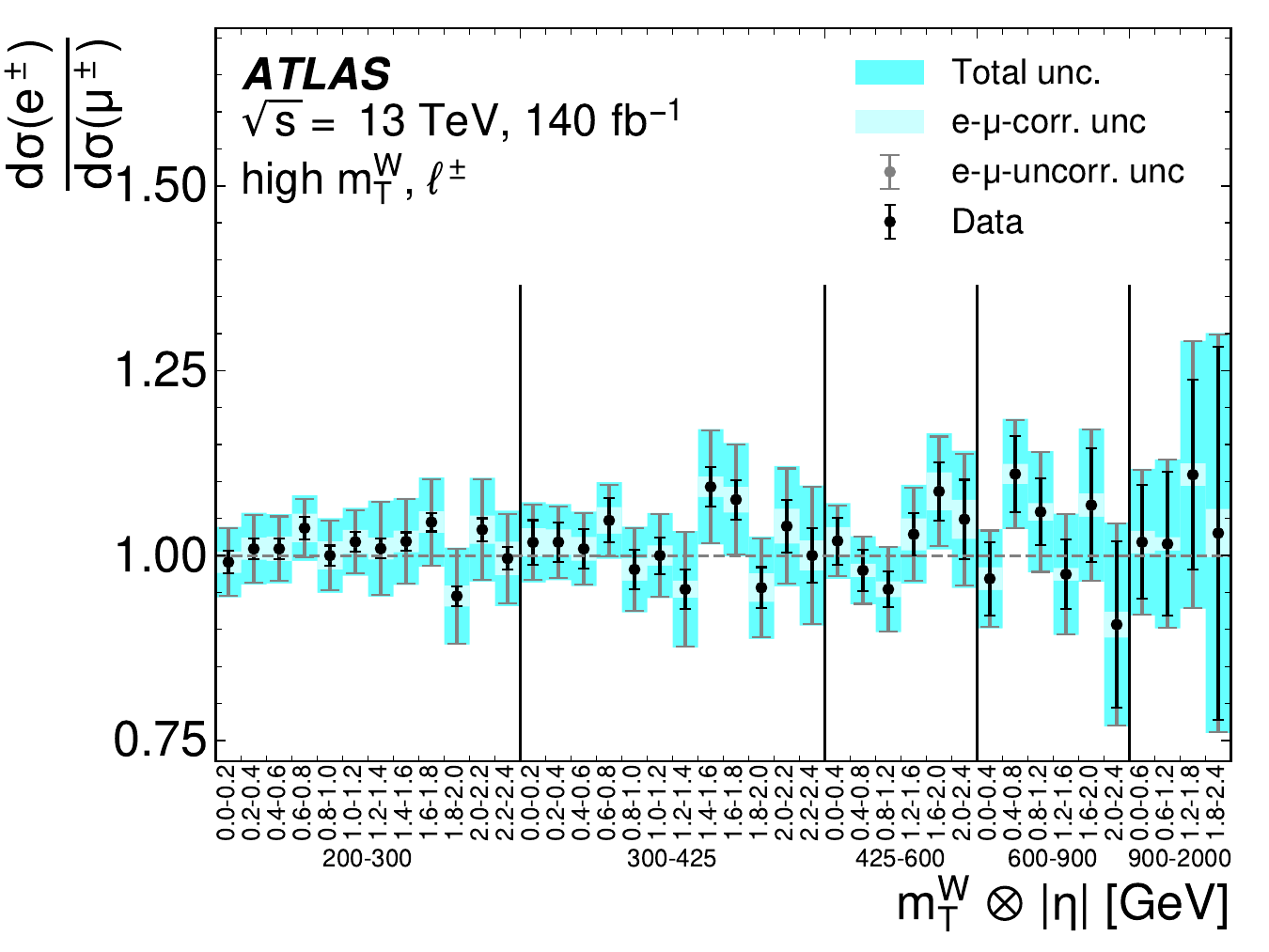}}
\caption{Ratio of the unfolded (a) single- and (b) double-differential electron- and muon-channel cross-sections in the combined $\ell^\pm$-channel. The error bars
represent the uncorrelated (systematic and statistical)
uncertainties between the electron and muon channels, while
the inner uncertainty band corresponds to the
correlated uncertainties. The total uncertainty in the ratio
is presented with the outer uncertainty band.}
\label{fig:emuRatio}
\end{figure}

Ratios of the charge-integrated
cross-sections for the $e^\pm$ to the $\mu^\pm$ final state
are shown in Figure~\ref{fig:emuRatio}.
Here all asymmetric uncertainties are symmetrized and the one-sided
uncertainties are mirrored.
This ratio provides a test of lepton flavour universality and is consistent with unity within the uncertainties.

\subsection{Combination of electron and muon channels}
\label{sec:combination}

The measurements in the $e^{\pm}$ and $\mu^{\pm}$ final states are
combined
using a $\chi^2$ minimisation procedure~\cite{Glazov:2005rn,H1:2009jxj,H1:2009pze} %
under the assumption that the unfolded measurements for the same charge
should agree.
The technique improves the statistical precision of the data as well as
the systematic uncertainty in case of uncertainties that are not (fully)
correlated between the lepton flavours.
The bin-to-bin correlated systematic uncertainties are
taken into account by introducing a nuisance parameter for each source
of uncertainty modelled as a unit Gaussian probability density
contributing to the $\chi^2$ definition. The systematic sources and
their correlations between channels are discussed in
Section~\ref{sec:systematics}. The nuisance parameter values are optimised
in the fit which minimises the $\chi^2$ difference between the
combined cross-section and the separate flavour channels, summed over
the measurement bins.
The combination provides new
orthogonal systematic uncertainty sources which are linear
combinations of the original sources, and therefore lose their
association to specific experimental causes. In order to aid
interpretation, the covariance matrix is rotated back to the (approximately) original
physical source decomposition. Both versions are available in \textsc{Hepdata}.            %
The combination is first performed for the double-differential cross-sections, and
simultaneously for both lepton charges,
which allows optimal systematic contraints from the separate charge
combinations and from the additional information encoded in the \abseta
variable.
The resulting shifts and constraints on the nuisance parameters are then transferred
to the combination of the single-differential cross-sections.

\begin{figure}
\centering
\subfloat[\label{fig:2Dcombplus}]{\includegraphics[width=.48\textwidth]{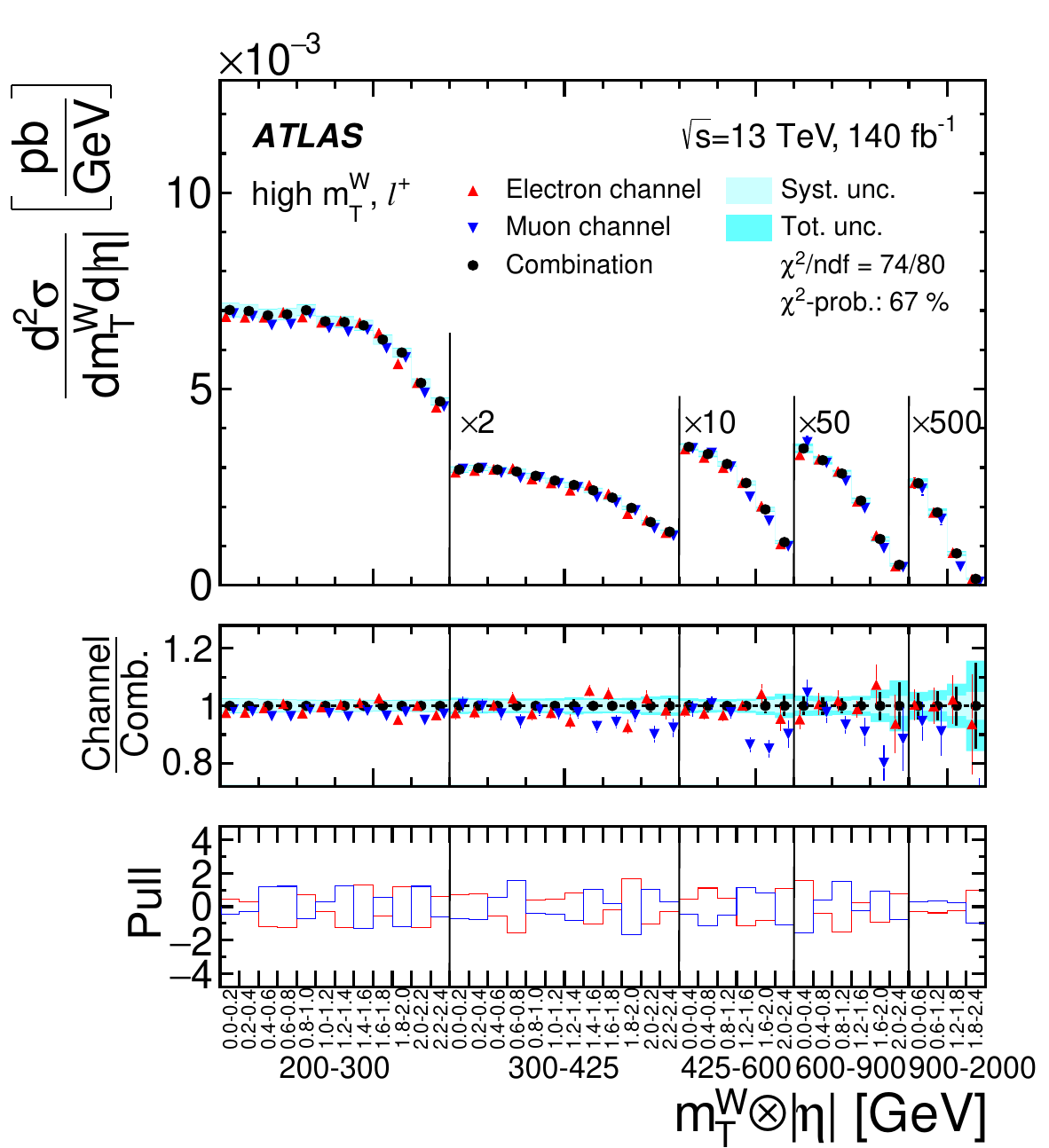}}
\subfloat[\label{fig:2Dcombminus}]{\includegraphics[width=.48\textwidth]{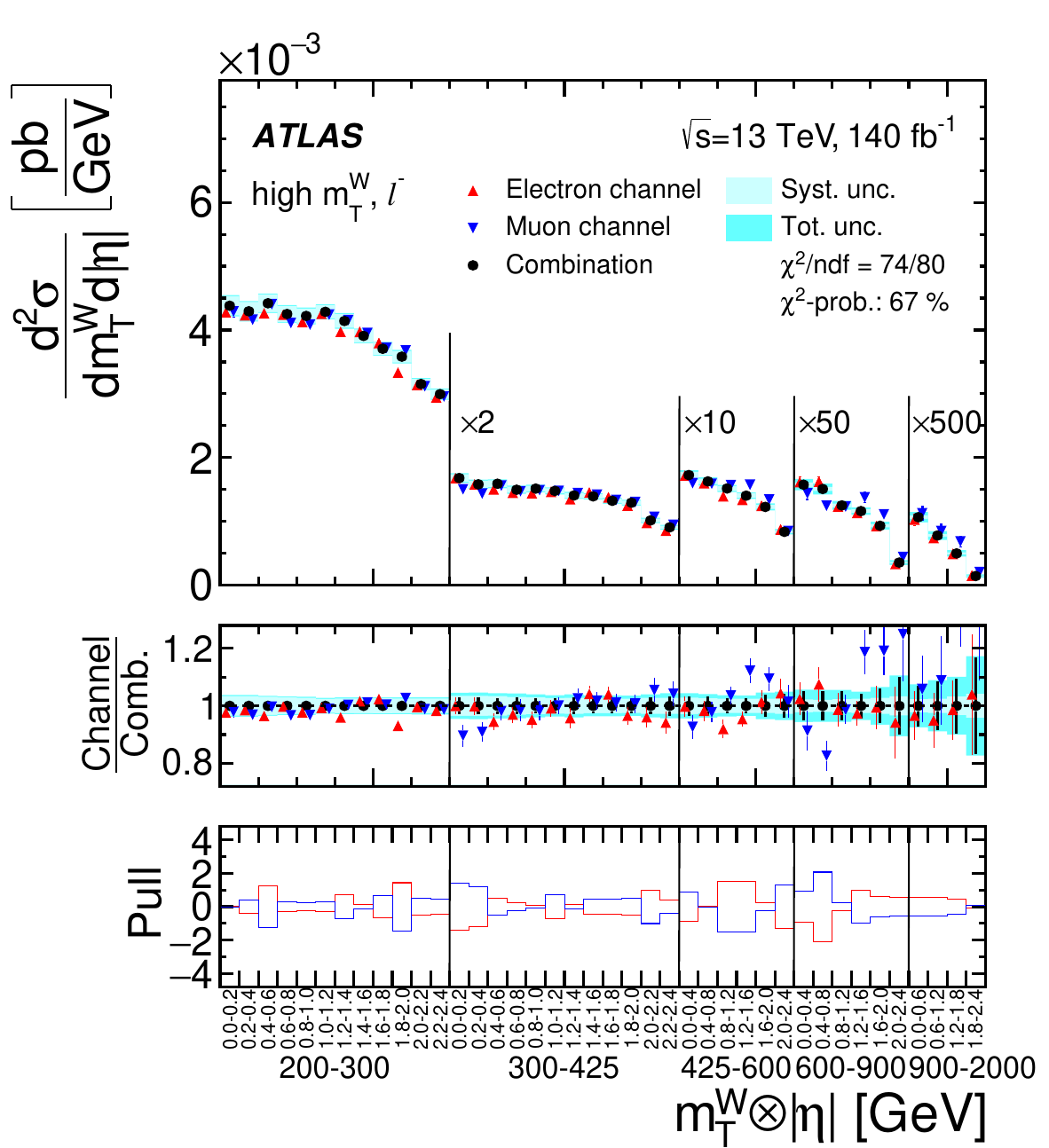}}
\caption{Electron, muon and combined fiducial Born-level cross-sections
binned in \mtw and \abseta ,
for the (a) positive and (b) negative lepton charge.
The error bars represent the statistical uncertainty. The inner shaded band represents the systematic uncertainty in the combined cross-sections,
and the outer shaded band represents the total measurement uncertainty (excluding the luminosity uncertainty).
For better readability, the cross-sections in the last four \mtw bins are scaled up by factors of 2, 10, 50 and 500, respectively.
The central panel shows the ratio of each measurement channel to the combined data, and the lower panel shows the pull of
the post-fit electron and muon channel measurements with respect to the combined data.
}
\label{fig:2Dcomb}
\end{figure}

The combined cross-sections $\textrm{d}^2\sigma/ \textrm{d}\mtw\textrm{d}|\eta|$ for the
$\ell^{+}$ and $\ell^{-}$ final states
are shown in Figure~\ref{fig:2Dcomb}.
The double-differential cross-section combination shows good agreement
between the lepton flavours.  The fit has $80$ degrees of freedom, and
the total $\chi^2$ is found to be $74$ yielding $\chi^2/\text{dof} = 0.92$.
In each figure the upper panels show the measured Born-level cross-sections for the electron channel, muon channel and the
combination. The ratio of the pre-fit individual channels to the combined
measurement is shown in the middle panel. The lower panel displays the
pulls of the two channels, defined as the difference between the
post-fit single-channel measurement and the combined result in units
of the bin-to-bin uncorrelated uncertainty. No coherent trends between the
measurements are observed and the pulls are found to be below two
standard deviations everywhere.

\begin{figure}
\centering
\subfloat[]{\includegraphics[width=.48\textwidth]{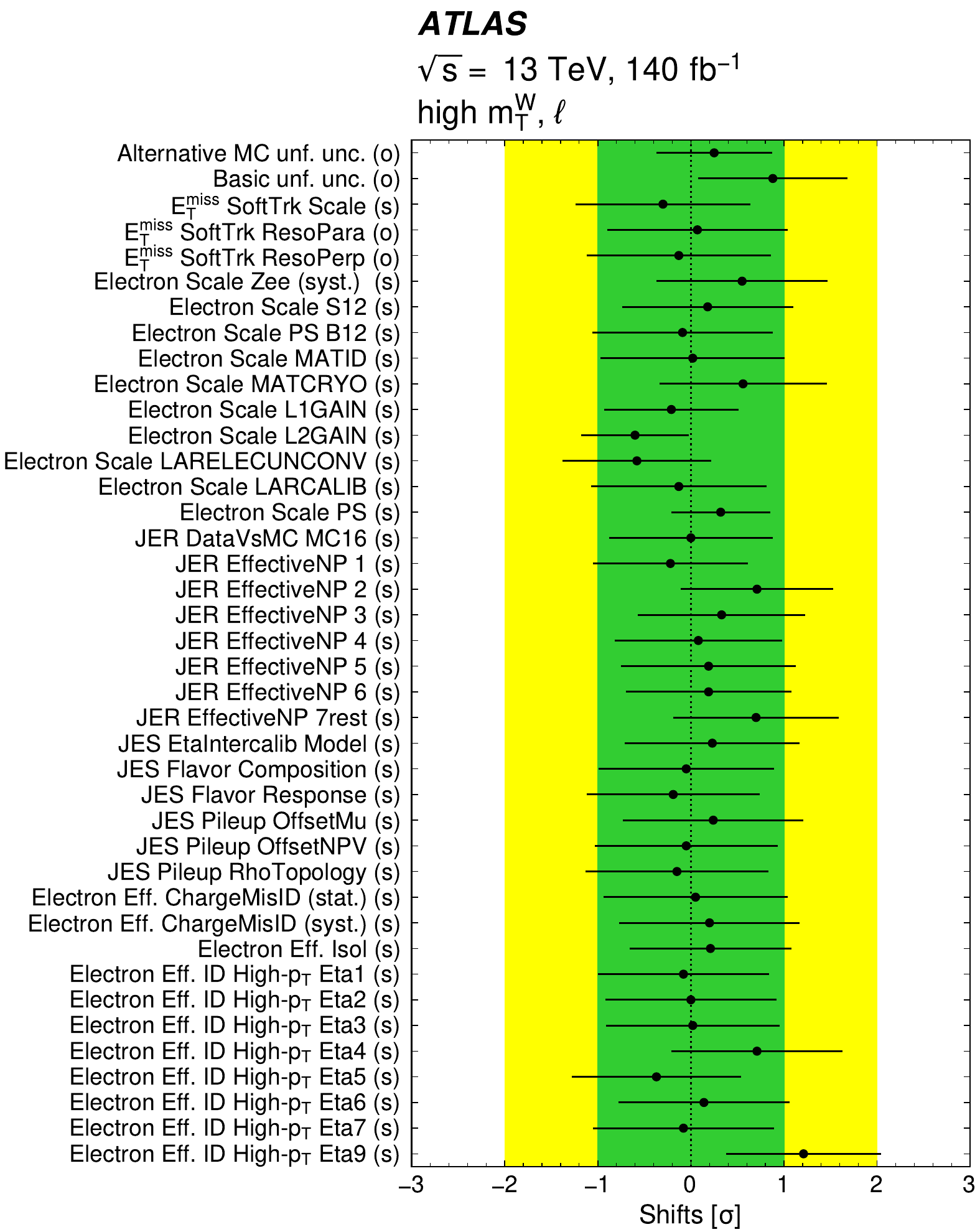}}
\subfloat[]{\includegraphics[width=.48\textwidth]{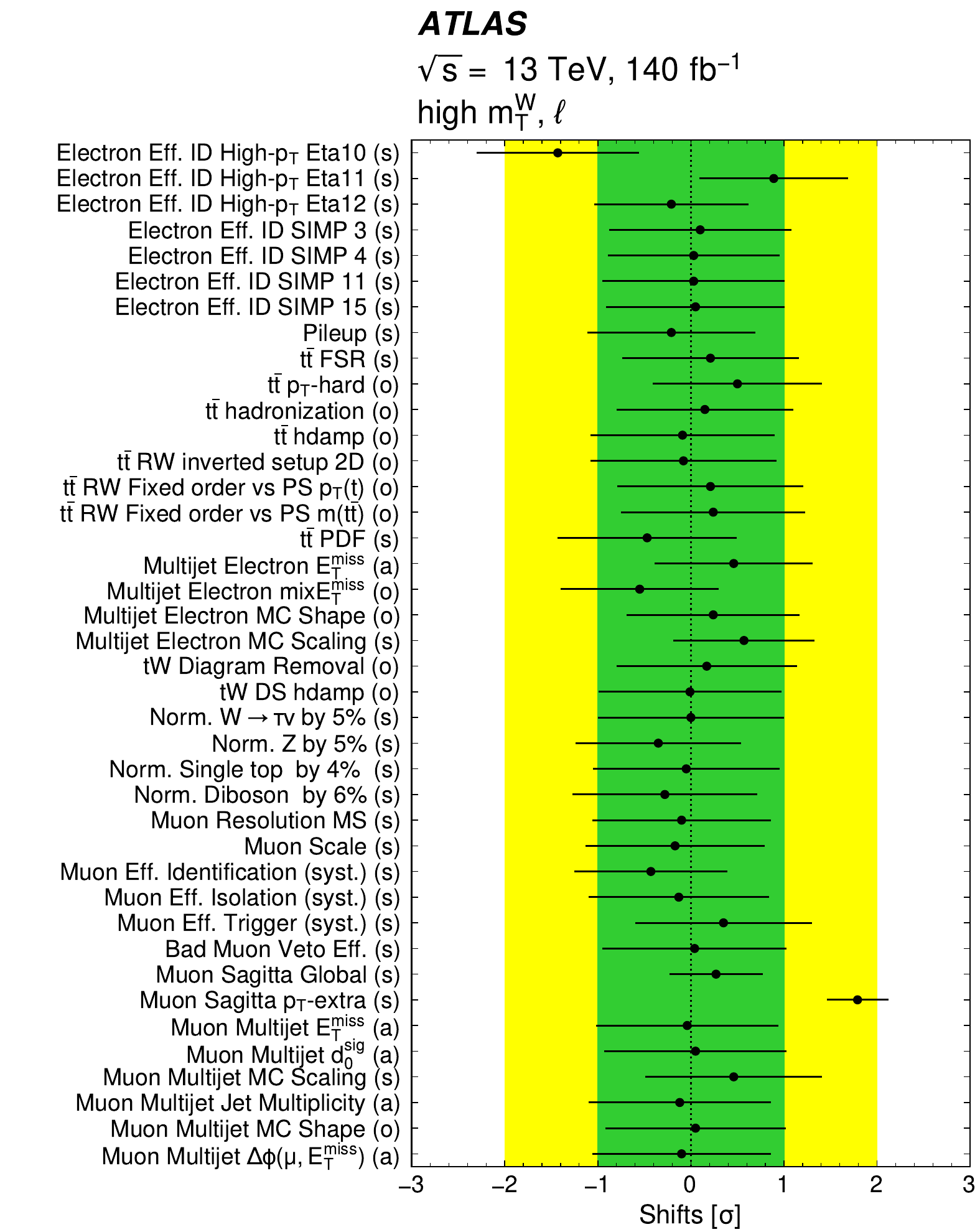}}
\caption{Shifts of the systematic uncertainty nuisance parameters after the
combination of the double-differential electron and muon cross-sections.
The systematic uncertainties are related to the unfolding procedure (``unf.''), the jet energy resolution/scale (``JER/JES''),  the \met scale and resolution, the electron and muon scale, resolution and efficiency (``Eff.''), the multijet and $t\bar{t}$ (where RW refers to a reweighting to NNLO) background estimates and normalization of small background processes (``Norm.'').
The uncertainties are marked whether they
are symmetric (``s''), asymmetric (``a'') or one-sided (``o''). The
systematic uncertainties labeled with ``\text{Electron}'' or
``\text{Muon}'' are not correlated between the lepton flavours.}
\label{fig:Combshifts}
\end{figure}

The post-fit nuisance parameters shown in Figure~\ref{fig:Combshifts}
fluctuate around a mean of zero with shifts typically below
one standard deviation. The exception to this is the high-\pt sagitta
bias parameter with a shift of $1.8$ standard deviations indicating a
small residual undercalibration. This systematic source only affects the region $\abseta>1.1$.  The
parameter is also highly constrained by the combination indicating the
calibration potential of this measurement. In addition, two of the twelve
$\eta$ binned electron identification systematics have pulls greater than one
standard deviation. These sources arise from the limited sample sizes
of the $Z\to e^+e^-$ data and are therefore
statistical in nature.

\subsection{Charge asymmetry of combined cross-sections}
\label{sec:asymmetry}

An asymmetry in the production of $W$ bosons arises from the purely
weak coupling to the quarks, the relative sizes of the production
helicity amplitudes, and the PDFs.
As the rapidity of the $W$ boson is not directly accessible,
measurements of the lepton charge
asymmetry, $A_{\ell}$, have been performed at the
LHC~\cite{STDM-2012-20,STDM-2017-13,CMS-SMP-18-012} at $7$, $8$ and
$13$~\TeV centre-of-mass energies in the resonant production region of
the $W$ boson.

The measurements presented here extend the experimental
determination of the asymmetry to large \mtw for the first time.
The single- and double-differential lepton charge asymmetries are defined as
\begin{equation*}
A_{\ell} = \frac{\textrm{d}\sigma_{+}  - \textrm{d}\sigma_{-} }{\textrm{d}\sigma_{+}  + \textrm{d}\sigma_{-} }\,\,\, ,
\end{equation*}
where $\textrm{d}\sigma_{\pm}$ represents the cross-section for the $\ell^+$ or the $\ell^-$ final
state respectively. The charge asymmetry uses the Born-level
flavour-combined measurements with orthogonal uncertainty sources
provided by the post-fit combination. They include correlations
between
the cross-sections for both charges
as the flavour combination is performed over both charges simultaneously as described
in Section~\ref{sec:combination}.

\begin{figure}
\centering
\subfloat[\label{fig:1Dlepasym}]{\includegraphics[width=.48\textwidth]{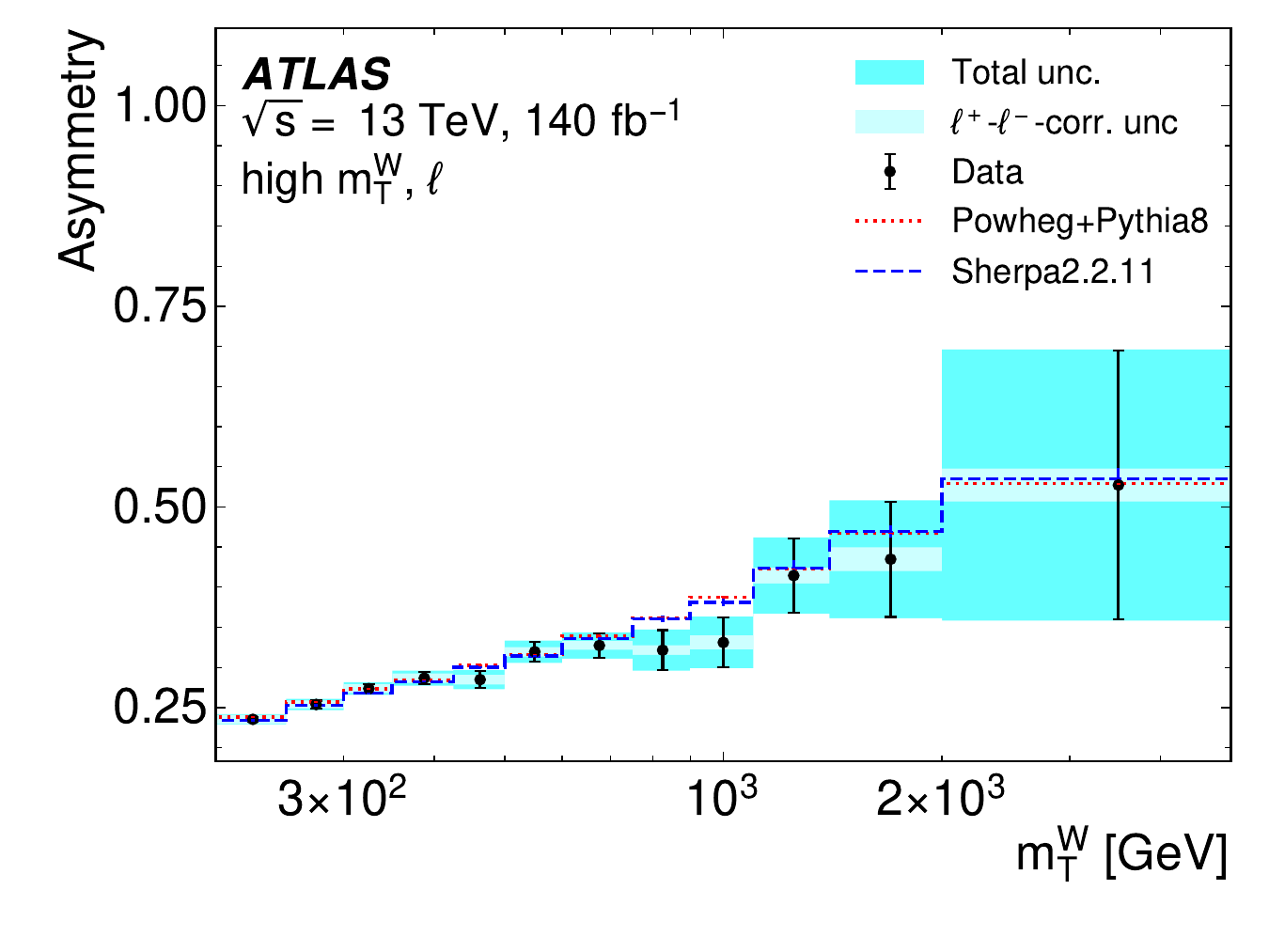}}
\subfloat[\label{fig:2Dlepasym}]{\includegraphics[width=.48\textwidth]{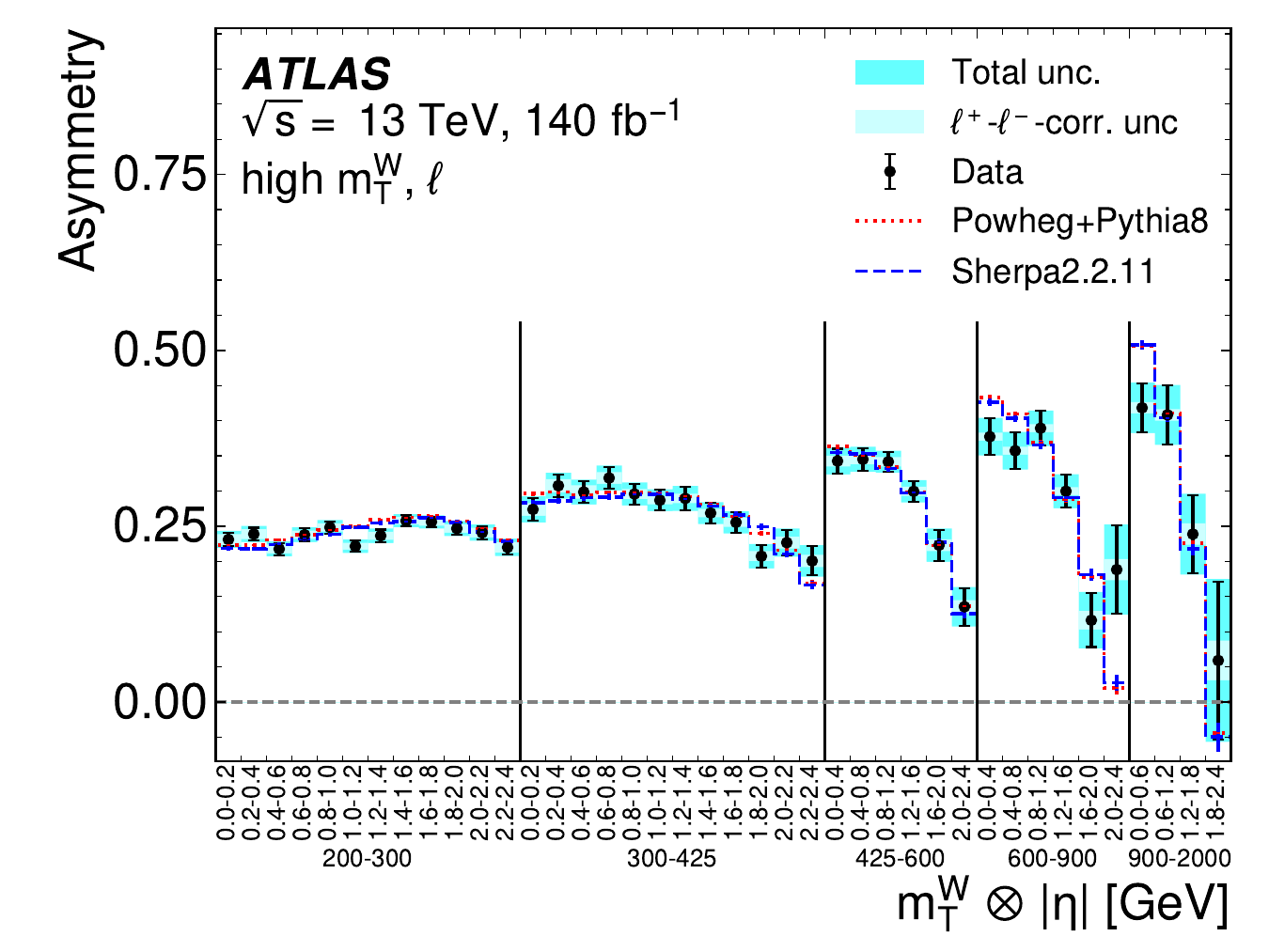}}
\caption{Charge asymmetry of the (a) single- and (b) double-differential
electron/muon combined
cross-section. The
error bars represent the statistical uncertainties.
The inner shaded band represents the systematic uncertainties that are correlated between the charges,
and the outer shaded band represents the total measurement uncertainty.
The data are compared to predictions from
\textsc{Sherpa} and \textsc{Powheg+Pythia}, both using the CT18\textsc{nnlo} PDF set.
}
\label{fig:lepasym}
\end{figure}

The single- and double-differential determinations of $A_{\ell}$ are shown in
Figure~\ref{fig:lepasym}.
The asymmetry is
found to be large and positive, increasing with \mtw, as expected from
the increasing contribution of the $u$-valence PDF at larger Bjorken
$x$.
The \abseta dependence of $A_{\ell}$ is observed to be
approximately constant for $\abseta \lesssim 1.2$
and to fall rapidly at higher values of \abseta for larger values of $\mtw$.
The \textsc{Powheg+Pythia} predictions are here updated to
the  \CT[18nnlo] PDF set, which is already used for \Sherpa.
Both predictions provide a good overall description of the measurements. In the highest \mtw and
\abseta bins the predicted asymmetry from \Sherpa and \textsc{Powheg+Pythia}
becomes slightly negative. However the measurements do not have the statistical precision to confirm this behaviour.

%

%
%

%
%
%
%
%
%

%
%

%


%
\section{Interpretation and discussion}
%

\subsection{Comparison to QCD predictions}
\label{sec:Comp_qcd}

The measured cross-sections are compared to a selection of QCD
predictions. In particular the MC predictions from
\textsc{Powheg+Pythia} and \textsc{Sherpa} are presented together with
a fixed-order calculation performed at NNLO with
\textsc{DYTurbo}~\cite{Camarda:2019zyx,Camarda:2021jsw}.
The NLO EW corrections applied to \textsc{Powheg+Pythia}  (see Section~\ref{sec:MC}) are also added
to \textsc{DYTurbo}, whereas \textsc{Sherpa} includes these contributions
natively.
The predictions from  \textsc{Powheg+Pythia}, \Sherpa and \textsc{DYTurbo} are shown for  the  \CT[18nnlo] PDF set.
In addition the \textsc{DYTurbo} prediction is shown
for four alternative PDF sets -
\textsc{MSHT20nnlo}~\cite{Thorne:2022abv},
\NNPDF[4.0nnlo]~\cite{NNPDF:2021njg},
\textsc{ATLASPDF21nnlo} ($T=3$)~\cite{STDM-2020-32},
and \CT[18qed]~\cite{Xie:2021ajm}.

\begin{figure}
\centering
\subfloat[]{ \includegraphics[width=.48\textwidth]{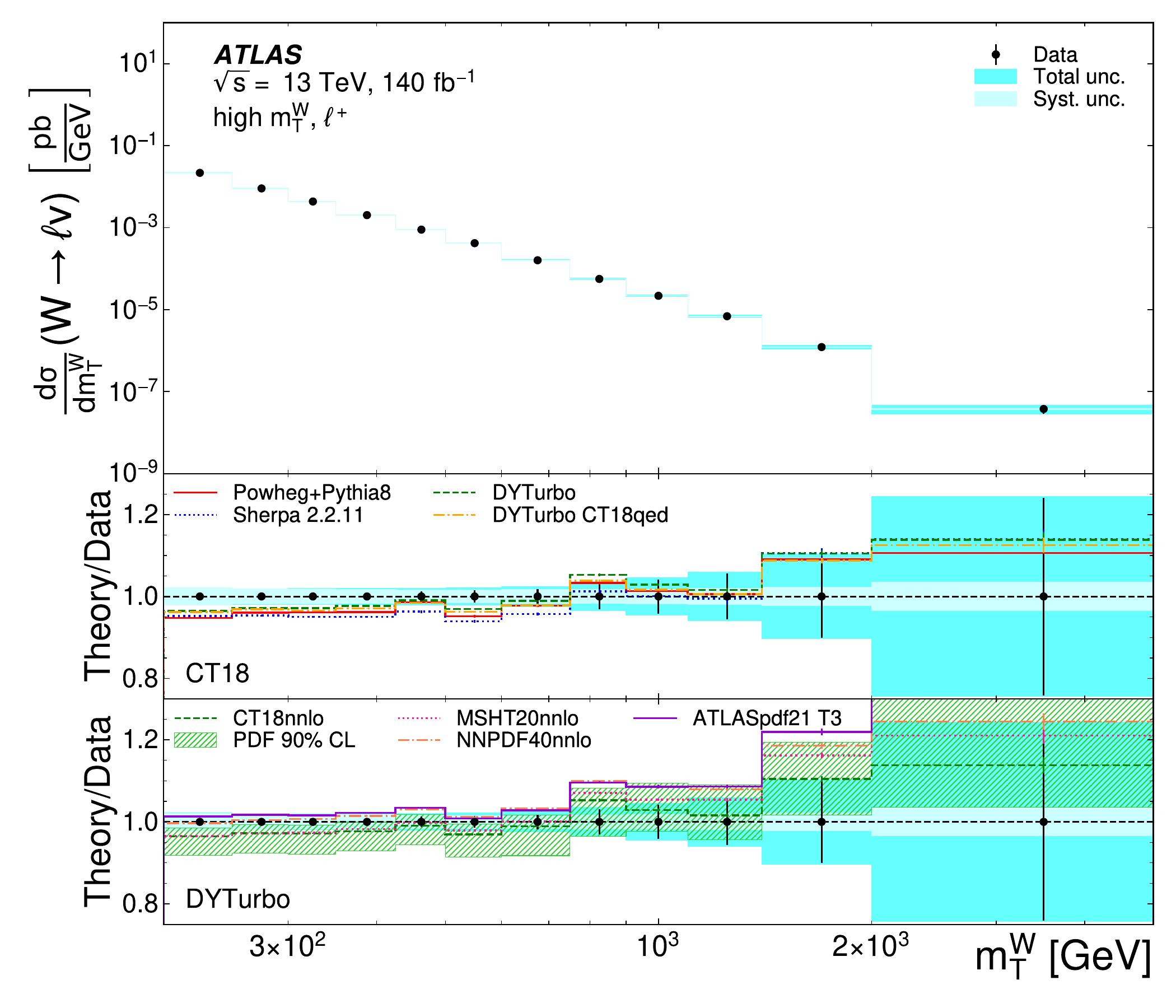}
}	\subfloat[]{         \includegraphics[width=.48\textwidth]{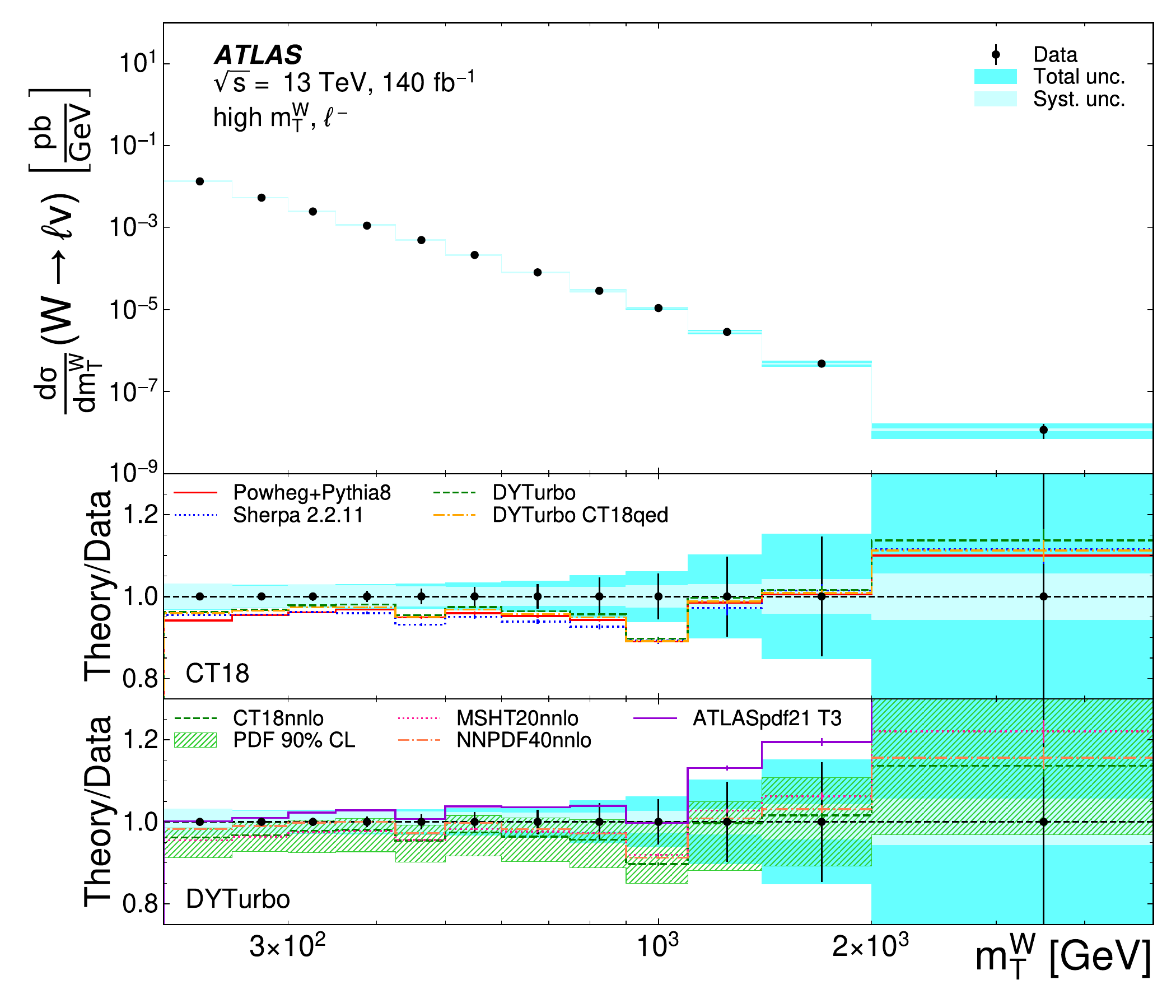} }
\caption{The combined Born-level cross-sections for the
single-differential distributions are shown for the (a) $\ell^+$ and (b)
$\ell^-$ final state. The middle panels show a comparison to the
predictions from \textsc{Sherpa}, \textsc{Powheg+Pythia} and the
fixed-order calculation from \textsc{DYTurbo}, each using the
CT\textsc{18nnlo} PDF. In addition, a fixed-order calculation using
\textsc{DYTurbo} and the  CT\textsc{18qed} proton PDF is shown. A comparison
to predictions with different PDFs using \textsc{DYTurbo} is
displayed in the lower panel. The 90\% CL PDF uncertainty is shown for CT\textsc{18nnlo}
prediction.
The statistical uncertainty of the
combination is displayed with the error bars. The inner uncertainty
band indicates the systematic uncertainty while the outer band
corresponds to the total measurement uncertainty (excluding the
luminosity uncertainty).}  \label{fig:theoComp1D}
\end{figure}

\begin{figure}
\centering
\includegraphics[width=0.95\textwidth]{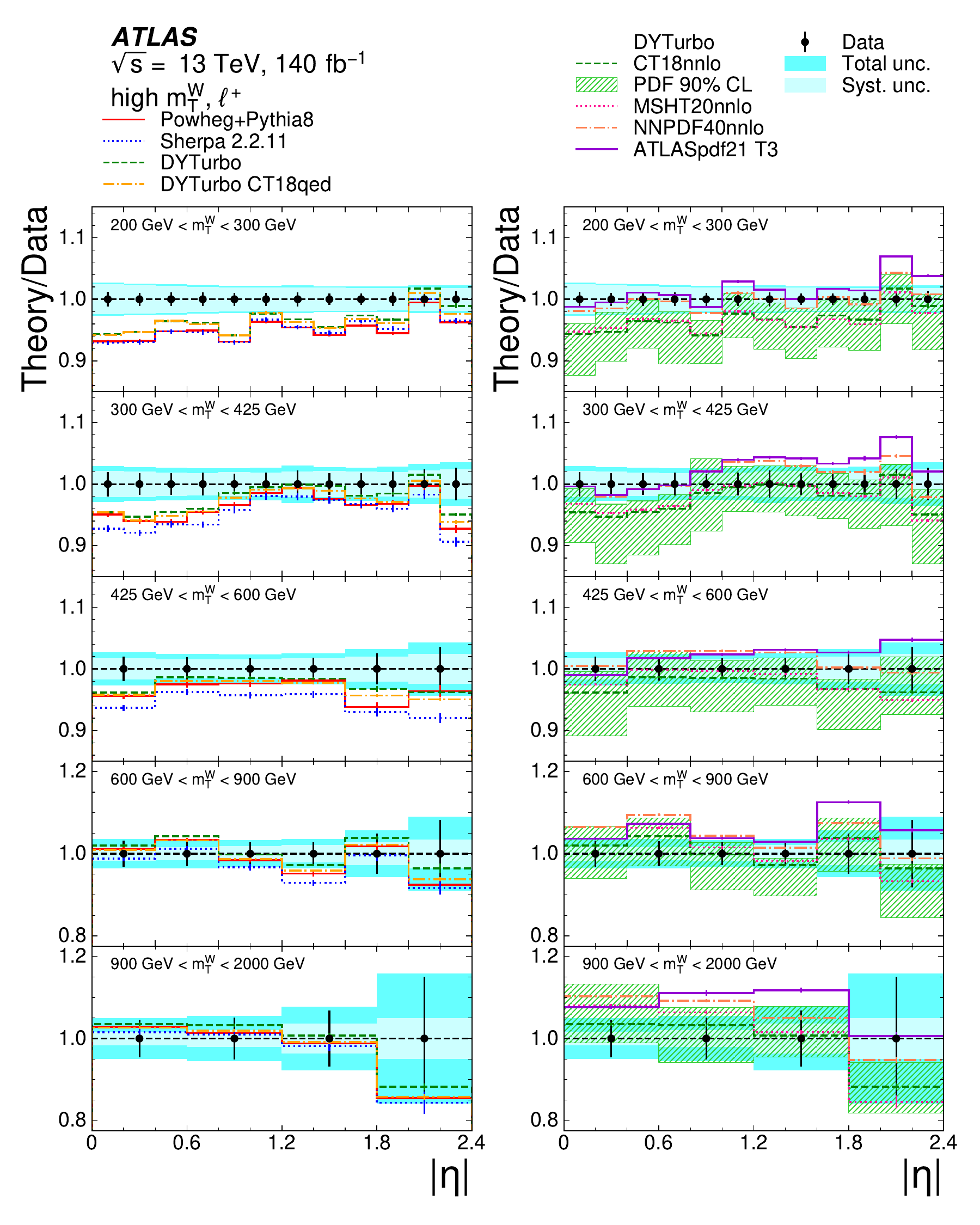}
\caption{The combined Born-level cross-sections for the
double-differential distributions are shown for the $\ell^+$
final state. The left panel shows a comparison to the
predictions from \textsc{Sherpa}, \textsc{Powheg+Pythia} and
the fixed-order calculation from \textsc{DYTurbo}, each
using the CT\textsc{18nnlo} PDF.  In addition, a fixed-order
calculation using \textsc{DYTurbo} and the CT\textsc{18qed} proton
PDF is shown. A comparison to predictions with different
PDFs using \textsc{DYTurbo} is displayed in the right
panel. The CT\textsc{18nnlo} 90\% CL PDF uncertainty is shown for the
prediction. The statistical uncertainty of the combination
is displayed with the error bars. The inner uncertainty band
indicates the systematic uncertainty while the outer band
corresponds to the total measurement uncertainty (excluding
the luminosity uncertainty).}
\label{fig:theoComp2Dplus}
\end{figure}

\begin{figure}
\centering
\includegraphics[width=0.95\textwidth]{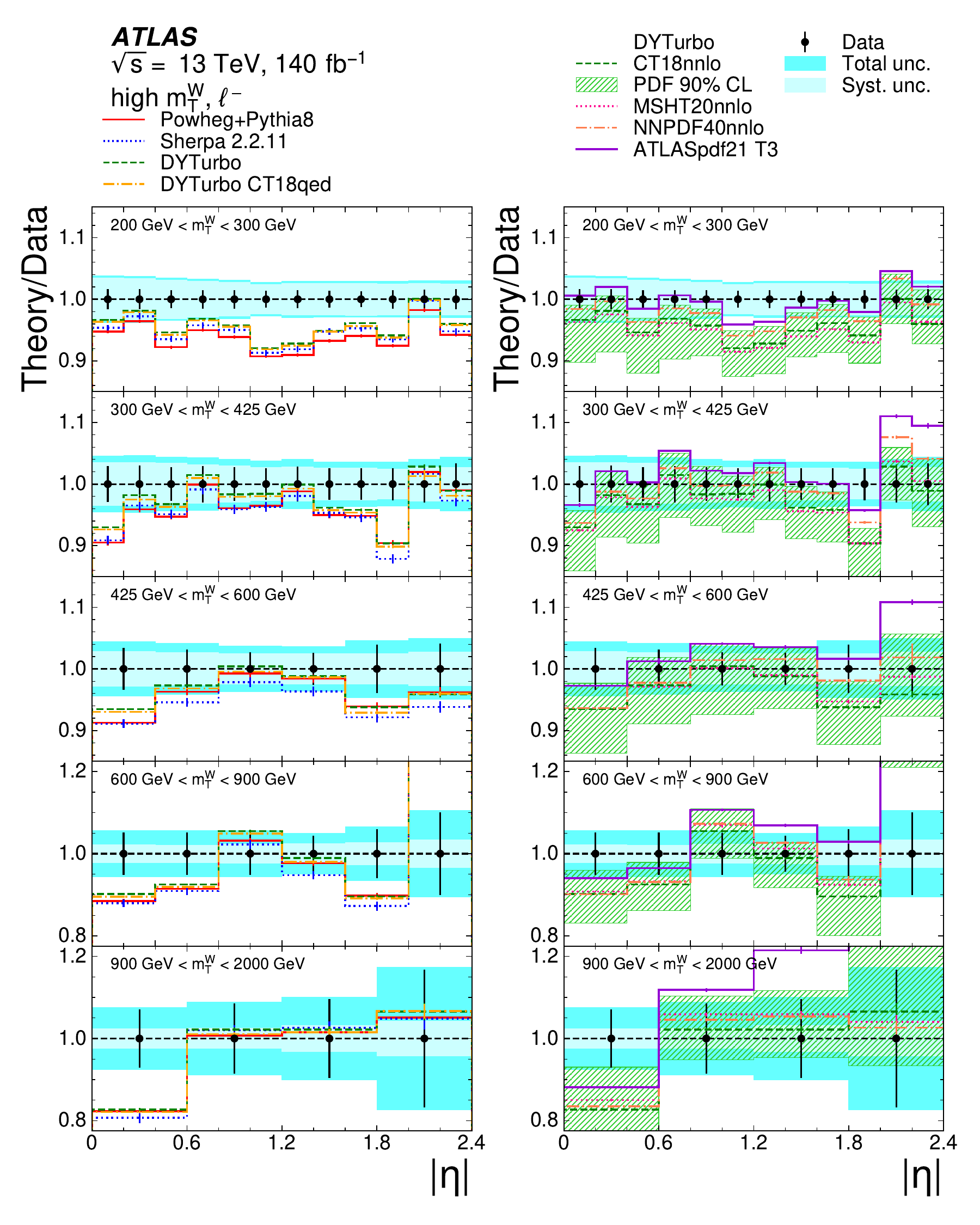}
\caption{
The combined Born-level cross-sections for the
double-differential distributions are shown for the $\ell^-$
final state. The left panel shows a comparison to the
predictions from \textsc{Sherpa}, \textsc{Powheg+Pythia} and
the fixed-order calculation from \textsc{DYTurbo}, each
using the CT\textsc{18nnlo} PDF.  In addition, a fixed-order
calculation using \textsc{DYTurbo} and the CT\textsc{18qed} proton
PDF is shown. A comparison to predictions with different
PDFs using \textsc{DYTurbo} is displayed in the right
panel. The CT\textsc{18nnlo} 90\% CL PDF uncertainty is shown for the
prediction. The statistical uncertainty of the combination
is displayed with the error bars. The inner uncertainty band
indicates the systematic uncertainty while the outer band
corresponds to the total measurement uncertainty (excluding
the luminosity uncertainty).}
\label{fig:theoComp2Dminus}
\end{figure}

The flavour-combined single- and double-differential cross-sections for the $\ell^+$ and $\ell^-$ final
states are compared to the predictions in Figures~\ref{fig:theoComp1D},~\ref{fig:theoComp2Dplus}, and~\ref{fig:theoComp2Dminus}.
The middle panels of Figure~\ref{fig:theoComp1D} and the left panels of Figures~\ref{fig:theoComp2Dplus} and~\ref{fig:theoComp2Dminus}
show that the \textsc{Powheg+Pythia}, \textsc{Sherpa} and \textsc{DYTurbo} predictions using the
\CT[18nnlo] PDF set
agree very well for both lepton charges. They are consistent with the measurements, although a moderate undershoot of the predictions is
observed for low \mtw. The panels also display the \textsc{DYTurbo} calculation
using the \CT[18qed] PDF which exhibits a consistently lower cross-section than from the
\CT[18nnlo] PDFs. This difference is however small and increases with increasing \mtw and increasing
$\abseta$ to up to $3\%$.

\begin{figure}
\centering
\subfloat[\label{fig:1DlepasymPDFComp}]{\includegraphics[width=.48\textwidth]{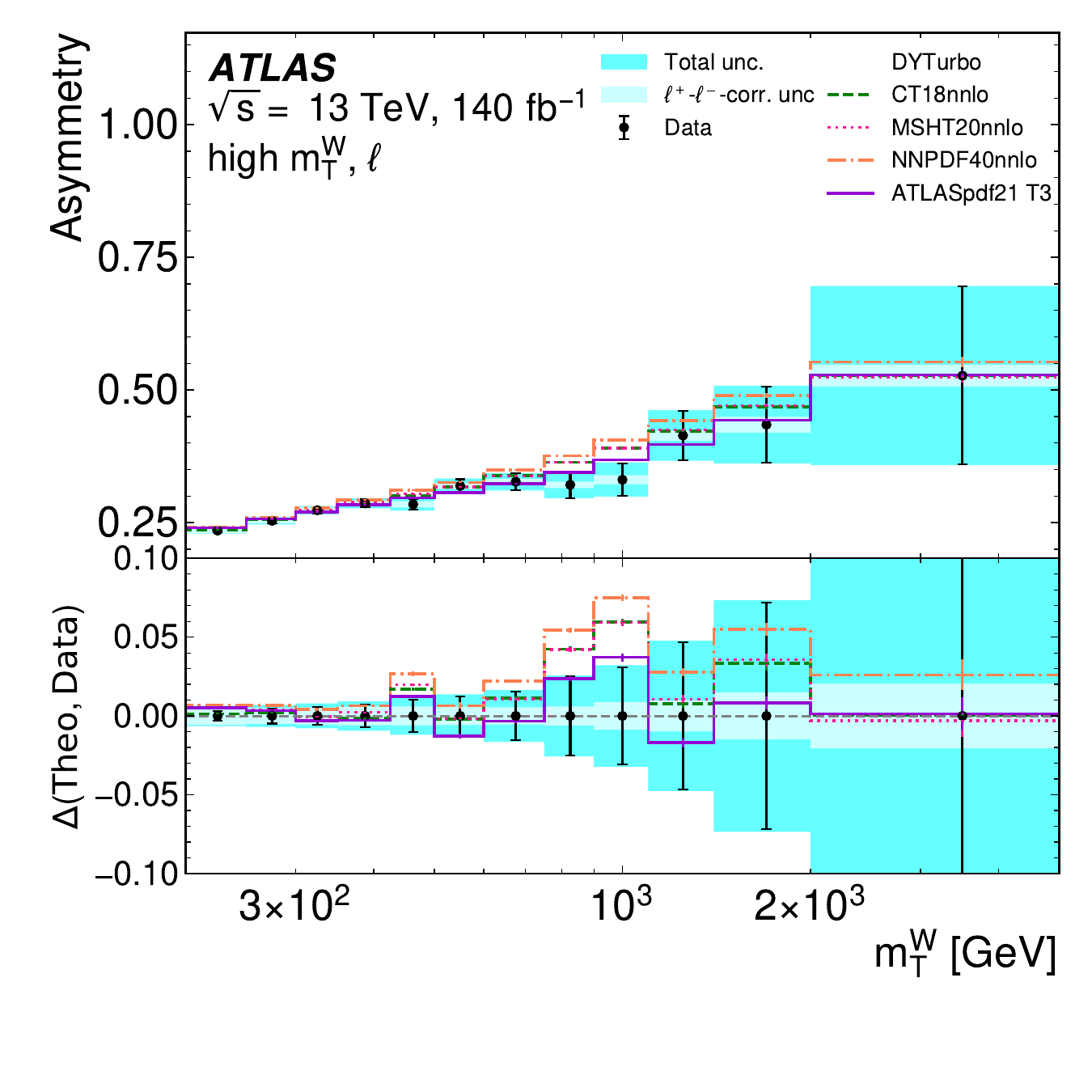}}
\subfloat[\label{fig:2DlepasymPDFComp}]{\includegraphics[width=.48\textwidth]{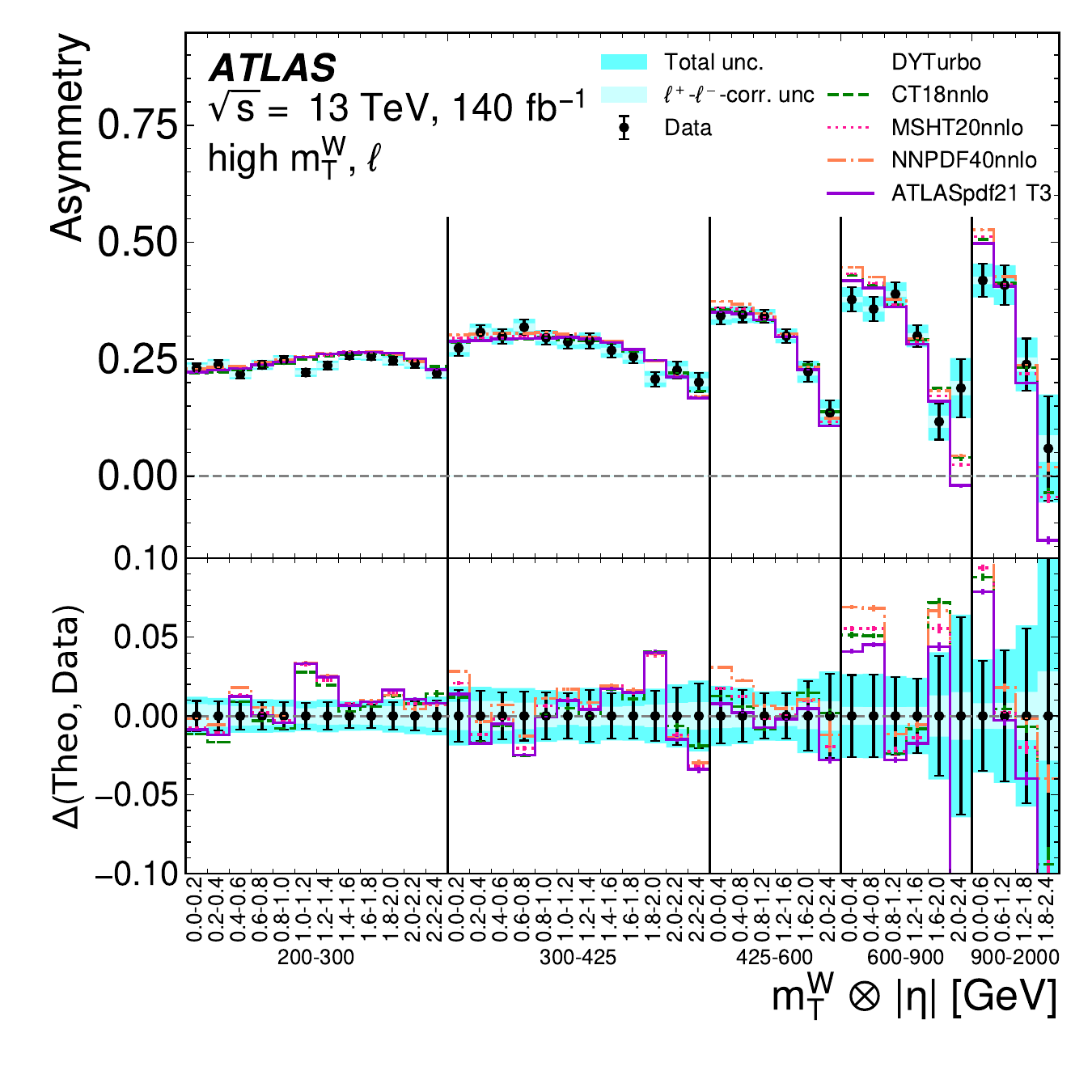}}
\caption{The Born-level lepton charge asymmetry of the
combined cross-sections are shown for the (a) single- and
(b) double-differential measurements. The error bars represent
the statistical and uncorrelated systematic uncertainties
between the $\ell^+$- and $\ell^-$- channels, while the inner
uncertainty band corresponds to the correlated
uncertainties. The total uncertainty in the asymmetry is
represented with the outer band. In addition, theoretical
predictions using \textsc{DYTurbo} for different PDFs are
displayed. The lower panels show the absolute difference
between each theory calculation and the measured asymmetry.}
\label{fig:lepasymPDFComp}
\end{figure}

The middle panels of Figure~\ref{fig:theoComp1D} and the left panels of Figures~\ref{fig:theoComp2Dplus} and~\ref{fig:theoComp2Dminus}
show \textsc{DYTurbo} predictions using different PDF sets
as well as the $90\%$ confidence level
PDF uncertainty bands from the \CT[18nnlo] PDF.
The prediction using \textsc{MSHT20nnlo} is typically rather similar to the one from the \CT[18nnlo] PDF set,
while the one using \NNPDF[4.0nnlo] is above the \CT[18nnlo] uncertainty band in the $\ell^+$ channel in the region $\mtw< 900$~\GeV.
The calculations with the \textsc{ATLASPDF21nnlo} are typically above the \CT[18nnlo] uncertainty band.
The \textsc{DYTurbo} predictions using different PDF sets are also compared to
the measured single- and double-differential charge asymmetry $A_{\ell}$ in Figure~\ref{fig:lepasymPDFComp}.
For the asymmetries the predictions agree well with the measurements and also for the separate $\ell^+$ and $\ell^-$ final
states the predictions are consistent with the measurements taking into account the sizeable  uncertainties due to the PDFs.
These uncertainties %
are found to be larger than the experimental
uncertainties for $\mtw \lesssim 1$~\TeV, indicating the potential of
these measurements to improve PDF determinations.

\subsection{Effective Field Theory constraints}

The measurements presented here may be sensitive to potential new
physics beyond the direct energy reach of the LHC.
EFTs are a useful tool for describing the physics below a
defined energy cutoff scale $\Lambda$.  The SMEFT is a generalised extension of the Standard Model,
offering a broad and largely model-independent approach to search for
new physics~\cite{Brivio:2017vri}.  This approach is adopted here to
interpret the cross-section measurements in terms of indirect
contributions from physics beyond the Standard Model.

The SMEFT Lagrangian includes all possible
operators constructed out of the SM field content:
\begin{equation*}
\mathcal{L}_\text{SMEFT} = \mathcal{L}_\text{SM} + \sum_{d>4} \mathcal{L}_{d} =  \mathcal{L}_\text{SM} + \sum_{d>4} \sum_{i} \frac{c_{i,d}}{{\Lambda}^{d-4}} \mathcal O_{i,d},
\end{equation*}
where $\mathcal{L}_\text{SM}$ is the SM Lagrangian, and $\mathcal{O}_{i,d}$ are SMEFT operators of dimension $d>4$.
Each operator is weighted by a dimensionless parameter, the Wilson coefficient $c_{i,d}$,
and is additionally suppressed by powers of the energy cutoff scale $\Lambda$.

It is conventional in SMEFT analyses to set $\Lambda$ to $1$~\TeV,
with higher dimensional operators having increasingly suppressed Wilson
coefficients.  For this reason, it is common to truncate the SMEFT to
dimension $d \leq 6$, as operators with $d>6$ will likely have reduced
impact on physical observables.  Additionally, all odd mass-dimension
operators in the SMEFT violate at least one of baryon or lepton number
conservation, which are believed to be strong symmetries.  All
odd-dimension operators can therefore also be neglected,
which justifies only considering  the dimension-$6$ operators of the SMEFT.

A complete set of dimension-6 operators of the SMEFT is given by
the Warsaw basis \cite{Grzadkowski:2010es}.  The Warsaw basis provides a
minimal basis with all redundant operators removed,  containing $2499$ dimension-$6$ operators, each associated with
a Wilson coefficient.  The Wilson coefficients are unknown free
parameters, whose values must be determined experimentally.  The number
of free parameters can be reduced by imposing symmetries amongst the
flavour structure of the fermion generations.  The most restrictive
flavour symmetry is known as the $U(3)^5$ symmetry, and reduces the
number of operators to $59$, considering only those that
conserve baryon and lepton number \cite{Brivio:2017vri}.

Physical observables receive corrections from the presence of
dimension-6 SMEFT operators. The amplitude-squared for a general
physics process is given by
\begin{equation*}
|\mathcal{A}^2| = |\mathcal{A}_{\text{SM}}|^2 + \sum_{i} \frac{c_i}{\Lambda^2}\mathcal{A}^\dag_{SM} \mathcal{A}_i + \sum_{i} \frac{|c_i|^2}{\Lambda^4}|\mathcal{A}_i|^2 + \sum_{i,j, i \neq j} \frac{c_i c_j}{\Lambda^4}\mathcal{A}^\dag_{i} \mathcal{A}_j,
\end{equation*}
where $\mathcal{A}_{\text{SM}}$ is the SM amplitude and
$\mathcal{A}_i$ is the contribution from dimension-6 operator $\mathcal{O}_i$ to
the amplitude $\mathcal{A}$.  The modifications to physical
observables can be separated into three terms:
\begin{itemize}
\item Linear interference between the SM and EFT amplitudes.
\item Pure quadratic EFT contribution.
\item Interference cross terms between EFT amplitudes.
\end{itemize}

The charged-current Drell--Yan process receives corrections from four
different dimension-6 operators, within the $U(3)^5$ symmetry scheme
of the Warsaw basis.  Those operators and their Wilson coefficients are
shown
in Table~\ref{tab:ccdyop}.

\begin{table}[b]
\caption{Dimension-6 SMEFT operators and their corresponding Wilson
coefficients affecting the charged current Drell--Yan, in the $U(3)^5$
symmetry of the Warsaw basis. Here, $H$ is the Higgs doublet, and
$q$, $l$ represent the left-handed quark and lepton doublets respectively.
The matrices $\tau^I$, $I={1,2,3}$ represent Pauli matrices, and
${D}^I_\mu$ is the usual Standard Model covariant derivative.
The subscript of each operator and Wilson coefficient denotes the fields
involved in the operator, and the superscript (1) or (3) denotes whether the operator
transforms as a singlet or triplet, respectively.}
\label{tab:ccdyop}
\centering
\begin{tabular}{llll}
\toprule{Wilson Coefficient} & {Operator}  \\
\midrule
$c^{(3)}_{lq}$ & $\mathcal{O}^{(3)}_{lq} = (\bar{l} \tau^I \gamma_\mu l) (\bar{q} \tau^I \gamma^\mu q)$\\
$c^{(1)}_{ll}$ & $\mathcal{O}^{(1)}_{ll} = (\bar{l} \gamma_\mu l)(\bar{l} \gamma^\mu l)$ \\
$c^{(3)}_{Hl}$ & $\mathcal{O}^{(3)}_{Hl} = (H^\dagger i \overleftrightarrow{D}^I_\mu H)(\bar{l} \tau^I \gamma^\mu l)$  \\
$c^{(3)}_{Hq}$ & $\mathcal{O}^{(3)}_{Hq} = (H^\dagger i \overleftrightarrow{D}^I_\mu H)(\bar{q} \tau^I \gamma^\mu q)$ \\
\bottomrule
\end{tabular}
\end{table}

The sensitivity of the charged-current Drell--Yan process to each
operator is studied by generating MC predictions within
the SMEFT and comparing against SM predictions.  Both SMEFT and SM
predictions at LO in QCD are produced using \MGNLO[2.9.9] \cite{Alwall:2011uj}
and the \textsc{SMEFTsim} 3.0 package \cite{Brivio:2020onw}. The ${m_W, m_Z, G_F}$ electroweak input parameter scheme is used.
Events are interfaced with  \Pythia[8.3] \cite{Bierlich:2022pfr} for parton
showering and hadronisation. The cutoff scale is set to $\Lambda = 1 $~\TeV.

The effects of each operator on the single-differential cross-section are
studied by setting the Wilson coefficient of each of the four
operators to unity and setting all
others to zero in turn.  It is observed that the quark-lepton contact operator
$\mathcal{O}^{(3)}_{lq}$ gives significant enhancements to the cross
section at high \mtw, whilst the other three operators
introduce a constant scaling in the cross-section with no mass
dependence. A similar analysis of the double-differential cross-sections
shows there is no additional SMEFT operator sensitivity associated
with the $\abseta$ dependence of the cross-sections.
Similarly, the charge-separated results do not yield additional sensitivity.
Therefore
the SMEFT analysis is performed on the single-differential charge-integrated cross
sections for the electron, muon and combined channels only.
The statistical model employed is a likelihood function with multivariate
Gaussian probability density functions representing the
measurement uncertainties. The likelihood is
written as
\begin{equation*}
L = \frac{1}{\sqrt{(2\pi)^{N_\text{bins}} |\Sigma|}} \exp \left\{ -\frac{1}{2} \left[ \vec{x} - \vec{\mu}(\vec{\theta}) \right]^\text{T} \Sigma^{-1} \left[ \vec{x} - \vec{\mu}(\vec{\theta}) \right] \right\} \times \prod_i \theta_i.
\end{equation*}
Here, $\Sigma$ is the covariance matrix,
which includes statistical uncertainties only and is therefore diagonal.
The vectors $\vec{x}$ and $\vec{\mu}$ represent the measured distribution and
predicted distributions, respectively.  Both the experimental and theoretical systematic
uncertainties are included as nuisance parameters $\vec{\theta}$, where a
shift of $\vec{\theta}_i = 1$ corresponds to a shift of the source $i$
by one standard deviation.

The EFT dependence of the cross-section is parametrised as a quadratic
function. In particular, the cross-section in bin $i$ is given by a
quadratic function of the Wilson coefficient $c$:
\begin{equation*}
\sigma_i = \sigma_{SM,i} +c \, \sigma_{\textrm{lin}, i} + c^2 \, \sigma_{\textrm{quad}, i},
\end{equation*}
where $\sigma_i$ is the SM prediction, and $\sigma_{\textrm{lin}, i}$ and $\sigma_{\textrm{quad}, i}$
are constants describing the dependence of the cross-section on $c$
and $c^2$ respectively.  These two coefficients, $\sigma_{\textrm{lin}, i}$ and
$\sigma_{\textrm{quad}, i}$, as well as $b_i$ must be determined from MC simulation in
order to infer information about the Wilson coefficients.

As the SMEFT and SM predictions are only computed at LO in QCD, the predictions are modified to account
for higher-order QCD and EW  corrections via
\begin{equation*} \label{renorm}
\sigma_i = \sigma^{\textrm{best}}_i  \left( 1 + c \frac{\sigma_{\textrm{lin}, i}}{ \sigma^{\textrm{LO}}_i } + c^2 \frac{\sigma_{\textrm{quad}, i}}{ \sigma^{\textrm{LO}}_i } \right),
\end{equation*}
where $\sigma^{\textrm{best}}_i$ represents a higher-order SM prediction,
and $\sigma^{\textrm{LO}}_i$ represents the LO SM prediction.
It is assumed that the relative effect of higher-order predictions does not
change in the presence of dimension-six operators.  The higher-order
SM prediction is generated using \textsc{DYTurbo}, as introduced Section~\ref{sec:Comp_qcd}. The
prediction uses the \CT[18nnlo] PDF set,
with $\alpha_s = 0.118$ and additive EW corrections.

Theoretical uncertainties arising from variations of
$\alpha_s$, PDFs, EW corrections, and the renormalisation ($\mu_R$) and
factorisation scale ($\mu_F$) variations are implemented
as follows:
the value of $\alpha_s(M_z)$ at the
$Z$ pole is varied by $\pm 0.001$;
$\mu_R$ and $\mu_F$ are
varied independently by factors of $2$ avoiding extreme variations;
the EW corrections are included as multiplicative factors
on the predictions, rather than additive; and the PDF eigenvector variations for the
\CT[18nnlo] set are used at the 90\% CL.

\begin{figure}
\centering
\subfloat[]{\includegraphics[width=.48\linewidth, trim={11mm 6mm 30mm 3mm},clip]{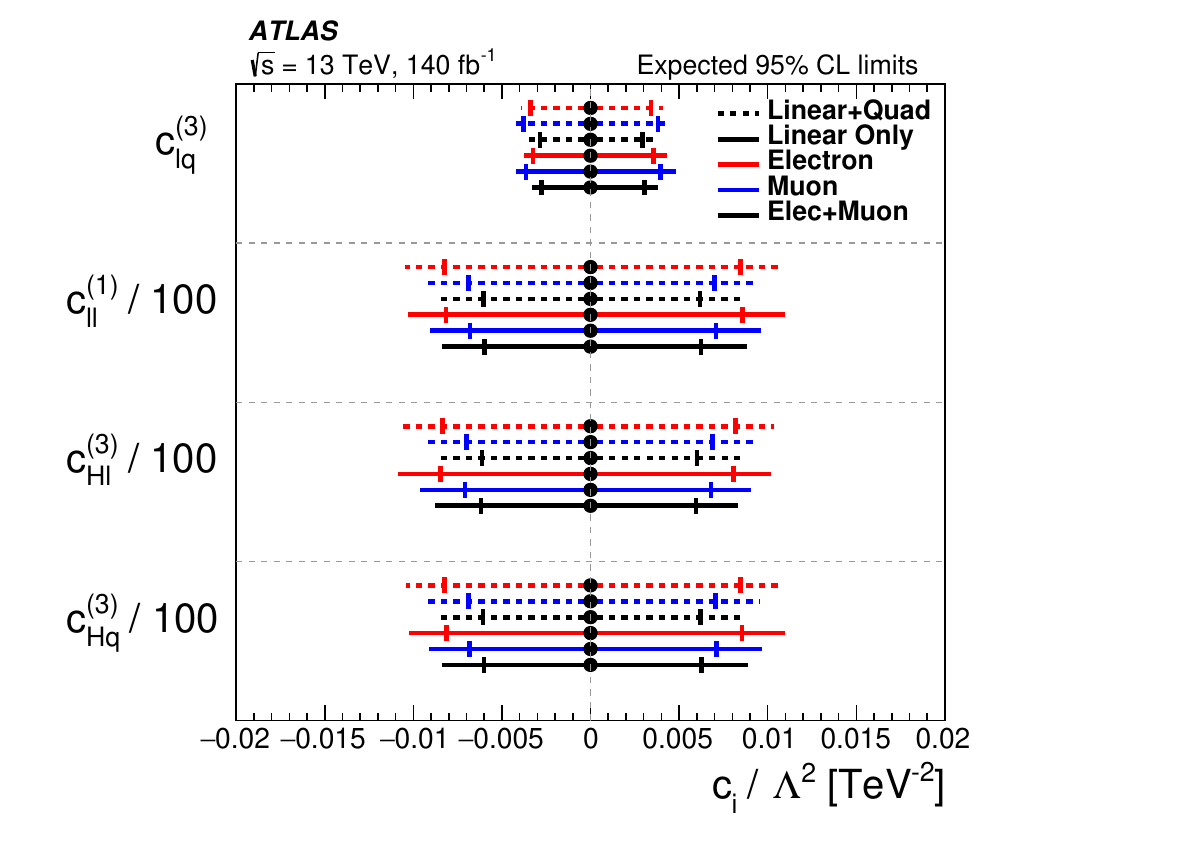}}
\subfloat[]{\includegraphics[width=.48\textwidth, trim={11mm 6mm 30mm 3mm},clip]{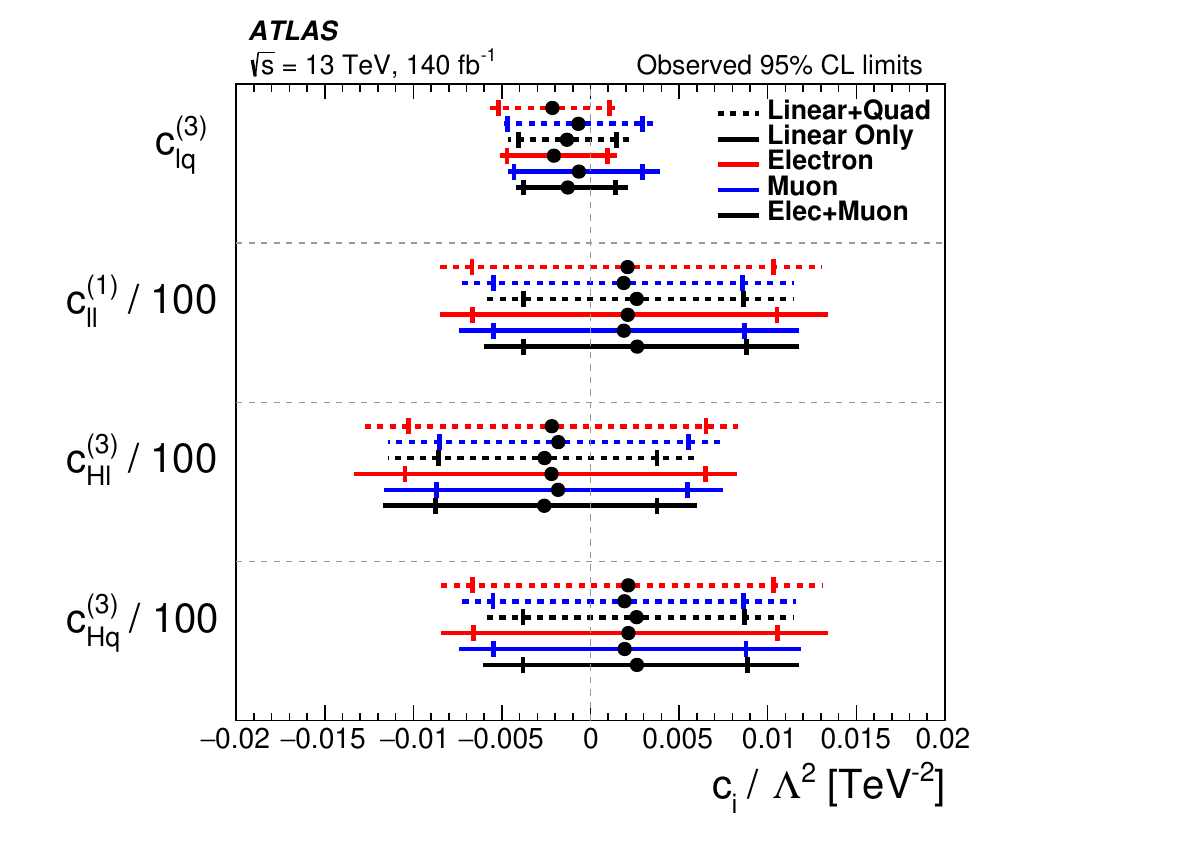}}
\caption{The (a) expected and (b) observed limits on Wilson coefficients at 95\%
CL. The results are shown for linear-only and linear+quadratic
fits, for both the electron and muon channels, as well as
their combination.
The inner (outer) error bars indicate the limits when using the 68\% (90\%) CL for the PDF uncertainty.
}
\label{fig:eft_constraints}
\end{figure}

Expected limits on the Wilson coefficients are derived by
using the \textsc{DYTurbo} prediction as the measurement pseudo-data
instead of using the actual measurement as for the observed limits.
The fits are performed first by using only the linear SM and EFT interference
terms, then by including the quadratic pure EFT contributions as well.
They are obtained by allowing only one coefficient to be non-zero at a time.

The expected and observed limits on the Wilson coefficients are given in Figure
\ref{fig:eft_constraints} and are also shown in Table \ref{tab:EFT_limits}.
All observed limits are compatible with
the $c_i=0$ assumption at 95\% CL, indicating no
significant deviations from the SM are observed.
The strongest limits are derived by using the data from the combined channel
due to the reduced statistical and systematic uncertainties after the electron-muon combination.
Differences between the constraints obtained using linear+quadratic and linear-only terms are negligible,
indicating the constraints are driven mostly by the linear-interference terms.
This suggests that the measurement is largely insensitive to the $1/\Lambda^4$ contributions
arising from interference between SM and dimension-8 operators,
implying the constraints are robust despite truncating the SMEFT to dimension-6.

The  limits on the EFT Wilson coefficients improve by a factor 1.4 to 2.6 when the fits are performed using only experimental uncertainties. This indicates the importance of increasing the precision of the theoretical predictions, in particular for the PDFs, in future measurements.
Figure \ref{fig:eft_constraints} shows that using the 68\% CL PDF uncertainty instead of the 90\% CL uncertainty improves the limits on the Wilson
coefficients by a factor of 1.1-1.4.
The limit on the Wilson coefficient $c^{(3)}_{lq}$ exceeds previous four-fermion limits using ATLAS and CMS data~\cite{Greljo:2017vvb,Boughezal:2023nhe} as well as limits from low-energy data~\cite{Falkowski:2017pss} and global analyses~\cite{Allwicher:2023shc,Bartocci:2023nvp}.

%
%
%


\begin{table}[t]
\fontsize{8.5}{8}\selectfont\setlength{\tabcolsep}{3pt}
\centering
\caption{The expected and observed EFT limits at 95\% CL, shown for the linear-only and linear+quadratic electron, muon, and combined fits.}
\begin{tabular}{lcccccc}
\toprule
& \multicolumn{3}{c}{{Expected (Linear + Quadratic) $[\text{TeV}^{-2}]$}} & \multicolumn{3}{c}{{Observed (Linear + Quadratic) $[\text{TeV}^{-2}]$}} \\
\midrule
& {Electron} & {Muon} & {Combined} & {Electron} & {Muon} & {Combined} \\
\midrule\midrule
\( c^{(3)}_{lq} / \Lambda^2 \) & \([-0.0039, 0.0041]\) & \([-0.0044, 0.0046]\) & \([-0.0035, 0.0037]\) & \([-0.0058, 0.0016]\) & \([-0.0051, 0.0038]\) & \([-0.0047, 0.0021]\) \\
\( c^{(1)}_{ll} / \Lambda^2 \) & \([-1.05, 1.08]\) & \([-0.92, 0.95]\) & \([-0.85, 0.87]\) & \([-0.86, 1.31]\) & \([-0.75, 1.15]\) & \([-0.60, 1.15]\) \\
\( c^{(3)}_{Hl} / \Lambda^2 \) & \([-1.07, 1.03]\) & \([-0.95, 0.92]\) & \([-0.87, 0.84]\) & \([-1.30, 0.83]\) & \([-1.14, 0.75]\) & \([-1.14, 0.60]\) \\
\( c^{(3)}_{Hq} / \Lambda^2 \) & \([-1.04, 1.08]\) & \([-0.93, 0.95]\) & \([-0.85, 0.88]\) & \([-0.85, 1.31]\) & \([-0.75, 1.16]\) & \([-0.61, 1.15]\) \\
\bottomrule
\end{tabular}\\
\vspace{0.5cm}
\begin{tabular}{lcccccc}
\toprule
& \multicolumn{3}{c}{{Expected (Linear) $[\text{TeV}^{-2}]$}} & \multicolumn{3}{c}{{Observed (Linear) $[\text{TeV}^{-2}]$}} \\
\midrule
& {Electron} & {Muon} & {Combined} & {Electron} & {Muon} & {Combined} \\
\midrule\midrule
\( c^{(3)}_{lq} / \Lambda^2 \) & \([-0.0038, 0.0043]\) & \([-0.0042, 0.0048]\) & \([-0.0033, 0.0038]\) & \([-0.0051, 0.0015]\) & \([-0.0047, 0.0039]\) & \([-0.0042, 0.0021]\) \\
\( c^{(1)}_{ll} / \Lambda^2 \) & \([-1.03, 1.10]\) & \([-0.91, 0.96]\) & \([-0.84, 0.89]\) & \([-0.85, 1.34]\) & \([-0.74, 1.18]\) & \([-0.60, 1.18]\) \\
\( c^{(3)}_{Hl} / \Lambda^2 \) & \([-1.08, 1.02]\) & \([-0.96, 0.91]\) & \([-0.88, 0.83]\) & \([-1.34, 0.83]\) & \([-1.17, 0.75]\) & \([-1.17, 0.60]\) \\
\( c^{(3)}_{Hq} / \Lambda^2 \) & \([-1.03, 1.10]\) & \([-0.91, 0.97]\) & \([-0.84, 0.89]\) & \([-0.84, 1.34]\) & \([-0.74, 1.19]\) & \([-0.61, 1.18]\) \\
\bottomrule
\end{tabular}
\label{tab:EFT_limits}
\end{table}



%
%
%
%

%
%
\FloatBarrier

\section{Conclusions}
%

\label{sec:conclusion}

This paper present a first measurement of the $W^{\pm}$
production cross-section above the resonant production region.
The single-differential cross-sections,
$\textrm{d}\sigma/\textrm{d}\mtw$, are measured in the region $200
\leq \mtw \leq 5000$~\GeV. The measurements are also presented as
double-differential cross-sections, $\textrm{d}^2\sigma/\textrm{d}\mtw\textrm{d}|\eta|$,
in the region $200 \leq \mtw \leq 2000$~\GeV and $0 \leq |\eta| \leq
2.4$. These
fiducial measurements use 140 fb$^{-1}$ of LHC $pp$ collision data
collected by ATLAS at a centre-of-mass energy of $13$~\TeV, and are performed separately for both charges
and for the electron and muon final states. They are unfolded for detector effects
to the Born level. Corrections to the dressed particle level are also
provided.

Combinations of the charge-separated cross-sections for the electron and muon final states
are performed to improve the statistical and systematic uncertainties
in the measurements.
These combined data compare well to state-of-the-art
theoretical predictions at NNLO in QCD including NLO EW effects.
The uncertainties in the predictions arising from the PDFs are found to be typically larger
than the experimental uncertainties, indicating the potential of these measurements to improve PDF determinations.
The single- and double-differential lepton charge
asymmetry, $A_{\ell}$, is presented and shown to be well described by
the predictions.

A test of lepton flavour universality in the range $200~\GeV<\mtw<5000$~\GeV is performed by determining the
ratio of the charge-combined cross-sections for the electron
to the muon final state. No deviations from the Standard
Model are observed.

Finally, the single-differential cross-sections are used to search for
signals of new physics in an effective field theory approach within the
SMEFT $U(3)^5$ symmetry model. The data constrain the Wilson
coefficients of four operators in the Warsaw basis. In
particular the measurements provide the world-leading constraints on
the Wilson coefficient $c^{(3)}_{lq}$.


%
%
%
%
%
%
%
%
%
%
%
%
%
%

%
%
%
%
%
\section*{Acknowledgements}
%

%

%
%

%
%

We thank CERN for the very successful operation of the LHC and its injectors, as well as the support staff at
CERN and at our institutions worldwide without whom ATLAS could not be operated efficiently.

The crucial computing support from all WLCG partners is acknowledged gratefully, in particular from CERN, the ATLAS Tier-1 facilities at TRIUMF/SFU (Canada), NDGF (Denmark, Norway, Sweden), CC-IN2P3 (France), KIT/GridKA (Germany), INFN-CNAF (Italy), NL-T1 (Netherlands), PIC (Spain), RAL (UK) and BNL (USA), the Tier-2 facilities worldwide and large non-WLCG resource providers. Major contributors of computing resources are listed in Ref.~\cite{ATL-SOFT-PUB-2025-001}.

We gratefully acknowledge the support of ANPCyT, Argentina; YerPhI, Armenia; ARC, Australia; BMWFW and FWF, Austria; ANAS, Azerbaijan; CNPq and FAPESP, Brazil; NSERC, NRC and CFI, Canada; CERN; ANID, Chile; CAS, MOST and NSFC, China; Minciencias, Colombia; MEYS CR, Czech Republic; DNRF and DNSRC, Denmark; IN2P3-CNRS and CEA-DRF/IRFU, France; SRNSFG, Georgia; BMBF, HGF and MPG, Germany; GSRI, Greece; RGC and Hong Kong SAR, China; ICHEP and Academy of Sciences and Humanities, Israel; INFN, Italy; MEXT and JSPS, Japan; CNRST, Morocco; NWO, Netherlands; RCN, Norway; MNiSW, Poland; FCT, Portugal; MNE/IFA, Romania; MSTDI, Serbia; MSSR, Slovakia; ARIS and MVZI, Slovenia; DSI/NRF, South Africa; MICIU/AEI, Spain; SRC and Wallenberg Foundation, Sweden; SERI, SNSF and Cantons of Bern and Geneva, Switzerland; NSTC, Taipei; TENMAK, T\"urkiye; STFC/UKRI, United Kingdom; DOE and NSF, United States of America.

Individual groups and members have received support from BCKDF, CANARIE, CRC and DRAC, Canada; CERN-CZ, FORTE and PRIMUS, Czech Republic; COST, ERC, ERDF, Horizon 2020, ICSC-NextGenerationEU and Marie Sk{\l}odowska-Curie Actions, European Union; Investissements d'Avenir Labex, Investissements d'Avenir Idex and ANR, France; DFG and AvH Foundation, Germany; Herakleitos, Thales and Aristeia programmes co-financed by EU-ESF and the Greek NSRF, Greece; BSF-NSF and MINERVA, Israel; NCN and NAWA, Poland; La Caixa Banking Foundation, CERCA Programme Generalitat de Catalunya and PROMETEO and GenT Programmes Generalitat Valenciana, Spain; G\"{o}ran Gustafssons Stiftelse, Sweden; The Royal Society and Leverhulme Trust, United Kingdom.

In addition, individual members wish to acknowledge support from Armenia: Yerevan Physics Institute (FAPERJ); CERN: European Organization for Nuclear Research (CERN DOCT); Chile: Agencia Nacional de Investigaci\'on y Desarrollo (FONDECYT 1230812, FONDECYT 1230987, FONDECYT 1240864); China: Chinese Ministry of Science and Technology (MOST-2023YFA1605700, MOST-2023YFA1609300), National Natural Science Foundation of China (NSFC - 12175119, NSFC 12275265); Czech Republic: Czech Science Foundation (GACR - 24-11373S), Ministry of Education Youth and Sports (ERC-CZ-LL2327, FORTE CZ.02.01.01/00/22\_008/0004632), PRIMUS Research Programme (PRIMUS/21/SCI/017); EU: H2020 European Research Council (ERC - 101002463); European Union: European Research Council (ERC - 948254, ERC 101089007, ERC, BARD, 101116429), European Regional Development Fund (SMASH COFUND 101081355, SLO ERDF), Horizon 2020 Framework Programme (MUCCA - CHIST-ERA-19-XAI-00), European Union, Future Artificial Intelligence Research (FAIR-NextGenerationEU PE00000013), Horizon 2020 (EuroHPC - EHPC-DEV-2024D11-051), Italian Center for High Performance Computing, Big Data and Quantum Computing (ICSC, NextGenerationEU); France: Agence Nationale de la Recherche (ANR-21-CE31-0013, ANR-21-CE31-0022, ANR-22-EDIR-0002); Germany: Baden-Württemberg Stiftung (BW Stiftung-Postdoc Eliteprogramme), Deutsche Forschungsgemeinschaft (DFG - 469666862, DFG - CR 312/5-2); China: Research Grants Council (GRF); Italy: Istituto Nazionale di Fisica Nucleare (ICSC, NextGenerationEU), Ministero dell'Università e della Ricerca (NextGenEU I53D23000820006 M4C2.1.1); Japan: Japan Society for the Promotion of Science (JSPS KAKENHI JP22H01227, JSPS KAKENHI JP22H04944, JSPS KAKENHI JP22KK0227, JSPS KAKENHI JP23KK0245); Norway: Research Council of Norway (RCN-314472); Poland: Ministry of Science and Higher Education (IDUB AGH, POB8, D4 no 9722), Polish National Science Centre (NCN 2021/42/E/ST2/00350, NCN OPUS 2023/51/B/ST2/02507, NCN OPUS nr 2022/47/B/ST2/03059, NCN UMO-2019/34/E/ST2/00393, UMO-2020/37/B/ST2/01043, UMO-2022/47/O/ST2/00148, UMO-2023/49/B/ST2/04085, UMO-2023/51/B/ST2/00920, UMO-2024/53/N/ST2/00869); Portugal: Foundation for Science and Technology (FCT); Spain: Generalitat Valenciana (Artemisa, FEDER, IDIFEDER/2018/048), Ministry of Science and Innovation (MCIN \& NextGenEU PCI2022-135018-2, MICIN \& FEDER PID2021-125273NB, RYC2019-028510-I, RYC2020-030254-I, RYC2021-031273-I, RYC2022-038164-I); Sweden: Carl Trygger Foundation (Carl Trygger Foundation CTS 22:2312), Swedish Research Council (Swedish Research Council 2023-04654, VR 2021-03651, VR 2022-03845, VR 2022-04683, VR 2023-03403, VR 2024-05451), Knut and Alice Wallenberg Foundation (KAW 2018.0458, KAW 2022.0358, KAW 2023.0366); Switzerland: Swiss National Science Foundation (SNSF - PCEFP2\_194658); United Kingdom: Leverhulme Trust (Leverhulme Trust RPG-2020-004), Royal Society (NIF-R1-231091); United States of America: U.S. Department of Energy (ECA DE-AC02-76SF00515), Neubauer Family Foundation.

%
%


%
%
%
%
%
%
%
%

%
%

%
%
%
\printbibliography
\clearpage
 
\begin{flushleft}
\hypersetup{urlcolor=black}
{\Large The ATLAS Collaboration}

\bigskip

\AtlasOrcid[0000-0002-6665-4934]{G.~Aad}$^\textrm{\scriptsize 104}$,
\AtlasOrcid[0000-0001-7616-1554]{E.~Aakvaag}$^\textrm{\scriptsize 17}$,
\AtlasOrcid[0000-0002-5888-2734]{B.~Abbott}$^\textrm{\scriptsize 123}$,
\AtlasOrcid[0000-0002-0287-5869]{S.~Abdelhameed}$^\textrm{\scriptsize 119a}$,
\AtlasOrcid[0000-0002-1002-1652]{K.~Abeling}$^\textrm{\scriptsize 55}$,
\AtlasOrcid[0000-0001-5763-2760]{N.J.~Abicht}$^\textrm{\scriptsize 49}$,
\AtlasOrcid[0000-0002-8496-9294]{S.H.~Abidi}$^\textrm{\scriptsize 30}$,
\AtlasOrcid[0009-0003-6578-220X]{M.~Aboelela}$^\textrm{\scriptsize 45}$,
\AtlasOrcid[0000-0002-9987-2292]{A.~Aboulhorma}$^\textrm{\scriptsize 36e}$,
\AtlasOrcid[0000-0001-5329-6640]{H.~Abramowicz}$^\textrm{\scriptsize 157}$
% ...
\bigskip
\\

% ...
$^{17}$Department for Physics and Technology, University of Bergen, Bergen; Norway.\\
% ...
$^{30}$Physics Department, Brookhaven National Laboratory, Upton NY; United States of America.\\
% ...
$^{36}$$^{(a)}$Facult\'e des Sciences Ain Chock, Universit\'e Hassan II de Casablanca;$^{(b)}$Facult\'{e} des Sciences, Universit\'{e} Ibn-Tofail, K\'{e}nitra;$^{(c)}$Facult\'e des Sciences Semlalia, Universit\'e Cadi Ayyad, LPHEA-Marrakech;$^{(d)}$LPMR, Facult\'e des Sciences, Universit\'e Mohamed Premier, Oujda;$^{(e)}$Facult\'e des sciences, Universit\'e Mohammed V, Rabat;$^{(f)}$Institute of Applied Physics, Mohammed VI Polytechnic University, Ben Guerir; Morocco.\\
% ...
$^{45}$Physics Department, Southern Methodist University, Dallas TX; United States of America.\\
% ...
$^{49}$Fakult\"{a}t Physik , Technische Universit{\"a}t Dortmund, Dortmund; Germany.\\
% ...
$^{55}$II. Physikalisches Institut, Georg-August-Universit\"{a}t G\"ottingen, G\"ottingen; Germany.\\
% ...
$^{104}$CPPM, Aix-Marseille Universit\'e, CNRS/IN2P3, Marseille; France.\\
% ...
$^{119}$$^{(a)}$New York University Abu Dhabi, Abu Dhabi;$^{(b)}$United Arab Emirates University, Al Ain; United Arab Emirates.\\
% ...
$^{123}$Homer L. Dodge Department of Physics and Astronomy, University of Oklahoma, Norman OK; United States of America.\\
% ...
$^{157}$Raymond and Beverly Sackler School of Physics and Astronomy, Tel Aviv University, Tel Aviv; Israel.\\
% ...
$^{a}$ Also at Affiliated with an institute formerly covered by a cooperation agreement with CERN.\\
$^{b}$ Also at An-Najah National University, Nablus; Palestine.\\
$^{c}$ Also at Borough of Manhattan Community College, City University of New York, New York NY; United States of America.\\
$^{d}$ Also at Center for Interdisciplinary Research and Innovation (CIRI-AUTH), Thessaloniki; Greece.\\
$^{e}$ Also at Centre of Physics of the Universities of Minho and Porto (CF-UM-UP); Portugal.\\
$^{f}$ Also at CERN, Geneva; Switzerland.\\
% ...

\end{flushleft}


%

%
%

%

%
%
%
%

\end{document}